\documentclass[a4paper,11pt] {article}
\usepackage[utf8]{inputenc}
\usepackage[left=2.5cm,right=2.5cm,top=1cm,bottom=4cm]{geometry}
\usepackage{color}
\usepackage{colortbl}
\usepackage{fancyhdr}
\usepackage{graphicx}
\usepackage{eurosym}
\usepackage{enumitem}
\usepackage{lastpage}
\usepackage{booktabs}
\usepackage[yyyymmdd]{datetime}
\usepackage{multirow}
\usepackage{pifont}
\usepackage[colorlinks=true, urlcolor=blue, linkcolor=black, citecolor=red]{hyperref}
\usepackage{amssymb} 
\usepackage[color]{changebar}
\usepackage[dvipsnames]{xcolor}
\usepackage{listings}
\usepackage{threeparttable}
\usepackage[acronym, toc]{glossaries}

\makeglossaries

\newacronym{1pa}{1PA}{one-post-adiabatic}

\newacronym{a/d}{A/D}{Analog-To-Digital Converter}
\newacronym{aam}{AAM}{Active Aperture Mechanism}
\newacronym{ab}{AB}{Actuator Block}
\newacronym{abcl}{ABCL}{As-Built Configuration List}
\newacronym{ac}{AC}{Alternating Current}
\newacronym{acm}{ACM}{Agency Controlled Milestones}
\newacronym{acs}{ACS}{Attitude Control System}
\newacronym{ad}{AD}{Applicable Document}
\newacronym{adc}{ADC}{Analog-To-Digital Converter}
\newacronym{adpll}{ADPLL}{all-digital phase-locked loop}
\newacronym{ads}{ADS}{Airbus Deutschland (Friedrichshafen)}
\newacronym{advoreach}{ADVOREACH}{Advocacy \&  Outreach Committee}
\newacronym{ae}{AE}{Arianespace}
\newacronym{aei}{AEI}{Albert Einstein Institute}
\newacronym{aetd}{AETD}{Applied Engineering and Technology Directorate}
\newacronym{agis}{AGIS}{Atomic Gravitational-wave Interferometer Sensor}
\newacronym{agn}{AGN}{Active Galactic Nuclei}
\newacronym{ai}{AI}{Atom Interferometry}
\newacronym{ail}{AIL}{Action Item List}
\newacronym{ait}{AIT}{"Assembly}
\newacronym{aiv}{AIV}{"Assembly}
\newacronym{aivt}{AIVT}{"Assembly}
\newacronym{ak}{AK}{Analytic Kludge}
\newacronym{ake}{AKE}{attitude absolute knowledge}
\newacronym{alma}{ALMA}{Atacama Large Millimeter/submillimeter Array}
\newacronym{am cvn}{AM CVN}{AM Canum Venaticorum star (cataclysmic variable)}
\newacronym{amr}{AMR}{Anisotropic Magnetoresistors}
\newacronym{amu}{AMU}{Accelerometer Measurement Unit}
\newacronym{ao}{AO}{Announcement Of Opportunity}
\newacronym{aocs}{AOCS}{Attitude \&  Orbit Control System}
\newacronym{aom}{AOM}{acousto-optic modulator}
\newacronym{ap}{AP}{(NASA's) Astrophysics Division}
\newacronym{apd}{APD}{(NASA's) Astrophysics Division}
\newacronym{ape}{APE}{Absolute Pointing Error}
\newacronym{api}{API}{Application Programming Interface}
\newacronym{apid}{APID}{Application Process Identification}
\newacronym{apm}{APM}{Antenna Pointing Mechanism}
\newacronym{asd}{ASD}{Amplitude Spectral Density}
\newacronym{asi}{ASI}{Agenzia Spaziale Italiana}
\newacronym{asm}{ASM}{Analogue Signal Monitor}
\newacronym{ast}{AST}{Autonomous Star Tracker}
\newacronym{astrowg}{ASTROWG}{Astrophysics Working Group}
\newacronym{asw}{ASW}{Application Software}
\newacronym{ata}{ATA}{Allen Telescope Array}
\newacronym{atc}{ATC}{UK Astronomy Technology Centre}
\newacronym{athena}{ATHENA}{Advanced Telescope for High-ENergy Astrophysics}
\newacronym{atlo}{ATLO}{"Assembly}
\newacronym{atm}{ATM}{Acoustic Telemetry Modem}
\newacronym{atm_}{ATM}{Augmented Telemetry}
\newacronym{atp}{ATP}{Authorisation To Proceed}
\newacronym{au}{AU}{Astronomical Unit}
\newacronym{au_}{AU}{gold}
\newacronym{avm}{AVM}{Avionic Model}
\newacronym{awg}{AWG}{American Wire Gauge}
\newacronym{awg_}{AWG}{Astronomy Working Group}

\newacronym{b}{B}{Black Hole Binary}
\newacronym{bam}{BAM}{Beam Alignment Mechanism}
\newacronym{bao}{BAO}{Baryonic Acoustic Oscillation}
\newacronym{bb}{BB}{Breadboard}
\newacronym{bbdr}{BBDR}{Breadboard Design Review}
\newacronym{bbn}{BBN}{Big Bang Nucleosynthesis}
\newacronym{bcr}{BCR}{Battery Charge Regulator}
\newacronym{bcrs}{BCRS}{Barycentric Celestial Reference System}
\newacronym{bdm}{BDM}{Bi-Level Discrete Monitor}
\newacronym{bee}{BEE}{Back End Electronics}
\newacronym{belspo}{BELSPO}{Belgian Science Policy}
\newacronym{ber}{BER}{Bit Error Rate}
\newacronym{bh}{BH}{Black Hole}
\newacronym{bhb}{BHB}{Black Hole Binary}
\newacronym{bipr}{BIPR}{Background Intellectual Properties Rights}
\newacronym{bm}{BM}{Balance Mass}
\newacronym{bme}{BME}{Barycentric Mean Ecliptic}
\newacronym{bob}{BOB}{Break Out Box}
\newacronym{bol}{BOL}{Beginning-Of-Life}
\newacronym{bsm}{BSM}{Physics beyond the Standard Model}
\newacronym{bsw}{BSW}{Basic Software}

\newacronym{cab}{CAB}{Change Appeal Board}
\newacronym{cac}{CAC}{Cost at Completion}
\newacronym{cad}{CAD}{Computer Aided Design}
\newacronym{cadm}{CADM}{Configuration And Data Management}
\newacronym{cam}{CAM}{Constellation Acquisition Mode}
\newacronym{car}{CAR}{Contamination Analysis Report}
\newacronym{cas}{CAS}{Constellation Acquisition Sensor}
\newacronym{cas-eu}{CAS-EU}{CAS Electronic Unit}
\newacronym{cas-oh}{CAS-OH}{CAS Optical Head}
\newacronym{catwp}{CATWP}{Catalogues Work Package}
\newacronym{cbe}{CBE}{Current Best Estimate}
\newacronym{cbod}{CBOD}{Clamp Band Opening Device}
\newacronym{ccc}{CCC}{Consortium Constituent Council}
\newacronym{cccp}{CCCP}{"Cleanliness}
\newacronym{ccd}{CCD}{Charge-Coupled Device}
\newacronym{ccn}{CCN}{Contract Change Notice}
\newacronym{ccpm}{CCPM}{Consortium Constellation Performance Model}
\newacronym{ccs}{CCS}{Central Checkout System}
\newacronym{ccsds}{CCSDS}{Consultative Committee for Space Data Systems}
\newacronym{ccu}{CCU}{Carging Control Unit}
\newacronym{ccu_}{CCU}{Charging Control Unit}
\newacronym{ccw}{CCW}{Counterclockwise}
\newacronym{cdf}{CDF}{Concurrent Design Facility}
\newacronym{cdm}{CDM}{Cold Dark Matter}
\newacronym{cdms}{CDMS}{Central Data Management System}
\newacronym{cdmu}{CDMU}{Control and Data Management Unit}
\newacronym{cdr}{CDR}{Critical Design Review}
\newacronym{ce}{CE}{Conducted Emission}
\newacronym{ce_}{CE}{Cosmic Explorer}
\newacronym{ce__}{CE}{European safety standard}
\newacronym{cel}{CEL}{Critical Event Log}
\newacronym{cf}{CF}{Communication Frame}
\newacronym{cfi}{CFI}{Customer-Furnished Items}
\newacronym{cfrp}{CFRP}{Carbon Fibre Reinforced Plastic}
\newacronym{cg}{CG}{Cold Gas}
\newacronym{cgmps}{CGMPS}{Cold Gas Micro Propulsion System}
\newacronym{cidl}{CIDL}{Configuration Item Data List}
\newacronym{cil}{CIL}{Critical Items List}
\newacronym{cip}{CIP}{Common Interface Pedestal}
\newacronym{cip_}{CIP}{Continuous Improvement Program}
\newacronym{cl}{CL}{Confidence Level}
\newacronym{cla}{CLA}{Coupled Loads Analysis}
\newacronym{clht}{CLHT}{Can LISA Hear This? (Detectability Calculator)}
\newacronym{cm}{CM}{Caging Mechanism}
\newacronym{cm_}{CM}{Charging Mechanism}
\newacronym{cm__}{CM}{Common Mode}
\newacronym{cm___}{CM}{Configuration Management}
\newacronym{cmb}{CMB}{Cosmic Microwave Background}
\newacronym{cmd}{CMD}{Charge Management Device}
\newacronym{cmm}{CMM}{Coordinate Measuring Machine}
\newacronym{cmnt}{CMNT}{Colloid Micro-Newton Thruster}
\newacronym{cmrr}{CMRR}{Common Mode Rejection Ratio}
\newacronym{cms}{CMS}{Charge Management System}
\newacronym{cmu}{CMU}{Charge Management Unit}
\newacronym{cnes}{CNES}{Centre National d'Études Spatiales}
\newacronym{cobe}{COBE}{Cosmic Background Explorer}
\newacronym{coc}{COC}{Certificate of Conformance}
\newacronym{cog}{COG}{Center of Gravity}
\newacronym{com}{COM}{Centre Of Mass}
\newacronym{combo}{COMBO}{Classifying Objects By Medium-Band Observations}
\newacronym{coms}{COMS}{Communications System (telemetry)}
\newacronym{cosmos}{COSMOS}{Cosmic Evolution Survey}
\newacronym{coswg}{COSWG}{Cosmology Working Group}
\newacronym{cote}{COTE}{Check-Out Test Equipment}
\newacronym{cots}{COTS}{Commercial off the Shelf}
\newacronym{cpc}{CPC}{Composite Processing Component}
\newacronym{cpis}{CPIS}{Consortium Provided Items}
\newacronym{cppa}{CPPA}{Coordinated Parts Procurement Agency}
\newacronym{cpta}{CPTA}{Chinese Pulsar Timing Array}
\newacronym{cpu}{CPU}{Central Processing Unit}
\newacronym{cqp}{CQP}{Calibrated Quadrant Photodiode Pair}
\newacronym{cr}{CR}{Change Request}
\newacronym{crb}{CRB}{Change Review Board}
\newacronym{crema}{CREMA}{Consolidated Report on Mission Analysis}
\newacronym{crs}{CRS}{Contamination requirements specification}
\newacronym{cs}{CS}{Conducted Susceptibility}
\newacronym{cs_}{CS}{Cosmic String}
\newacronym{csar}{CSAR}{Configuration Status Accounting Report}
\newacronym{csd}{CSD}{Cross Spectral Density}
\newacronym{cst}{CST}{Community Science Team}
\newacronym{cte}{CTE}{Coefficient Of Thermal Expansion}
\newacronym{ctp}{CTP}{Core Technology Program}
\newacronym{cu}{CU}{Coordination Unit}
\newacronym{cvcm}{CVCM}{Collected Volatile Condensable Material}
\newacronym{cvm}{CVM}{Caging and Venting Mechanism}
\newacronym{cw}{CW}{ClockWise}
\newacronym{cw_}{CW}{Continuous Wave}
\newacronym{cwd}{CWD}{Compact White Dwarf binaries}

\newacronym{d/a}{D/A}{Digital -To-Analogue Converter}
\newacronym{d/sre}{D/SRE}{Director of Science \&  Robotic Exploration}
\newacronym{da}{DA}{Data Analysis}
\newacronym{daap}{DAAP}{Deep Analysis Alert Pipeline}
\newacronym{dac}{DAC}{Digital-to-Analog Converter}
\newacronym{dacu}{DACU}{Diagnostic Acquisition and Control Unit}
\newacronym{daftwp}{DAFTWP}{Data Analysis Framework and Tools Work Packages}
\newacronym{daughter-s}{DAUGHTER-S}{Daughter spacecraft}
\newacronym{db}{DB}{Data Base}
\newacronym{dc}{DC}{Direct Current}
\newacronym{dcbh}{DCBH}{Direct Collapse Black Hole}
\newacronym{dcc}{DCC}{Data Computing Centre}
\newacronym{dccs}{DCCS}{Data Computing Centers}
\newacronym{dci}{DCI}{Dossier de Contrôle des Interfaces}
\newacronym{dciu}{DCIU}{Digital Control Interface Unit}
\newacronym{dcl}{DCL}{Declared Components List}
\newacronym{dcp}{DCP}{De-Commissioning Phase}
\newacronym{dcr}{DCR}{Document Change Request}
\newacronym{dd}{DD}{Displacement Damage}
\newacronym{dde}{DDE}{Diagnostics Drive Electronics}
\newacronym{ddf}{DDF}{Design Definition File}
\newacronym{ddpc}{DDPC}{Distributed Data Processing Centre}
\newacronym{dds}{DDS}{Data and Diagnostic Subsystem}
\newacronym{ddvp}{DDVP}{"Design}
\newacronym{deep2}{DEEP2}{Deep Extragalactic Evolutionary Probe 2}
\newacronym{dei}{DEI}{"Diversity}
\newacronym{df}{DF}{drag-free}
\newacronym{dfacs}{DFACS}{Drag-Free Attitude Control System}
\newacronym{dgmm}{DGMM}{Detailed Geometrical Mathematical Model}
\newacronym{dhlc}{DHLC}{Data Handling And Laser Control}
\newacronym{djf}{DJF}{Design Justification File}
\newacronym{dll}{DLL}{Delay-Locked Loop}
\newacronym{dlr}{DLR}{Deutsches Zentrum für Luft- und Raumfahrt}
\newacronym{dm}{DM}{Dark Matter}
\newacronym{dm_}{DM}{Development Model}
\newacronym{dm__}{DM}{Differential Mode}
\newacronym{dml}{DML}{Declared Material List}
\newacronym{dmpl}{DMPL}{Declared Mechanical Part List}
\newacronym{dms}{DMS}{Document Management System}
\newacronym{dmu}{DMU}{Data Management Unit}
\newacronym{dnel}{DNEL}{Disconnection of Non-Essential Loads}
\newacronym{dod}{DOD}{Depth Of Discharge}
\newacronym{dof}{DOF}{Degrees Of Freedom}
\newacronym{dp}{DP}{Diagnostic Package}
\newacronym{dpa}{DPA}{Destructive Physical Analysis}
\newacronym{dpc}{DPC}{Data Processing Centre}
\newacronym{dpewg}{DPEWG}{Detection and Parameter Estimation Work Packages}
\newacronym{dpl}{DPL}{Daily Pipeline}
\newacronym{dpl_}{DPL}{Declared Process List}
\newacronym{dpll}{DPLL}{Digital Phase Locked Loop}
\newacronym{dps}{DPS}{Differential Power Sensing}
\newacronym{dr}{DR}{data release}
\newacronym{drb}{DRB}{Delivery Review Board}
\newacronym{drd}{DRD}{Document Requirements Definition}
\newacronym{drl}{DRL}{Document Requirements List}
\newacronym{drpt}{DRPT}{Data Release Product Team}
\newacronym{drs}{DRS}{Disturbance Reduction System}
\newacronym{ds}{DS}{Diagnostics Subsystem}
\newacronym{dsa}{DSA}{Deep Space Antenna}
\newacronym{dsc}{DSC}{Daughter Space Craft}
\newacronym{dscu}{DSCU}{Diagnostics Signal Conditioning Unit}
\newacronym{dsn}{DSN}{Deep Space Network}
\newacronym{dsp}{DSP}{Data Signal Processing}
\newacronym{dst}{DST}{Deep Space Transponder}
\newacronym{dtcp}{DTCP}{Daily Telemetry Communications Period}
\newacronym{dtm}{DTM}{Deterministic Transfer Manoeuvre}
\newacronym{dtmm}{DTMM}{Detailed Thermal Mathematical Model}
\newacronym{dtu}{DTU}{Danmarks Tekniske Universitet}
\newacronym{dut}{DUT}{Device under test}
\newacronym{dwd}{DWD}{Double White Dwarf (binary)}
\newacronym{dws}{DWS}{Differential Wavefront Sensing}

\newacronym{e2e}{E2E}{End-to-End}
\newacronym{ear}{EAR}{Export Authorisation Regulation}
\newacronym{ebb}{EBB}{Engineering (or Elegant) Bread Board}
\newacronym{ebm}{EBM}{External Balance Mass}
\newacronym{ec}{EC}{Executive Committee}
\newacronym{ec_}{EC}{Export Control}
\newacronym{ecdl}{ECDL}{Extended Cavity Diode Laser}
\newacronym{ecp}{ECP}{Export Control Plan}
\newacronym{ecss}{ECSS}{European Cooperation for Space Standardization}
\newacronym{edu}{EDU}{Engineering Development Unit}
\newacronym{eed}{EED}{Electro-Explosive Device}
\newacronym{eee}{EEE}{"Electrical}
\newacronym{eee_}{EEE}{Electronic/Electrical Engineering Equipment}
\newacronym{eelv}{EELV}{Evolved Expendable Launch Vehicle}
\newacronym{eeprom}{EEPROM}{Electrically Erasable Programmable Read-Only Memory}
\newacronym{efc2}{EFC2}{ESA/ESTEC Frame Contract 2}
\newacronym{egaps}{EGAPS}{European Galactic Plane Surveys}
\newacronym{egse}{EGSE}{Electrical Ground Support Equipment}
\newacronym{eh}{EH}{Electrode Housing}
\newacronym{eid}{EID}{Experiment Interface Document}
\newacronym{eidp}{EIDP}{End Item Data Package}
\newacronym{eirp}{EIRP}{Effective Isotropic Radiative Power}
\newacronym{elisa}{ELISA}{Evolved Laser Interferometer Space Antenna}
\newacronym{elite}{ELITE}{European LISA Technology Experiment}
\newacronym{elt}{ELT}{Extremely Large Telescope}
\newacronym{elv}{ELV}{Expendable Launch Vehicle}
\newacronym{em}{EM}{Engineering Model}
\newacronym{ema}{EMA}{Electro-Magnetic}
\newacronym{emc}{EMC}{Earth-Mean Ecliptic}
\newacronym{emc_}{EMC}{Electro-Magnetic Compatibility}
\newacronym{eme}{EME}{Earth Mean Equator of J2000}
\newacronym{emi}{EMI}{Electro-Magnetic Interference}
\newacronym{emrb}{EMRB}{Extreme Mass-Ratio Bursts}
\newacronym{emri}{EMRI}{Extreme Mass-Ratio Inspiral}
\newacronym{encts}{ENCTS}{Nonconformance tracking}
\newacronym{eob}{EOB}{effective one-body}
\newacronym{eoc}{EOC}{End Of Charge}
\newacronym{eod}{EOD}{End Of Discharge}
\newacronym{eol}{EOL}{End-Of-Life}
\newacronym{eom}{EOM}{Electro-Optical Modulator}
\newacronym{eom_}{EOM}{Equations of Motion}
\newacronym{epms}{EPMS}{Extended Phase Measuring System}
\newacronym{eppl}{EPPL}{European Preferred Part List}
\newacronym{eps}{EPS}{Electrical Power System}
\newacronym{eps_}{EPS}{Extended Press-Schechter Formalism}
\newacronym{epta}{EPTA}{European Pulsar Timing Array}
\newacronym{eqm}{EQM}{Engineering Qualification Model}
\newacronym{erst}{ERST}{Early Release Science Time}
\newacronym{esa}{ESA}{European Space Agency}
\newacronym{esac}{ESAC}{European Space Astronomy Centre}
\newacronym{escc}{ESCC}{European Space Components Coordination}
\newacronym{esd}{ESD}{Electrostatic Discharge}
\newacronym{esdc}{ESDC}{ESAC Science Data Centre}
\newacronym{esoc}{ESOC}{European Space Operations Centre}
\newacronym{esp}{ESP}{Extended Science Phase}
\newacronym{estec}{ESTEC}{European Space Research and Technology Centre}
\newacronym{estrack}{ESTRACK}{European Space TRACKing}
\newacronym{et}{ET}{Einstein Telescope}
\newacronym{eth}{ETH}{Eidgenössische Technische Hochschule Zürich}
\newacronym{ethz}{ETHZ}{Eidgenössische Technische Hochschule Zürich}
\newacronym{etu}{ETU}{Engineering Thermal Unit}
\newacronym{ew}{EW}{Electroweak}

\newacronym{fa}{FA}{Fiber Amplifier}
\newacronym{faq}{FAQ}{Frequently Asked Question(s)}
\newacronym{fbd}{FBD}{Functional Block Diagram}
\newacronym{fcc}{FCC}{Future Circular Collider}
\newacronym{fcl}{FCL}{Fold-back Current Limiter}
\newacronym{fcv}{FCV}{Flow Control Valve}
\newacronym{fcv_}{FCV}{Functional Configuration Verification}
\newacronym{fd}{FD}{Flight Dynamics}
\newacronym{fddb}{FDDB}{Flight Dynamics Data Base}
\newacronym{fdir}{FDIR}{"Failure Detection}
\newacronym{fdm}{FDM}{Frequency Distribution Module}
\newacronym{fds}{FDS}{Frequency Distribution System}
\newacronym{fdv}{FDV}{Fill and Drain Valve}
\newacronym{fe}{FE}{Finite Element}
\newacronym{fee}{FEE}{Front-End Electronics}
\newacronym{fee sau}{FEE SAU}{Front-End Electronics Sensing And Actuation Unit}
\newacronym{feep}{FEEP}{Field-Emission Electric Propulsion}
\newacronym{fem}{FEM}{Finite Element Model}
\newacronym{fer}{FER}{Frame Error Rate}
\newacronym{ff ogse}{FF OGSE}{Far-Field Optical Ground Support Equipment}
\newacronym{ff-ogse}{FF-OGSE}{Far-Field Optical Ground Support Equipment}
\newacronym{ffogse}{FFOGSE}{Far Field Optical Ground Support Equipment}
\newacronym{ffpt}{FFPT}{Full Functional Performance Test}
\newacronym{fft}{FFT}{Fast Fourier Transform}
\newacronym{fft_}{FFT}{Full Functional Test}
\newacronym{fios}{FIOS}{Fibre Injector Optical Subassembly}
\newacronym{fit}{FIT}{Failure In Time (per billion cycles)}
\newacronym{fits}{FITS}{Flexible Image Transport System}
\newacronym{fm}{FM}{Flight Model}
\newacronym{fmea}{FMEA}{"Failure Mode}
\newacronym{fmeca}{FMECA}{Failure Mode Effects and Criticality Analysis}
\newacronym{fmt}{FMT}{Formulation Management Team}
\newacronym{fo}{FO}{Fibre optics}
\newacronym{foh}{FOH}{Fibre Optical Harness}
\newacronym{fom}{FOM}{Figures of Merit}
\newacronym{fop}{FOP}{Flight Operations Plan}
\newacronym{fopt}{FOPT}{First-Order Phase Transition}
\newacronym{fos}{FOS}{Factor of Safety}
\newacronym{fov}{FOV}{Field of View}
\newacronym{fpag}{FPAG}{Fundamental Physics Advisory Group}
\newacronym{fpga}{FPGA}{Field-Programmable Gate Array}
\newacronym{fpwg}{FPWG}{Fundamental Physics Working Group}
\newacronym{fr}{FR}{Laser Frequency Reference}
\newacronym{frs}{FRS}{Frequency Reference System}
\newacronym{frs-e}{FRS-E}{Frequency Reference System- Electrical}
\newacronym{frs-o}{FRS-O}{Frequency Reference System- Optical}
\newacronym{fs}{FS}{Flight Segment}
\newacronym{fs_}{FS}{Flight Spare}
\newacronym{fsm}{FSM}{Finite State Machine}
\newacronym{fsu}{FSU}{Fibre Switching Unit}
\newacronym{fsua}{FSUA}{Fiber Switch Unit Actuator}
\newacronym{fte}{FTE}{Full-Time Equivalent}
\newacronym{fumo}{FUMO}{Functional Model}
\newacronym{fvv}{FVV}{Fill and Vent Valve}
\newacronym{fwhm}{FWHM}{Full Width Half Maximum}
\newacronym{fzu}{FZU}{Institute of Physics of the Czech Academy of Sciences}

\newacronym{gb}{GB}{Galactic Binary}
\newacronym{gbs}{GBS}{Galactic Binaries}
\newacronym{gcr}{GCR}{Galactic Cosmic Ray}
\newacronym{gcrs}{GCRS}{Geocentric Celestial Reference System}
\newacronym{gnd}{GND}{Ground}
\newacronym{goat}{GOAT}{Gravitational Observatory Advisory Team}
\newacronym{gprm}{GPRM}{"Grabbing}
\newacronym{gps}{GPS}{Global Positioning System}
\newacronym{gr}{GR}{General Theory Of Relativity}
\newacronym{gr740}{GR740}{LEON4 SPARCv8 microprocessor}
\newacronym{grace}{GRACE}{Gravity Recovery and Climate Experiment}
\newacronym{grace-fo}{GRACE-FO}{GRACE Follow-On}
\newacronym{grb}{GRB}{Gamma-Ray Burst}
\newacronym{grs}{GRS}{Gravitational Reference (Sub)system}
\newacronym{grs-fee}{GRS-FEE}{Gravitational Reference (Sub)system Front End Electronics}
\newacronym{grs-tme}{GRS-TME}{Gravitational Reference (Sub)system Test Mass Emulator}
\newacronym{grsh}{GRSH}{Gravitational Reference (Sub)system Head}
\newacronym{gs}{GS}{Ground Segment}
\newacronym{gs_}{GS}{Ground Station}
\newacronym{gse}{GSE}{Ground Support Equipment}
\newacronym{gsf}{GSF}{Gravitational Self-Force}
\newacronym{gsfc}{GSFC}{Goddard Space Flight Center}
\newacronym{gsir}{GSIR}{Ground Segment Implementation Review}
\newacronym{gsorr}{GSORR}{Ground Segment Operations Readiness Review}
\newacronym{gsrr}{GSRR}{Ground Segment Readiness Review}
\newacronym{gto}{GTO}{Geostationary Transfer Orbit}
\newacronym{gut}{GUT}{Grand Unified Theory}
\newacronym{gw}{GW}{Gravitational Wave}
\newacronym{gw-wg}{GW-WG}{Gravitational-wave Observatory Working Group}
\newacronym{gwb}{GWB}{Gravitational Wave Background}
\newacronym{gwic}{GWIC}{Gravitational Wave International Committee}

\newacronym{h/w}{H/W}{Hardware}
\newacronym{hbr}{HBR}{High Bit Rate}
\newacronym{hc-hpc}{HC-HPC}{High Current-High Power Command}
\newacronym{hdf}{HDF}{Hierarchical Data Format}
\newacronym{hdrm}{HDRM}{Hold Down and Release Mechanism}
\newacronym{hdsw}{HDSW}{Hardware-Dependent Software}
\newacronym{hemp}{HEMP}{High-Efficiency Micropropulsion}
\newacronym{heo}{HEO}{High Earth Orbit}
\newacronym{heto}{HETO}{Heliocentric Earth Trailing Orbit}
\newacronym{hg}{HG}{Mercury}
\newacronym{hga}{HGA}{High-Gain Antenna}
\newacronym{hk}{HK}{Housekeeping}
\newacronym{hl-lhc}{HL-LHC}{High Luminosity Large Hadron Collider}
\newacronym{hpc}{HPC}{High Power Command}
\newacronym{hr}{HR}{High Resolution}
\newacronym{hr-la}{HR-LA}{High Resolution Low Authority}
\newacronym{hr-ma}{HR-MA}{High Resolution Maximum Authority}
\newacronym{hsia}{HSIA}{Hardware Software Interaction Analysis}
\newacronym{hst}{HST}{Hubble Space Telescope}
\newacronym{hsu}{HSU}{Harness Splicing Unit}
\newacronym{hv-hpc}{HV-HPC}{High Voltage-High Power Command}
\newacronym{hw}{HW}{Hardware}

\newacronym{iandt}{I\& T}{Integration and Test(ing)}
\newacronym{i/f}{I/F}{interface}
\newacronym{ia}{IA}{Instrument Amplifier}
\newacronym{iaas}{IAAS}{Infrastructure As A Service}
\newacronym{iau}{IAU}{International Astronomical Union}
\newacronym{ibm}{IBM}{Internal Balance Mass}
\newacronym{icc}{ICC}{Instrument Control Computer}
\newacronym{icd}{ICD}{Interface Control Document}
\newacronym{icrf}{ICRF}{International Celestial Reference Frame}
\newacronym{icrs}{ICRS}{International Celestial Reference System}
\newacronym{icso}{ICSO}{Innermost Stable Circular Orbit}
\newacronym{ida}{IDA}{Initial Displacement Angle}
\newacronym{idl}{IDL}{Interferometer Data Log}
\newacronym{idm}{IDM}{Information Documentation Management}
\newacronym{ids}{IDS}{Interferometric Detection (Sub)system}
\newacronym{ieec}{IEEC}{Institut d’Estudis Espacials de Catalunya}
\newacronym{if}{IF}{Interface}
\newacronym{ifo}{IFO}{InterFerOmeter}
\newacronym{ifp}{IFP}{In-Field Pointing}
\newacronym{igm}{IGM}{Inter-Galactic Medium}
\newacronym{ils}{ILS}{Instrument Line of Sight}
\newacronym{ima}{IMA}{Integrated Modular Avionics}
\newacronym{imbh}{IMBH}{Intermediate-Mass Black Hole}
\newacronym{imbhb}{IMBHB}{Intermediate Mass Black Hole Binary}
\newacronym{imf}{IMF}{Initial Mass Function}
\newacronym{imr}{IMR}{Inspiral-Merger-Ringdown}
\newacronym{imri}{IMRI}{Intermediate Mass-Ratio Inspiral}
\newacronym{ims}{IMS}{Interferometric Measurement System}
\newacronym{in2p3}{IN2P3}{National Institute of Nuclear and Particle Physics}
\newacronym{infn}{INFN}{Instituto Nazionale di Fisica Nucleare (National Institute for Nuclear Physics)}
\newacronym{ingaas}{INGAAS}{Indium-Gallium Arsenide}
\newacronym{inpta}{INPTA}{Indian pulsar Timing Array}
\newacronym{inrep}{INREP}{Initial Noise Reduction Pipeline}
\newacronym{insprl}{INSPRL}{Interferometer in Space for Detecting Gravity-wave Radiation using Lasers}
\newacronym{iocr}{IOCR}{In-Orbit Commissioning Review}
\newacronym{iot}{IOT}{Instruments Operations Team}
\newacronym{ipc}{IPC}{Industrial Policy Committee}
\newacronym{ipn}{IPN}{Input Parameters and Noises}
\newacronym{ipt}{IPT}{Integrated Product Team}
\newacronym{iptf}{IPTF}{intermediate Palomar Transient Factory}
\newacronym{ipu}{IPU}{Interferometric Processing Unit}
\newacronym{ir}{IR}{Infra-red}
\newacronym{ird}{IRD}{Interface Requirement Document}
\newacronym{isc}{ISC}{Inter-spacecraft Laser Communication}
\newacronym{ish}{ISH}{Inertial Sensor Head}
\newacronym{isi}{ISI}{Inter-Satellite Interferometer}
\newacronym{ism}{ISM}{Instrument Sensitivity Model}
\newacronym{iso}{ISO}{International Organization for Standardization}
\newacronym{ispe}{ISPE}{Instrument Scientists \&  Payload Experts}
\newacronym{isuk}{ISUK}{Inertial Sensor UV Kit}
\newacronym{isvv}{ISVV}{Independent Software Verification and Validation}
\newacronym{it}{IT}{Information Technology}
\newacronym{it_}{IT}{Integration and Test(ing)}
\newacronym{itar}{ITAR}{International Traffic in Arms Regulations}
\newacronym{itat}{ITAT}{Integrated Technical Advisory Team(s)}
\newacronym{itrf}{ITRF}{International Terrestrial Reference Frame}
\newacronym{itt}{ITT}{Invitation To Tender}

\newacronym{jila}{JILA}{Joint Institute For Laboratory Astrophysics}
\newacronym{jpip}{JPIP}{Joint Project Implementation Plan}
\newacronym{jpl}{JPL}{Jet Propulsion Laboratory}
\newacronym{jwst}{JWST}{James Webb Space Telescope}

\newacronym{kagra}{KAGRA}{Kamioka Gravitational-wave Detector}
\newacronym{kip}{KIP}{Key Inspection Point}
\newacronym{ko}{KO}{Kick-off}
\newacronym{kpi}{KPI}{Key Performance Indicator}
\newacronym{kru}{KRU}{Kourou}
\newacronym{ksc}{KSC}{Kennedy Space Center}
\newacronym{ku}{KU}{Katholieke Universiteit Leuven}
\newacronym{kul}{KUL}{Katholieke Universiteit Leuven}

\newacronym{l0}{L0}{Level 0 data}
\newacronym{l0.5}{L0.5}{Level 0.5 data}
\newacronym{l1}{L1}{Level 1 data}
\newacronym{l2}{L2}{Level 2 data}
\newacronym{l3}{L3}{Level 3 data}
\newacronym{l3st}{L3ST}{L3 Study Team}
\newacronym{la}{LA}{Laser Assembly}
\newacronym{la_}{LA}{Launch Adapter}
\newacronym{lad}{LAD}{LISA Acronym Dictionary}
\newacronym{lagos}{LAGOS}{Laser Antenna For Gravitational-Radiation Observation In Space}
\newacronym{lat}{LAT}{Lot Acceptance Test}
\newacronym{lbr}{LBR}{Low Bit Rate}
\newacronym{lca}{LCA}{LISA Core Assembly}
\newacronym{lcc}{LCC}{Life Cycle Cost}
\newacronym{lcdm}{LCDM}{Lambda Cold Dark Matter}
\newacronym{lcl}{LCL}{LISA Consortium Lead}
\newacronym{lcl_}{LCL}{Latching Current Limiter}
\newacronym{lcm}{LCM}{Spacecraft Composite Stack}
\newacronym{lcpm}{LCPM}{LISA Constellation Performance Model}
\newacronym{ld}{LD}{Laser Diode}
\newacronym{ldb}{LDB}{LISA Mission Database}
\newacronym{ldc}{LDC}{LISA Data Challenge}
\newacronym{ldcwg}{LDCWG}{LISA Data Challenge Working Group}
\newacronym{ldp}{LDP}{LISA Data Processing}
\newacronym{ldpc}{LDPC}{Low-Density Parity Check}
\newacronym{ldpg}{LDPG}{LISA Data Progressing Group}
\newacronym{lecs}{LECS}{LISA Early Career Scientist Group}
\newacronym{led}{LED}{Light-Emitting Diode}
\newacronym{lem}{LEM}{Laser Electronics Module}
\newacronym{leop}{LEOP}{Launch and Early Operations Phase}
\newacronym{lfn}{LFN}{Laser Frequency Noise}
\newacronym{lfs}{LFS}{Laser Frequency Stabilisation}
\newacronym{lga}{LGA}{Low-Gain Antenna}
\newacronym{lh}{LH}{Laser Head}
\newacronym{lhc}{LHC}{Large Hadron Collider}
\newacronym{lhcp}{LHCP}{Left Hand Circular Polarization}
\newacronym{li}{LI}{List}
\newacronym{lic}{LIC}{Lisa Instrument Consortium}
\newacronym{lid/lic}{LID/LIC}{Laser Induced Damage (LID) and Laser Induced Contamination (LIC)}
\newacronym{lig}{LIG}{LISA Instrument Group}
\newacronym{ligo}{LIGO}{Laser Interferometer Gravitational Wave Observatory}
\newacronym{limas}{LIMAS}{LISA Instrument and Metrology Avionics System}
\newacronym{lincs}{LINCS}{LISA Internal Networking Committee for Science}
\newacronym{lisa}{LISA}{Laser Interferometer Space Antenna}
\newacronym{lisa-mrd-0}{LISA-MRD-0}{Mission Requirement Document}
\newacronym{lisn}{LISN}{Line Impedance Stabilization Network}
\newacronym{lispeg}{LISPEG}{LISA Instrument Scientists and Payload Experts Group}
\newacronym{list}{LIST}{Lisa International Science Team}
\newacronym{llap}{LLAP}{Low Latency Alert Pipeline}
\newacronym{lld}{LLD}{Launch Lock Device}
\newacronym{lli}{LLI}{Long Lead Item}
\newacronym{lm}{LM}{Life Model}
\newacronym{lmc}{LMC}{Large Magellanic Cloud}
\newacronym{lmf}{LMF}{LISA Mission Formulation}
\newacronym{lmt}{LMT}{Large Momentum Transfer}
\newacronym{lo}{LO}{Local Oscillator}
\newacronym{loa}{LOA}{Letter of Agreement}
\newacronym{locs}{LOCS}{LISA Opto-mechanical Core System}
\newacronym{lolipp}{LOLIPP}{Level 0 To Levl 1 Pipeline Procedure}
\newacronym{lom}{LOM}{Laser Optical Module}
\newacronym{los}{LOS}{Line Of Sight}
\newacronym{lpf}{LPF}{Lisa Pathfinder}
\newacronym{lps}{LPS}{Laser Pre-stabilization System}
\newacronym{lps_}{LPS}{Launch Power Supply}
\newacronym{lri}{LRI}{Laser-Ranging Interferometer}
\newacronym{lrr}{LRR}{Launch Readiness Review}
\newacronym{ls}{LS}{Laser (Sub)system}
\newacronym{lsa}{LSA}{Launch Service Agreement}
\newacronym{lsb}{LSB}{Least Significant Bit}
\newacronym{lsg}{LSG}{Lisa Science Group}
\newacronym{lsgcore}{LSGCORE}{LISA Science Group Core Team}
\newacronym{lso}{LSO}{Last Stable Orbit}
\newacronym{lss}{LSS}{Laser Systems}
\newacronym{lsst}{LSST}{Large Synoptic Survey Telescope}
\newacronym{lst}{LST}{LISA Science Team}
\newacronym{ltp}{LTP}{Lisa Pathfinder Technology Package}
\newacronym{ltpda}{LTPDA}{LISA Technology Package Data Analysis}
\newacronym{ltt}{LTT}{Light Travel Time}
\newacronym{lut}{LUT}{Look-Up Table}
\newacronym{lv}{LV}{Latch Valve}
\newacronym{lv_}{LV}{Launch vehicle}
\newacronym{lva}{LVA}{Launch Vehicle Adaptor}
\newacronym{lvds}{LVDS}{Low Voltage Differential Signalling}
\newacronym{lvk}{LVK}{"LIGO}

\newacronym{ma}{MA}{Mission Assurance}
\newacronym{mac}{MAC}{Mass Acceleration Curve}
\newacronym{mag}{MAG}{Mission Analysis Guidelines}
\newacronym{mait}{MAIT}{Manufacturing Assembly Integration Tolerances}
\newacronym{mar}{MAR}{Mission Adoption Review}
\newacronym{maxi}{MAXI}{Monitor Of All-Sky X-Ray Image}
\newacronym{mbe}{MBE}{Model Best Estimate}
\newacronym{mbh}{MBH}{Massive Black Hole}
\newacronym[longplural=massive black hole binaries]{mbhb}{MBHB}{Massive Black Hole Binary}
\newacronym{mbw}{MBW}{Measurement Bandwidth}
\newacronym{mci}{MCI}{"Mass}
\newacronym{mclk}{MCLK}{Master Clock}
\newacronym{mcmc}{MCMC}{Markov-Chain Monte Carlo}
\newacronym{mcr}{MCR}{Mission Consolidation Review}
\newacronym{mcu}{MCU}{Mechanism Control Unit}
\newacronym{mdr}{MDR}{Main Data Repository}
\newacronym{mecc}{MECC}{Merger Event Coordination Committee}
\newacronym{meda}{MEDA}{Mean Earth Displacement Angle}
\newacronym{mel}{MEL}{Master Equipment List}
\newacronym{meop}{MEOP}{Maximum Expected Operating Pressure}
\newacronym{mep}{MEP}{Maximum Expected Pressure}
\newacronym{mfr}{MFR}{Mission Formulation Review}
\newacronym{mfs}{MFS}{Main Frequency Stabilisation}
\newacronym{mgse}{MGSE}{Mechanical Ground Support Equipment}
\newacronym{mib}{MIB}{Mission Information Base}
\newacronym{micd}{MICD}{Mechanical Interface Control Document}
\newacronym{micinn}{MICINN}{Ministerio de Ciencia e Innovación del Reino de España}
\newacronym{mida}{MIDA}{Mean (Earth) Initial Displacement Angle}
\newacronym{mifo}{MIFO}{Metallic Interferometer}
\newacronym{mil}{MIL}{(Spec) Specification Of The US Department Of Defence}
\newacronym{min}{MIN}{Minimum}
\newacronym{mip}{MIP}{Mandatory Inspection Point}
\newacronym{mird}{MIRD}{Mission Implementation Requirement Document}
\newacronym{mla}{MLA}{Multi-Lateral Agreement}
\newacronym{mlb}{MLB}{Motorised Light Band}
\newacronym{mldc}{MLDC}{Mock Lisa Data Challenge}
\newacronym{mli}{MLI}{Multi Layer Insulation}
\newacronym{mm}{MM}{Mission Manager}
\newacronym{mmawp}{MMAWP}{Multi-Messenger Astrophysics Work Package}
\newacronym{mmh}{MMH}{Monomethyl Hydrazine}
\newacronym{mmic}{MMIC}{Monolithic Microwave Integrated Circuit}
\newacronym{mo}{MO}{Maser Oscillator}
\newacronym{mo_}{MO}{molybdenum}
\newacronym{moad}{MOAD}{Mission Operations Assumptions Document}
\newacronym{moc}{MOC}{Mission Operations Centre}
\newacronym{mofpa}{MOFPA}{Master Oscillator Fibre Power Amplifier}
\newacronym{moi}{MOI}{Moment of Inertia}
\newacronym{mon-3}{MON-3}{Mixed Oxides Of Nitrogen With Si3 Nitric Oxide}
\newacronym{mopa}{MOPA}{Master Oscillator Power Amplifier}
\newacronym{moq}{MOQ}{Minimum Order Quantity}
\newacronym{mos}{MOS}{Margin of Safety}
\newacronym[longplural=moveable optical sub-assemblies]{mosa}{MOSA}{Moving Optical Sub-Assembly}
\newacronym{mot}{MOT}{Magneto-Optical Trap}
\newacronym{mother-s/c}{MOTHER-S/C}{Mother spacecraft}
\newacronym{mou}{MOU}{Memorandum of Understanding}
\newacronym{mpa}{MPA}{Micro Propulsion Assembly}
\newacronym{mpe}{MPE}{Max Planck Institute for Extraterrestrial Physics}
\newacronym{mpe_}{MPE}{Micro Propulsion Electronics}
\newacronym{mpr}{MPR}{Mission Performance Requirement}
\newacronym{mps}{MPS}{Micro Propulsion System}
\newacronym{mps cg}{MPS CG}{Cold Gas Micro Propulsion System}
\newacronym{mr}{MR}{Measurement Requirement}
\newacronym{mr_}{MR}{Mission Requirement}
\newacronym{mrb}{MRB}{Material Review Board}
\newacronym{mrd}{MRD}{Mission Requirements Document}
\newacronym{mrf}{MRF}{Mechanical Reference Frame}
\newacronym{mrl}{MRL}{Manufacturing readiness Level}
\newacronym{msb}{MSB}{Material Review Board}
\newacronym{msc}{MSC}{Mother Spacecraft}
\newacronym{msem}{MSEM}{Mission Systems Engineering Managers}
\newacronym{mss}{MSS}{Mosa Support Structure}
\newacronym{mta}{MTA}{Micro Propulsion Thruster Assembly}
\newacronym{mtl}{MTL}{Mission Time Line}
\newacronym{mtr}{MTR}{Mid-Term Review}
\newacronym{mts}{MTS}{Metasurface (antennas)}

\newacronym{n/a}{N/A}{Not Applicable}
\newacronym{na}{NA}{Not Applicable}
\newacronym{na_}{NA}{Numerical Aperture}
\newacronym{nanograv}{NANOGRAV}{North American Nanohertz Observatory for Gravitational Waves}
\newacronym{nasa}{NASA}{National Aeronautic and Space Administration}
\newacronym{nco}{NCO}{numerically controlled oscillator}
\newacronym{ncr}{NCR}{Non Conformance Report}
\newacronym{ncts}{NCTS}{Non-Conformance Tracking System}
\newacronym{nda}{NDA}{Non-Destructive Analysis}
\newacronym{nda_}{NDA}{Non-Disclosure Agreement}
\newacronym{ndi}{NDI}{Non-Destructive Inspection}
\newacronym{nea}{NEA}{Non Explosive Actuator}
\newacronym{necp}{NECP}{Near Earth Commissioning Phase}
\newacronym{nep}{NEP}{Noise Equivalent Power}
\newacronym{nggm}{NGGM}{Next-Generation Gravity Mission}
\newacronym{ngmp}{NGMP}{Next Generation Micro Processor}
\newacronym{ngo}{NGO}{New Gravitational Wave Observatory}
\newacronym{ngrm}{NGRM}{Next Generation Radiation Monitor}
\newacronym{nicm}{NICM}{NASA Instrument Cost Model}
\newacronym{nls}{NLS}{NASA Launch Services}
\newacronym{nlso}{NLSO}{NASA LISA Study Office}
\newacronym{nlst}{NLST}{Nasa Lisa Study Team}
\newacronym{npmb}{NPMB}{National Project Manager Board}
\newacronym{npro}{NPRO}{Non-Planar Ring Oscillator}
\newacronym{nr}{NR}{Numerical Relativity}
\newacronym{nrac}{NRAC}{Non-Recurrent Agent Cost}
\newacronym{nrb}{NRB}{Nonconformance Review Board}
\newacronym{nrc}{NRC}{National Research Council}
\newacronym{nrc_}{NRC}{Non-Recurrent Cost}
\newacronym{nre}{NRE}{Non-Recurrent Expense}
\newacronym{nre_}{NRE}{Non-Recurring Engineering}
\newacronym{nrt}{NRT}{Near-Real-Time Pipeline}
\newacronym{nrvc}{NRVC}{Non-Recurrent Vendor Cost}
\newacronym{nrz-l}{NRZ-L}{Non Return To Zero Level}
\newacronym{ns}{NS}{Neutron Star}
\newacronym{nsgs}{NSGS}{NASA Science Ground Segment}
\newacronym{nsp}{NSP}{Nominal Science Phase}
\newacronym{nspires}{NSPIRES}{NASA Solicitation and Proposal Integrated Review and Evaluation System}
\newacronym{ntc}{NTC}{Negative Temperature Coefficient}
\newacronym{nwnh}{NWNH}{New Worlds New Horizons}

\newacronym{oam}{OAM}{Optical Assembly Mechanics}
\newacronym{oas}{OAS}{Optical Assembly Subsystem}
\newacronym{oatm}{OATM}{Optical Assembly Tracking Mechanism}
\newacronym{ob}{OB}{Optical Bench}
\newacronym{oba}{OBA}{Optical Bench Assembly}
\newacronym{obc}{OBC}{On-Board Computer}
\newacronym{obcp}{OBCP}{On-Board Control Procedure}
\newacronym{obdh}{OBDH}{On Board Data Handling}
\newacronym{obmcu}{OBMCU}{Optical Bench Mechanism Control Unit}
\newacronym{obs}{OBS}{Organisation Breakdown Structure}
\newacronym{obsw}{OBSW}{On-Board Software}
\newacronym{obt}{OBT}{On-Board Time}
\newacronym{ocp}{OCP}{Over Current Protection}
\newacronym{ocr}{OCR}{Observatory Commissioning Review}
\newacronym{ogs}{OGS}{Operational Ground Segment}
\newacronym{ogse}{OGSE}{Optical Ground Support Equipment}
\newacronym{oicd}{OICD}{Optical ICD}
\newacronym{oirf}{OIRF}{Optical Interface Reference Frame}
\newacronym{om}{OM}{Optical Model}
\newacronym{oms}{OMS}{Optical Metrology System}
\newacronym{op}{OP}{Optical Path}
\newacronym{opl}{OPL}{Optical Path Length}
\newacronym{or}{OR}{Operational Requirement}
\newacronym{oro}{ORO}{Optical Read-Out}
\newacronym{ot}{OT}{optical truss}
\newacronym{ots}{OTS}{Off The Shelf}
\newacronym{ots_}{OTS}{Optical Test (Sub)Systems}
\newacronym{ovp}{OVP}{Over Voltage Protection}

\newacronym{p/l}{P/L}{Payload}
\newacronym{p/m}{P/M}{Propulsion Module}
\newacronym{pa}{PA}{Power Amplifier}
\newacronym{pa_}{PA}{Product Assurance}
\newacronym{paands}{PA\& S}{Product Assurance and Safety}
\newacronym{paa}{PAA}{Point-Ahead Angle}
\newacronym{paam}{PAAM}{Point-Ahead Angle Mechanism}
\newacronym{pac}{PAC}{Particulate Contamination}
\newacronym{pad}{PAD}{Parts Approval Document}
\newacronym{pan-starrs}{PAN-STARRS}{The Panoramic Survey Telescope \&  Rapid Response System}
\newacronym{par}{PAR}{phase accumulation register}
\newacronym{pard}{PARD}{Product Assurance and Safety Requirement Document}
\newacronym{parx}{PARX}{public archive}
\newacronym{patp}{PATP}{Preliminary Authorisation To Proceed}
\newacronym{pbh}{PBH}{Primordial Black Hole}
\newacronym{pbs}{PBS}{Polarising Beam Splitter (might be an LPF acronym)}
\newacronym{pc}{PC}{Processing Component}
\newacronym{pcb}{PCB}{Parts Control Board}
\newacronym{pcb_}{PCB}{Printed Circuit Board}
\newacronym{pcc}{PCC}{Payload Control Computer}
\newacronym{pcdu}{PCDU}{Power Control And Distribution Unit}
\newacronym{pcos}{PCOS}{Physics of the Cosmos}
\newacronym{pcp}{PCP}{Payload Commanding and Processing}
\newacronym{pcp-gse}{PCP-GSE}{Payload Commanding and Processing Ground Support Equipment}
\newacronym{pcs}{PCS}{Payload Control Subsystem}
\newacronym{pcu}{PCU}{Phasemeter Control Unit}
\newacronym{pcu_}{PCU}{Power Conditioning Unit}
\newacronym{pd}{PD}{Photo Diode}
\newacronym{pdb}{PDB}{Parameter Data Base}
\newacronym{pdd}{PDD}{Payload Description Document}
\newacronym{pdf}{PDF}{Probability Density Function}
\newacronym{pdh}{PDH}{Probability Data Handling}
\newacronym{pdr}{PDR}{Preliminary Design Review}
\newacronym{pds}{PDS}{Photo Detector System}
\newacronym{pem}{PEM}{Payload Engineering Model}
\newacronym{pfci}{PFCI}{Potential Fracture Critical Item}
\newacronym{pfm}{PFM}{Proto-Flight Model}
\newacronym{pfo}{PFO}{Particle fallout}
\newacronym{pi}{PI}{Principal Investigator}
\newacronym{piar}{PIAR}{Performance and Impact Assessment Review}
\newacronym{pir}{PIR}{phase increment register}
\newacronym{pl}{PL}{Payload}
\newacronym{pl_}{PL}{Power-Law (model)}
\newacronym{pll}{PLL}{Phase-Locked Loop}
\newacronym{pls}{PLS}{Power Law Sensitivity}
\newacronym{pm}{PM}{Phase Meter}
\newacronym{pm_}{PM}{Progress Meeting/Management}
\newacronym{pm__}{PM}{Proof Mass}
\newacronym{pmb}{PMB}{Project Manager Board}
\newacronym{pmdsp}{PMDSP}{Phase Meter Digital Signal Processor}
\newacronym{pmf}{PMF}{Polarization-Maintaining Fibre}
\newacronym{pmfde}{PMFDE}{Phase Meter Frequency Distribution Electronics}
\newacronym{pmfee}{PMFEE}{Phase Meter Front-End Electronics}
\newacronym{pmon}{PMON}{Power Monitor(s)}
\newacronym{pmp}{PMP}{"Parts}
\newacronym{pmr}{PMR}{Payload Management Requirements}
\newacronym{pms}{PMS}{Phase Measurement Subsystem}
\newacronym{pn}{PN}{Post-Newtonian}
\newacronym{po}{PO}{Purchase Order}
\newacronym{poad}{POAD}{Performance Operations Assumptions Document}
\newacronym{por}{POR}{Payload Operations Request}
\newacronym{ppe}{PPE}{Parameterized Post-Einsteinian}
\newacronym{ppr}{PPR}{Proper Pseudo-Range}
\newacronym{pps}{PPS}{Pulse Per Second}
\newacronym{ppta}{PPTA}{Parkes Pulsar Timing Array}
\newacronym{ppu}{PPU}{Power Processing Unit}
\newacronym{prd}{PRD}{Payload Requirement Document}
\newacronym{prds}{PRDS}{Phase Reference Distribution System}
\newacronym{prds-ogse}{PRDS-OGSE}{Phase Reference Distribution System - Optical Ground Support Equipment}
\newacronym{prn}{PRN}{Pseudo-Random (Speudo Noise)}
\newacronym{prr}{PRR}{Preliminary Requirements Review}
\newacronym{prt}{PRT}{Platinum Resistance Thermometers}
\newacronym{ps}{PS}{Project Scientist}
\newacronym{pscu}{PSCU}{Power Signal Conditioning Unit}
\newacronym{psd}{PSD}{Power Spectral Density}
\newacronym{psf}{PSF}{Point-Spread Function}
\newacronym{psm}{PSM}{Power Supply Module}
\newacronym{psr}{PSR}{Pulsar}
\newacronym{pt}{PT}{Phase Transition}
\newacronym{pt_}{PT}{Pilot Tone}
\newacronym{pt__}{PT}{platinum}
\newacronym{pta}{PTA}{Pulsar Timing Array}
\newacronym{ptf}{PTF}{Palomar Transient Factory}
\newacronym{ptr}{PTR}{Post Test Review}
\newacronym{pus}{PUS}{Packet Utilisation Standard}
\newacronym{pva}{PVA}{Photovoltaic Assembly}
\newacronym{pvc}{PVC}{Polyvinyl Chloride}
\newacronym{py}{PY}{Person Years}

\newacronym{qa}{QA}{Quality Assurance}
\newacronym{qci}{QCI}{Quality Conformance Inspection}
\newacronym{qe}{QE}{Quantum Efficiency}
\newacronym{qm}{QM}{Qualification Model}
\newacronym{qml}{QML}{Qualified Manufacturer List}
\newacronym{qnm}{QNM}{Quasi-Normal Mode}
\newacronym{qpd}{QPD}{Quadrant Photodetector}
\newacronym{qpl}{QPL}{Qualified Product List}
\newacronym{qpr}{QPR}{Quadrant Photo-Receiver}
\newacronym{qr}{QR}{Qualification Review}
\newacronym{qsl}{QSL}{Qualification Status List}
\newacronym{qsl_}{QSL}{Quasi-Static Load}
\newacronym{qso}{QSO}{Quasi-Stellar Object}

\newacronym{r-lcl}{R-LCL}{Retriggerable Latching current limiter}
\newacronym{raan}{RAAN}{Right Ascension of the Ascending Node}
\newacronym{racs}{RACS}{Reaction and Attitude Control subsystem (combined RCS and MPS)}
\newacronym{rad}{RAD}{Radiation}
\newacronym{ram}{RAM}{Random Access Memory}
\newacronym{rams}{RAMS}{"Reliability}
\newacronym{rats}{RATS}{Rapid Time Survey}
\newacronym{rc}{RC}{Recurrent Cost}
\newacronym{rd}{RD}{Reference Document}
\newacronym{rdbms}{RDBMS}{Relational Database Management System}
\newacronym{re}{RE}{Radiated Emission}
\newacronym{ref}{REF}{Reference}
\newacronym{ref_}{REF}{Reference Interferometer}
\newacronym{rf}{RF}{Radio Frequency}
\newacronym{rfa}{RFA}{Request for Approval}
\newacronym{rfa_}{RFA}{Request for Frequency Assignment}
\newacronym{rfd}{RFD}{Request for Deviation}
\newacronym{rfdn}{RFDN}{Radio Frequency Distribution Network}
\newacronym{rfi}{RFI}{Reference Interferometer}
\newacronym{rfi_}{RFI}{Request for Information}
\newacronym{rfp}{RFP}{Request for Proposal}
\newacronym{rfq}{RFQ}{Request for Quotation}
\newacronym{rft}{RFT}{Reduced Functional Test}
\newacronym{rfw}{RFW}{Request for Waiver}
\newacronym{rgmm}{RGMM}{Reduced Geometrical Mathematical Model}
\newacronym{rh}{RH}{Reference Hole}
\newacronym{rh_}{RH}{Relative Humidity}
\newacronym{rhcp}{RHCP}{Right Hand Circular Polarization}
\newacronym{rid}{RID}{Review Item Discrepancy}
\newacronym{rin}{RIN}{Relative Intensity Noise}
\newacronym{rit}{RIT}{Radio-Frequency Ion Thruster}
\newacronym{riu}{RIU}{Remote Interface Unit}
\newacronym{rm}{RM}{Radiation Monitor}
\newacronym{rms}{RMS}{Root Mean Square}
\newacronym{rob}{ROB}{Read Out Box}
\newacronym{rom}{ROM}{Read Only Memory}
\newacronym{rom_}{ROM}{Rough Order of Magnitude}
\newacronym{rpe}{RPE}{"Relative Pointing/Performance Error (= jitter}
\newacronym{rs}{RS}{Radiated Susceptibility}
\newacronym{rs_}{RS}{Requirement Specification}
\newacronym{rss}{RSS}{Root Sum Square}
\newacronym{rt}{RT}{Remote Terminal}
\newacronym{rtb}{RTB}{Real-time Test Bed}
\newacronym{rtmm}{RTMM}{Reduced Thermal Mathematical Model}
\newacronym{rtos}{RTOS}{Real Time Operating System}
\newacronym{rtu}{RTU}{Remote Terminal Unit}
\newacronym{ru}{RU}{Reference Unit}
\newacronym{rvt}{RVT}{Radiation Verification Testing}
\newacronym{rx}{RX}{Received Laser (from far Spacecraft)}
\newacronym{rx_}{RX}{Receiver/Reception}
\newacronym{rxte}{RXTE}{Rossi X-Ray Timing Explorer}

\newacronym{s/c}{S/C}{Spacecraft}
\newacronym{s/c-p/m}{S/C-P/M}{Spacecraft/Propulsion-Module}
\newacronym{s/s}{S/S}{Sub-system}
\newacronym{s/w}{S/W}{Software}
\newacronym{sa}{SA}{Solar Array}
\newacronym{sa_}{SA}{Sub Address}
\newacronym{sard}{SARD}{System AIV Requirements Document}
\newacronym{sau}{SAU}{Sensing and Actuation Unit}
\newacronym{savoir}{SAVOIR}{Space AVionics Open Interface aRchitecture}
\newacronym{sbcc}{SBCC}{Single Board Computer Core}
\newacronym{sbdl}{SBDL}{Standard Balanced Digital Link}
\newacronym{sbh}{SBH}{stellar-mass Black Hole}
\newacronym[longplural=stellar-mass black hole binaries]{sbhb}{SBHB}{stellar-mass Black Hole binary}
\newacronym{sbs}{SBS}{Stimulated Brillouin Scattering}
\newacronym{sc}{SC}{Spacecraft}
\newacronym{scc}{SCC}{Stress Corrosion Cracking}
\newacronym{sccp}{SCCP}{System Commissioning and Calibration Phase}
\newacronym{scet}{SCET}{Spacecraft Elapsed Time}
\newacronym{scf}{SCF}{Software Configuration File}
\newacronym{sci}{SCI}{Science Directorate of ESA}
\newacronym{sci_}{SCI}{Science Interferometer}
\newacronym{sci-f}{SCI-F}{Future Missions Department of ESA Directorate of Science}
\newacronym{sci-m}{SCI-M}{"Project Control and Management Support Service}
\newacronym{sci-p}{SCI-P}{Projects Department of ESA Directorate of Science}
\newacronym{sci-s}{SCI-S}{Science and Operations Department of ESA Directorate of Science}
\newacronym{sci-sd}{SCI-SD}{Development Division of SCI-S}
\newacronym{scird}{SCIRD}{Science Requirements Document}
\newacronym{scmp}{SCMP}{Software Configuration Management Plan}
\newacronym{sco}{SCO}{Stellar-mass Compact Object}
\newacronym{scoc}{SCOC}{Science Operations Coordinator}
\newacronym{scoe}{SCOE}{Special Check Out Equipment}
\newacronym{scot}{SCOT}{Special Condition Of Tender}
\newacronym{scu}{SCU}{Signal Conditioning Unit}
\newacronym{sdb}{SDB}{System Data Base}
\newacronym{sdm}{SDM}{SOC Development Manager}
\newacronym{sdp}{SDP}{system data pool}
\newacronym{sds}{SDS}{Science Diagnostics Subsystem}
\newacronym{sdss}{SDSS}{Sloan Digital Sky Survey}
\newacronym{se}{SE}{System Engineer(ing)}
\newacronym{see}{SEE}{Single Event Effect}
\newacronym{sefi}{SEFI}{Single Event Functional Interruption}
\newacronym{sel}{SEL}{Single Event Latch-up}
\newacronym{seo}{SEO}{System Engineering Office}
\newacronym{sep}{SEP}{Solar Electric Propulsion}
\newacronym{sep_}{SEP}{Solar Energetic Particle}
\newacronym{sepd}{SEPD}{Single-Element Photo Diode}
\newacronym{set}{SET}{Single Event Transient}
\newacronym{set_}{SET}{Supplier Evaluation Tool}
\newacronym{seu}{SEU}{Single Event Upset}
\newacronym{sgicd}{SGICD}{Space to Ground ICD}
\newacronym{sgm}{SGM}{Safeguard Memory}
\newacronym{sgo}{SGO}{Space-based Gravitational-wave Observatory}
\newacronym{sgs}{SGS}{Science Ground Segment}
\newacronym{sgwb}{SGWB}{Stochastic Gravitational-wave Background}
\newacronym{si}{SI}{Science investigation}
\newacronym{sia}{SIA}{Sun Incident Angle}
\newacronym{sic}{SIC}{Silicon Carbide}
\newacronym{sigw}{SIGW}{Scalar-Induced Gravitational Waves}
\newacronym{sim}{SIM}{Space Interferometry Mission}
\newacronym{sip}{SIP}{Science Implementation Plan}
\newacronym{sips}{SIPS}{Science Implementation Plans}
\newacronym{sir}{SIR}{System Integration Review}
\newacronym{sird}{SIRD}{Science Implementation Requirements Document}
\newacronym{siso}{SISO}{Single Input/Single Output}
\newacronym{siwp}{SIWP}{Science Interpretation Work Package}
\newacronym{ska}{SKA}{Square Kilometre Array}
\newacronym{skm}{SKM}{Station Keeping Manoeuvre}
\newacronym{sl}{SL}{Scatter Light}
\newacronym{sli}{SLI}{Single Layer Insulation}
\newacronym{sm}{SM}{Structural Model}
\newacronym{smbh}{SMBH}{Super-Massive Black Hole}
\newacronym{smc}{SMC}{Small Magellanic Cloud}
\newacronym{smd}{SMD}{Surface mount device}
\newacronym{smf}{SMF}{Single Mode Fiber}
\newacronym{smm}{SMM}{Structural Mathematical Model}
\newacronym{smp}{SMP}{Science Management Plan}
\newacronym{smt}{SMT}{Surface mounted technology}
\newacronym{snr}{SNR}{Signal-To-Noise Ratio}
\newacronym{so}{SO}{Science objective}
\newacronym{soad}{SOAD}{Science Operations Assumptions Document}
\newacronym{sobh}{SOBH}{Stellar origin black hole}
\newacronym{sobhb}{SOBHB}{Stellar Origin Black Hole Binary}
\newacronym{soc}{SOC}{Science Operation Centre}
\newacronym{soc_}{SOC}{State of Charge}
\newacronym{socd}{SOCD}{Science Operations Concept Document}
\newacronym{sodpc}{SODPC}{Science Operations Data Processing Center}
\newacronym{sosl}{SOSL}{Science Operations Study Lead}
\newacronym{sovt}{SOVT}{System Operations Validation Tests}
\newacronym{sow}{SOW}{Statement Of Work}
\newacronym{sp}{SP}{Settable Parameters}
\newacronym{spa}{SPA}{Stationary Phase Approximation}
\newacronym{spacom}{SPACOM}{Space Control}
\newacronym{spb}{SPB}{Senior Procurement Board}
\newacronym{spc}{SPC}{Science Programme Committee}
\newacronym{spf}{SPF}{Single Point Failure}
\newacronym{spff}{SPFF}{Single Point Failure Free}
\newacronym{spr}{SPR}{Software Problem Report}
\newacronym{spw}{SPW}{SpaceWire}
\newacronym{squid}{SQUID}{Superconducting Quantum Interference Device}
\newacronym{srd}{SRD}{System Requirements Document}
\newacronym{srdb}{SRDB}{System Reference Database}
\newacronym{srf}{SRF}{Spacecraft Reference Frame}
\newacronym{sron}{SRON}{Netherlands Institute for Space Research}
\newacronym{srp}{SRP}{Solar Radiation Pressure}
\newacronym{srr}{SRR}{System Requirements Review}
\newacronym{srs}{SRS}{spacecraft reference system}
\newacronym{ss}{SS}{Sub-system}
\newacronym{ssac}{SSAC}{Space Science Advisory Committee}
\newacronym{ssb}{SSB}{Solar System Barycenter}
\newacronym{ssc}{SSC}{Senior Science Committee}
\newacronym{ssewg}{SSEWG}{Solar System and Exploration Working Group}
\newacronym{ssf}{SSF}{Structural Scaling Factor}
\newacronym{sso}{SSO}{Swiss Space Office}
\newacronym{ssrd}{SSRD}{Space Segment Requirements Document}
\newacronym{sss}{SSS}{SW System Specifications}
\newacronym{sst}{SST}{Science Study Team}
\newacronym{ste}{STE}{Service Time Equivalent}
\newacronym{stft}{STFT}{Short-Time Fourier Transform}
\newacronym{stm}{STM}{Structural Model}
\newacronym{stm_}{STM}{Structural Thermal Model}
\newacronym{stoc}{STOC}{Science Technology Operations Centre}
\newacronym{stom}{STOM}{Structural Optical Model}
\newacronym{stop}{STOP}{Structural Thermo-Optical Performance}
\newacronym{stp}{STP}{Science Topical Panel}
\newacronym{str}{STR}{Coarse Star Tracker}
\newacronym{str_}{STR}{Star Tracker}
\newacronym{sts}{STS}{IDS  (Interferometric Detection System) Stable Structure}
\newacronym{sum}{SUM}{SW User Manual}
\newacronym{susw}{SUSW}{Start-up SW}
\newacronym{svf}{SVF}{SW Validation Facility}
\newacronym{svt}{SVT}{System Verification Test}
\newacronym{sw}{SW}{Software}
\newacronym{swt}{SWT}{Science Working Team}

\newacronym{ta}{TA}{Telescope Assembly}
\newacronym{taa}{TAA}{Technical Assistance Agreement}
\newacronym{tas}{TAS}{Thales-Alenia Aerospace Systems}
\newacronym{tb(t)}{TB(T)}{Thermal Balance (Test)}
\newacronym{tbc}{TBC}{To Be Confirmed}
\newacronym{tbd}{TBD}{To Be Defined}
\newacronym{tbm}{TBM}{Trimming Balance Mass}
\newacronym{tbt}{TBT}{Thermal Balance Test}
\newacronym{tbw}{TBW}{To Be Written}
\newacronym{tc}{TC}{Telecommand}
\newacronym{tc/tm}{TC/TM}{Telecommand/Telemetry}
\newacronym{tcb}{TCB}{Barycentric Coordinate Time}
\newacronym{tcg}{TCG}{Geocentric Coordinate Time}
\newacronym{tcls}{TCLS}{Triple Core LockStep}
\newacronym{tcm}{TCM}{Trajectory Correction Manoeuvre}
\newacronym{tcs}{TCS}{Thermal Control Subsystem}
\newacronym{tcvr}{TCVR}{Transceiver}
\newacronym{tda}{TDA}{Technology Development Activities}
\newacronym{tdb}{TDB}{Barycentric Dynamical Time}
\newacronym{tdi}{TDI}{Time-Delay Interferometry}
\newacronym{tdi x}{TDI X}{LISA's 2-arm configuration}
\newacronym{tdir}{TDIR}{Time Delay Interferometry Ranging}
\newacronym{tecc}{TECC}{Transient Event Coordination Committee}
\newacronym{tel}{TEL}{Telescope}
\newacronym{the}{THE}{On-Board Clock Time}
\newacronym{ti}{TI}{titanium}
\newacronym{tia}{TIA}{Transimpedance Amplifiers}
\newacronym{tid}{TID}{Total Ionising Dose}
\newacronym{tl}{TL}{or TEL Telescope}
\newacronym{tm}{TM}{Telemetry}
\newacronym{tm_}{TM}{Test Mass (Often Proof Mass)}
\newacronym{tm-i/f}{TM-I/F}{test mass interface}
\newacronym{tm-ogse}{TM-OGSE}{Test-Mass Optical Ground Equipment}
\newacronym{tm/tc}{TM/TC}{Telemetry/Telecommand}
\newacronym{tme}{TME}{Test Mass Emulator}
\newacronym{tmi}{TMI}{test mass interferometer}
\newacronym{tml}{TML}{Total Mass Loss}
\newacronym{tmm}{TMM}{Thermal Mathematical Model}
\newacronym{tmt}{TMT}{Telemetry/Telecommand}
\newacronym{tmtc}{TMTC}{Telemetry/Telecommand}
\newacronym{tn}{TN}{Technical Note}
\newacronym{tnid}{TNID}{Total Non-Ionising Dose}
\newacronym{tno}{TNO}{Nederlandse Organisatie voor Toegepast Natuurwetenschappelijk Onderzoek}
\newacronym{toba}{TOBA}{Telescope \&  Optical Bench Assembly}
\newacronym{toga}{TOGA}{"Telescope}
\newacronym{tp}{TP}{Transfer Phase}
\newacronym{tpro}{TPRO}{Test Procedure}
\newacronym{tps}{TPS}{Spacecraft Proper Time}
\newacronym{tra}{TRA}{Technology Readiness Assessment}
\newacronym{trb}{TRB}{Test Review Board}
\newacronym{trip}{TRIP}{Technical Readiness and Implementation Plan}
\newacronym{trl}{TRL}{Technology Readiness Level}
\newacronym{trp}{TRP}{Temperature Reference Points}
\newacronym{trr}{TRR}{Test Readiness Review}
\newacronym{ts}{TS}{Telescope}
\newacronym{ts_}{TS}{Transmission Stage}
\newacronym{tsm}{TSM}{Temperature Sensor Monitor}
\newacronym{tsp}{TSP}{Temporal and Spatial Partitioning}
\newacronym{tspe}{TSPE}{Test specification}
\newacronym{ttandc}{TT\& C}{"Telemetry}
\newacronym{ttl}{TTL}{Tilt-to-Length}
\newacronym{tvac}{TVAC}{Thermal Vacuum}
\newacronym{tvc}{TVC}{Total Vendor Cost}
\newacronym{tvt}{TVT}{Thermal Vacuum Test}
\newacronym{twta}{TWTA}{Traveling-Wave Tube Amplifier}
\newacronym{tx}{TX}{Transmission}
\newacronym{tx_}{TX}{Transmitted Laser (main laser in OB)}

\newacronym{uart}{UART}{Universal Asynchronous Receiver Transmitter}
\newacronym{ucb}{UCB}{Ultra-Compact Binaries}
\newacronym{ufl}{UFL}{University of Florida - Gainesville}
\newacronym{ugl}{UGL}{University of Glasgow}
\newacronym{uic}{UIC}{Unit Identification Code}
\newacronym{uk atc}{UK ATC}{UK Astronomy Technology Centre}
\newacronym{uksa}{UKSA}{United Kingdom Space Agency}
\newacronym{ulu}{ULU}{UV Light Unit}
\newacronym{ulx}{ULX}{Ultra-Luminous X-ray source}
\newacronym{urf}{URF}{Unit Reference Frame}
\newacronym{url}{URL}{Uniform Resource Locator}
\newacronym{us}{US}{Ultra-Sonic}
\newacronym{uso}{USO}{Ultra-Stable Oscillator}
\newacronym{usr}{USR}{Ultra Slow-Roll Inflation Model}
\newacronym{ut}{UT}{University of Trento}
\newacronym{utc}{UTC}{Coordinated Universal Time}
\newacronym{uv}{UV}{Ultra-Violet}

\newacronym{vandv}{V\& V}{Verification and Validation}
\newacronym{vast}{VAST}{Variables And Slow Transients}
\newacronym{vb}{VB}{Verification Binary}
\newacronym{vc}{VC}{Vacuum Chamber}
\newacronym{vcd}{VCD}{Verification Control Document}
\newacronym{vcm}{VCM}{Verification Control Matrix}
\newacronym{vd}{VD}{Vent Duct}
\newacronym{ve}{VE}{Vacuum Enclosure}
\newacronym{vef}{VEF}{Vacuum Enclosure Frame}
\newacronym{vgb}{VGB}{Verification Galactic Binary}
\newacronym{virgo}{VIRGO}{Virgo}
\newacronym{vlbi}{VLBI}{Very-long-baseline interferometry}
\newacronym{vm}{VM}{Verification Method}
\newacronym{vms}{VMS}{Very Massive Star}
\newacronym{vp}{VP}{Verification Program}

\newacronym{w}{W}{tungsten}
\newacronym{wavwg}{WAVWG}{Waveform Working Group}
\newacronym{wbs}{WBS}{Work Breakdown Structure}
\newacronym{wca}{WCA}{Worst Case Analysis}
\newacronym{wd}{WD}{Watch Dog}
\newacronym{wd_}{WD}{White Dwarf}
\newacronym{wfe}{WFE}{Wave Front Error}
\newacronym{wg}{WG}{Working Group}
\newacronym{wmap}{WMAP}{Wilkison Microwave Anisotropy Probe}
\newacronym{wp}{WP}{Work Package}
\newacronym{wps}{WPS}{Work Packages}
\newacronym{wr}{WR}{Wide Range}
\newacronym{wsb}{WSB}{Weak Stability Boundary}

\newacronym{xda}{XDA}{X-band Downlink Assembly}
\newacronym{xgnsb}{XGNSB}{Extragalactic Neutron Star Binary}
\newacronym{xml}{XML}{Extensible Markup Language}
\newacronym{xmri}{XMRI}{extremely mass-ratio inspiral}

\newacronym{yb}{YB}{ytterbium}

\newacronym{zar}{ZAR}{Zemax Archive File}
\newacronym{zifo}{ZIFO}{Zerudur Interferomenter}
\newacronym{ztf}{ZTF}{Zwicky Transient Facility}

\usepackage{bm}
\usepackage{physics}
\usepackage{siunitx}
\usepackage{cleveref}
\usepackage{mathtools}
\usepackage{verbatim}
\usepackage{tikz}
\usepackage{tikz-3dplot}
\usetikzlibrary{3d,arrows.meta,positioning,calc}
\usetikzlibrary{angles, quotes}
\usetikzlibrary{positioning}
\usepackage[normalem]{ulem}
\usepackage{appendix}

\newcommand{\delay}{\mathbf{D}}
\newcommand{\adv}{\mathbf{A}}

\tdplotsetmaincoords{60}{110}
\pgfmathsetmacro{\rvec}{1.3}
\pgfmathsetmacro{\thetavec}{30}
\pgfmathsetmacro{\phivec}{60}

\definecolor{pink}{rgb}{0.55,0,0.52}
\definecolor{mygreen}{rgb}{0.19,0.55,0.11}
\definecolor{dkgreen}{rgb}{0,0.6,0}
\definecolor{gray}{rgb}{0.5,0.5,0.5}
\definecolor{mauve}{rgb}{0.58,0,0.82}
\definecolor{verbgray}{gray}{0.9}
\definecolor{lightblue}{rgb}{0.85,0.9,1}
\definecolor{lightgreen}{rgb}{0.85,1,0.85}
\definecolor{lightorange}{rgb}{1,0.94,0.8}
\definecolor{Darkgreen}{rgb}{0,0.4,0}
\definecolor{dodgerblue}{HTML}{1E90FF}
\definecolor{chromeyellow}{rgb}{1.0, 0.65, 0.0}

\newcommand{\bh}[1]{\bm{ \hat {#1}}}

\def\be{\begin{equation}}
\def\en{\end{equation}}
\def\bea{\begin{eqnarray}}
\def\ena{\end{eqnarray}}
\def\bsub{\begin{subequations}}
\def\esub{\end{subequations}}

\newcommand{\sYlm}{{}_{-2}Y_{\ell m}}
\newcommand{\sYlmstar}{{}_{-2}Y_{\ell, -m}^{*}}
\newcommand{\sYlminusmstar}{{}_{-2}Y_{\ell, -m}^{*}}

\renewcommand{\vu}{\mbox{\boldmath${u}$}}
\newcommand{\vv}{\mbox{\boldmath${v}$}}
\newcommand{\vp}{\mbox{\boldmath${p}$}}
\newcommand{\vq}{\mbox{\boldmath${q}$}}
\newcommand{\vx}{\mbox{\boldmath${x}$}}
\newcommand{\vy}{\mbox{\boldmath${y}$}}
\newcommand{\vz}{\mbox{\boldmath${z}$}}

\newcommand{\ReportTitle}{SGS Conventions Document}           
\newcommand{\ReportShortTitle}{SGS Conventions}     
\newcommand{\ReportIssue}{01}                                                                  
\newcommand{\ReportRevision}{20}                                                             
\newcommand{\ReportRef}{LISA-DDPC-SEG-TN-007}                                           
\newcommand{\ReportDate}{\today}
\newcommand{\ReportStatus}{}

\definecolor{pink}{rgb}{0.55,0,0.52}
\definecolor{mygreen}{rgb}{0.19,0.55,0.11}
\definecolor{dkgreen}{rgb}{0,0.6,0}
\definecolor{gray}{rgb}{0.5,0.5,0.5}
\definecolor{mauve}{rgb}{0.58,0,0.82}
\definecolor{verbgray}{gray}{0.9}
\definecolor{lightblue}{rgb}{0.85,0.9,1}
\definecolor{lightgreen}{rgb}{0.85,1,0.85}
\definecolor{lightorange}{rgb}{1,0.94,0.8}
\definecolor{forestgreen}{rgb}{0.1,0.49,0.07}

\begin{document}

\pagenumbering{arabic}

\begin{titlepage}
\thispagestyle{fancy}

\vspace{30mm}
\noindent
{\fontsize{32pt}{14pt}\selectfont \ReportTitle}

\vspace{5mm}
\noindent
{ \fontsize{18pt}{14pt}\selectfont \ReportRef }

\vspace{3mm}
\textit{\textsf{v\ReportIssue\_\ReportRevision} - \ReportDate}

\vspace{30mm}
\begin{center}
\includegraphics[scale=2]{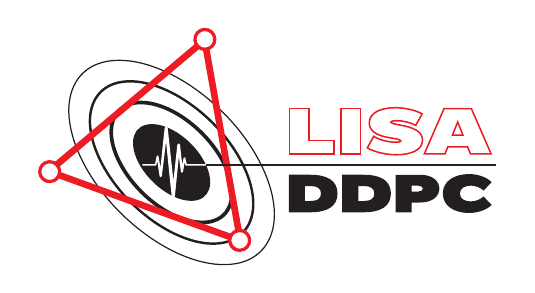}
\end{center}

\cfoot{For LISA SGS Use only}

\newpage

\lhead{\includegraphics[scale=0.30,trim={5mm 4mm 7mm 5mm},clip]{figures/2025-128_Logo_LISA_DDPC_def_Coul.pdf}}
\chead{}
\rhead{\begin{tabular}{r} 
\textsc{\ReportShortTitle} \\  
\textbf{\thepage / \pageref{LastPage}}
\end{tabular}}

\cfoot{
\textcolor{gray}{Reference: \ReportRef - Date of issue: \ReportDate} \\
\textcolor{gray}{Issue Revision: \ReportIssue\_\ReportRevision - Status: \ReportStatus} \\
\vspace{2mm}
For LISA SGS Use only
\vspace{2mm}
}


\begin{center}
{\Large \textbf{DESCRIPTION}}  
\end{center}

\vspace{0.5cm}

{
\renewcommand{\arraystretch}{1.8}
\begin{table}[htp]

  \begin{tabular}{ m{0.15\textwidth} m{0.85\textwidth}}
  
  N/Ref & \ReportRef \\
  \hline
  
  Issue\_Revision & \ReportIssue\_\ReportRevision \\
  \hline
  
  Date & \ReportDate \\
  \hline
  
  Title & \ReportTitle \\
  \hline
  
  Authors & Quentin Baghi (APC) \\
  \hline 
  
  Abstract & This document aims to provide a reference for conventions used in data simulations, waveforms, and analysis pipelines within the \gls{ddpc} of the \gls{lisa} mission. 
  It can also be considered as good practices for conventions in all publications related to \gls{lisa}.

  \end{tabular} 
\end{table}
}

\vspace{2cm}

\begin{center}
{\Large \textbf{APPROVAL}}  
\end{center}

\vspace{0.5cm}

{
\renewcommand{\arraystretch}{1.3}
\begin{table}[htp]

  \begin{tabular}{ m{0.2\textwidth} m{0.4\textwidth} m{0.4\textwidth}}
  
   & Names & Signatures \\
   & & (with the dates) \\
  \hline
  
  DDPC Manager & Hong-Nga Nguyen & \\
  & (on behalf of DDPC authors) & \\
  & & \\
  & & \\
  \hline
  
  DDPC Scientist & Antoine Petiteau & \\
  & & \\
  & & \\
  \hline
  
  DDPC QA & Brigitte Huynh & \\
  Manager & (on behalf of DDPC PAQA Team) & \\
  & & \\
  & & \\
  \hline

  \end{tabular} 
\end{table}
}



\newpage

\begin{center}
{\Large \textbf{DOCUMENT CHANGE RECORD}}\\  
\end{center}

\noindent
\begin{tabular}{|>{\centering\arraybackslash}p{0.05\textwidth}|>{\centering\arraybackslash}p{0.05\textwidth}|>{\centering\arraybackslash}p{0.12\textwidth}|m{0.28\textwidth}|m{0.4\textwidth}|}
\hline
{\bf Iss.} & {\bf Rev.} & {\bf Date} & {\bf Author} & {\bf Reason for change} \\
\hline
\hline
00 & 00 & 2024/06/01 &  Q. Baghi (APC)  & First Version  \\
01 & 00 & 2025/06/16 &  Q. Baghi (APC)  & After DDPC review \\
01 & 10 & 2026/03/20 & Q. Baghi (APC) & Minor revision for arXiv submission \\
01 & 20 & 2026/04/28 & Q. Baghi (APC) & Updates to align with v2.1 of AD1\\
\hline

\end{tabular}

\vspace{30mm}

\begin{center}
  {\Large \textbf{DISTRIBUTION LIST}}\\
\end{center}

\noindent
\begin{tabular}{|>{\centering\arraybackslash}p{0.52\textwidth}| >{\centering\arraybackslash}p{0.2\textwidth}|  >{\centering\arraybackslash}p{0.2\textwidth} |}
\hline
{\bf Recipient} & {\bf Restricted} & {\bf Not restricted} \\
\hline
\hline
DDPC &  & \ding{55} \\
\hline
SOC &  & \ding{55} \\
\hline
P\&O  &  & \ding{55} \\
\hline
NSGS  &  & \ding{55} \\
\hline
LST  &  & \ding{55} \\
\hline
\end{tabular}

\newpage
  
\begin{center}
  {\Large \textbf{CONTRIBUTOR LIST}}\\
\end{center}

\noindent
\begin{tabular}{|m{0.3\textwidth}| m{0.35\textwidth}| m{0.35\textwidth} |}
\hline
{\bf Author's name} & {\bf Institute} & {\bf Location } \\
\hline
Stanislas Babak   & APC & Paris (France) \\
Quentin Baghi     &  APC & Paris (France) \\
Leor Barack       & University of Southampton & Southampton (UK)\\
Jean-Baptiste Bayle & CEA/IRFU & Saclay (France) \\
Ollie Burke      & University of Glasgow & Glasgow (UK)\\
Raffi Enficiaud  & AEI & Potsdam (Germany)\\
Hector Estelles  & Institute of Space Sciences & Barcelona (Spain)\\
Cecilio Garc\'{i}a-Quir\'{o}s  & UZH & Z\"{u}rich (Switzerland)\\
Olaf Hartwig    & AEI & Hannover (Germany) \\
Aurelien Hees   & LTE & Paris (France)\\
Sascha Husa     & Institute of Space Sciences & Barcelona (Spain)\\
Henri Inchausp\'{e} & KU Leuven &  Leuven (Belgium)\\
Eric Joffre     & ESTEC & Noordwijk (Netherlands) \\
Antoine Klein  & University of Birmingham & Birmingham (UK) \\
Philip Lynch    & AEI & Potsdam (Germany)\\
Sylvain Marsat  & L2IT & Toulouse (France)\\
Jonathan Menu & KU Leuven &  Leuven (Belgium)\\
Zach Nasipak & University of Southampton & Southampton (UK)\\
Ramon Pardo De Santayana  & ESAC &  Villafranca del Castillo (Spain) \\
Harald Pfeiffer  & AEI & Potsdam (Germany)\\
Lorenzo Pompili & University of Nottingham / AEI & Nottingham (UK)\\
Adam Pound  & University of Southampton & Southampton (UK)\\
Geraint Pratten  & University of Birmingham & Birmingham (UK) \\
Antoni Ramos-Buades  & IAC3/UIB & Palma (Spain)\\
Carlos Sopuerta  & Institute of Space Sciences & Barcelona (Spain)\\
Niels Warburton & University College Dublin & Dublin (Ireland)\\
\hline
\end{tabular}

\begin{center}
{\Large \textbf{AKNOWLEDGMENTS OF FUNDING AGENCIES}}\\
\end{center}

\noindent
\begin{tabular}{|m{0.85\textwidth}| m{0.13\textwidth}|}
	\hline
	{\bf Name} & {\bf Acronym} \\
	\hline
	European Space Agency & ESA \\
	National Astronotics and Space Administration & NASA \\
	Agenzia Spaziale Italiana &ASI \\
	Belgian Science Policy Office & BELSPO\\
	Centre National d'\'{E}tudes Spatiales & CNES \\
	Deutsches Zentrum f\"{u}r Luft & DLR \\
	Ministerio de Ciencia e Innovaci\'{o}n & MICINN \\
	Research Ireland & RI\\
	Royal Society of London & \\
	Space Research Organisation Netherlands & SRON \\
	State Research Agency, Ministry for Science, Innovation and Universities of Spain & AEI\\
	Swiss National Science Foundation & SNSF\\
	UK Research and Innovation & UKRI\\
	UK Space Agency & UKSA\\
	\hline
\end{tabular}

\newpage
\tableofcontents

\newpage
\listoffigures

\newpage

\begin{center}
  {\Large \textbf{SCOPE}}\\
\end{center}

This document aims to provide a reference for conventions used in data simulations, waveforms, and analysis pipelines within the \gls{ddpc}. 
It can also be considered as good practices for conventions in all publications related to the \gls{lisa} mission.

\vspace{12mm}

\begin{center}
  {\Large \textbf{APPLICABLES AND REFERENCE DOCUMENTS}}\\
\end{center}

\vspace{8mm}
{\large \textbf{Applicable documents}}
\vspace{2mm}

\noindent
\begin{tabular}{|m{0.06\textwidth} |m{0.34\textwidth}| m{0.4\textwidth}| m{0.1\textwidth} |}
\hline
{\bf No.} & {\bf Reference} & {\bf Title} & {\bf Version} \\
\hline
\hline
\hypertarget{AD1}{AD1} & ESA-LISA-EST-MIS-LI-0001 & LISA Definitions, Conventions and Acronyms & 2.1  \\
\hline 
\hypertarget{AD2}{AD2} & ESA-LISA-EST-MIS-DD-0002 & LISA Performance Model Description & 0.1 \\
\hline
\hypertarget{AD3}{AD3} & ESA-LISA-ESOC-MAS-RP-0001 & LISA Consolidated Report on Mission Analysis & 1.2 \\
\hline
\hline
\end{tabular}

\vspace{8mm}


\newpage

\begin{center}
  {\Large \textbf{GLOSSARY}}\\
\end{center}

The acronyms definitions for the LISA DDPC  project are available in the following documents:
\begin{itemize}
  \item ESA-LISA-EST-MIS-LI-0001 - LISA Acronyms, Definitions and Conventions
  \item (in preparation) - LISA DDPC Acronyms and Definitions
\end{itemize}

\printglossary            
\printglossary[type=\acronymtype] 

\vspace{8mm}
\begin{center}
  {\Large \textbf{POSITION IN THE DOCUMENT TREE}}\\
\end{center}

-

\end{titlepage}


\lhead{\includegraphics[scale=0.30,trim={5mm 4mm 7mm 5mm},clip]{figures/2025-128_Logo_LISA_DDPC_def_Coul.pdf}}
\chead{}
\rhead{\begin{tabular}{r} 
\textsc{\ReportShortTitle} \\  
\textbf{\thepage / \pageref{LastPage}}
\end{tabular}}

\cfoot{
\textcolor{gray}{Reference: \ReportRef - Date of issue: \ReportDate} \\
\textcolor{gray}{Issue Revision: \ReportIssue\_\ReportRevision - Status: \ReportStatus} \\
\vspace{2mm}
For LISA SGS Use only
\vspace{2mm}
}

\clearpage
\glsresetall

\section*{Introduction}

The SGS Conventions document (formerly untitled LISA Rosetta Stone) aims at providing recommendations for standards and conventions to use in the \gls{sgs} related to 
\begin{itemize}
	\item Time-to-frequency transformation
    \item Gravitational wave source parametrization
    \item LISA instrument response to gravitational waves
    \item TDI definitions and labelling
    \item Definition of reference frames including inertial, constellation, and source frames
    \item Stochastic gravitational waves
\end{itemize}

This is meant to be a living document in the sense that modifications may be made in the course of the \gls{ddpc} project.

The document provides standards to exchange information among coordination units and  sub-groups within the \gls{sgs}. It lays common grounds for source populations, data simulations, data analysis, and catalogue makers. 
Note that we are aware that some conventions are more suitable for certain applications than for others. In some cases, instead of imposing one reference convention, we decided to provide several definitions (depending for example on the gravitational wave source type) along with transformations to go from one definition to the other.

The conventions explicited in this document either derive from \gls{esa} applicable documents listed below (especially \hyperlink{AD1}{AD1}) or are specific to the \gls{sgs} and are introduced only here.

\newpage

\section{Fourier transform and spectral densities}\label{sec:FT}

\subsection{Fourier transform conventions}
We define the Fourier transform (FT) operator $\mathcal{F}[\, \cdot\, ]$ on a time-domain function $F$ via $\mathcal{F}[F](f) = \widetilde{F}(f)$ as
\begin{equation}
    \mathcal{F}[F](f) = \widetilde{F}(f) = \int_{-\infty}^{+\infty}{F(t)e^{-i 2 \pi f t} \dd{t}}\,,
    \label{eq:defFourier}
\end{equation}
with associated Inverse Fourier transform (IFT) 
\begin{equation}
    \mathcal{F}^{-1}[\widetilde{F}](t) = F(t) = \int_{-\infty}^{+\infty}{\widetilde{F}(f) e^{i 2 \pi f t} \dd{f}}\,.
    \label{eq:IFTdefFourier}
\end{equation}
\subsection{From continuous to discrete signals}
\subsubsection{Time domain representation}
Let $F(t)$ be a continuous function (which may be real or complex valued). Its discrete-time representation is defined by sampling at times 
\begin{equation}\label{eq:def_sampling_interval}
t_{n} = n\Delta t\,, \qquad n = 0, 1, \ldots N -1
\end{equation}
such that 
\begin{equation}
    F_n = F(t_n) = F(n \Delta t)\,.
\end{equation}
The total duration of the signal is defined by $T = N\Delta t$ with sampling rate $f_{s}$
\begin{equation}\label{eq:def_sampling_rate}
f_{s} = \frac{1}{\Delta t}\,.
\end{equation}
\subsubsection{Frequency domain representation}

Discretising both Eq.~\eqref{eq:defFourier} and Eq.~\eqref{eq:IFTdefFourier} results in the discrete Fourier transform (DFT) and the inverse discrete Fourier transform (IDFT) respectively
\begin{align}\label{eq:DFT}
    \widetilde{F}_k &= \Delta t\sum_{n = 0}^{N-1}{F_n e^{-i 2 \pi f_{k} t_{n}}} = \Delta t\sum_{n = 0}^{N-1}{F_n e^{-i 2 \pi k n / N}}\,,\\
    F_n &= \Delta f \sum_{k = k_\mathrm{min}}^{k_\mathrm{max}}{\widetilde{F}_k e^{i 2 \pi f_{k} t_{n}}} = \Delta f \sum_{k = k_\mathrm{min}}^{k_\mathrm{max}}{\widetilde{F}_k e^{i 2 \pi k n / N}} \label{eq:DFT}\,.
\end{align}
The sampling frequencies $f_k = k\Delta f$ for frequency resolution $\Delta f =1/(N\Delta t)$ are associated with the DFT entries $\widetilde{F}_k$. The indices $k$ are defined on the set $k \in \{k_\mathrm{min}, \ldots, -1, 0, 1, \ldots, k_\mathrm{max}\}$ with $k_\mathrm{min}=-\bigl \lfloor N/2 \bigl \rfloor$ and $k_\mathrm{max}=\bigl \lfloor N/2 \bigl \rfloor$ if $N$ is odd and $N/2-1$ if $N$ is even. In the case of a real signal, we have the symmetry $\widetilde{F}(-f) = \widetilde{F}(f)^{\star}$ implying that the DFT is completely determined by one-side of the spectrum consisting of $\bigl \lfloor N/2 \bigl \rfloor +1 $ components with frequency indices are $k \in [0, k_{\text{max}}]$. 

\subsection{Power Spectral Density}
Let $F(t)$ now be a (weakly) stationary continuous stochastic process with some (for now arbitrary) distribution. The time-domain auto-covariance function $C_{F}$ of the process $F$ is then only determined by the lag $\tau = |t_{2} - t_{1}|$. 
\begin{equation}
\mathbb{E}_{F}[F(t)F(t+\tau)] = C_{F}(\tau)
\end{equation}
where the expectation is taken under the data generating process that determines $F$. We can relate the auto-covariance function $C_{F}$ of some process $F$ to the \gls{psd} of the process defined via $S_{F}(f)$~\cite{wiener1930generalized,khintchine1934korrelationstheorie}
\begin{equation}
\mathcal{F}(C_{F}(\tau)) = \int_{-\infty}^{+\infty}C_{F}(\tau)\exp(-2\pi i f\tau)\dd\tau = \frac{1}{2}S_{F}(f)\,,
\end{equation}
where the factor of 1/2 is a convention choice. Here we denote $S_{F}(f)$ as the \emph{one-sided} \gls{psd} defined over the positive frequency components $f\geq 0$. 

The amplitude spectral density (ASD) is defined as the square root of the \gls{psd}, i.e.,
\begin{equation}
S_{F}^{1/2}(f) = \sqrt{S_{F}(f)}.
\end{equation}


\subsection{Relation to some existing software}

The convention presented in Eq.~(\ref{eq:DFT}) agrees with the ones used in LALSuite~\cite{lal}\footnote{See \href{https://lscsoft.docs.ligo.org/lalsuite/lal/group___time_freq_f_f_t__h.html}{\texttt{https://lscsoft.docs.ligo.org/lalsuite/lal/group\_\_\_time\_freq\_f\_f\_t\_\_h.html}}.}.

Note that the \texttt{numpy.fft} implementation (with default normalization parameter) and \texttt{Matlab}'s \texttt{fft} function  define the DFT and IDFT conventions similarly to Eq.~(\ref{eq:DFT}) but without the $\Delta t$ factor for the DFT and by replacing the $\Delta f$ factor by $1/N$ for the IDFT. The ordering of indices for \texttt{numpy.fft} is worth mentioning, with the zeroth frequency bin the first index $k_{0} = 0$, the next half $k_{\text{pos}} \in \{1,\ldots,N/2 - 1\}$ the positive frequency spectrum and $k_{\text{neg}}\in\{N/2,\ldots, N-1\}$ the negative frequency spectrum (if required).

Note that \texttt{fftw} follows another convention, where the backward transform does not have the $1/N$ leading factor.

\section{Physical source parameters}

\subsection{Universe}\label{sec:universe}

The standard cosmology that will be assumed is Planck15 \cite{Planck:2015fie} as implemented in {\tt astropy} \cite{price2018astropy}.
The luminosity distance will be denoted as $D_L$.

 Choices for values of mission parameters and constants on the \gls{pdb} according to \gls{sird} requirements are defined via the {\tt lisaconstants} software package \cite{bayle_2022_6627346}.
Support is provided for the programming languages Python, C, and C++ (for the latter two as header files).
\texttt{LISA Constants} is intended to be used consistently by other pieces of software related to the simulation of the instrument, of gravitational wave signals, and others.

\subsection{Sky coordinates}\label{sec:sky_coordinates}

We define sky coordinates in the coordinate system attached to an inertial reference frame centred on the \gls{ssb}. Preferably, we adopt the coordinate system attached to the (equatorial) \gls{icrf}, as described in Sec.~\ref{sec:equatorial_frame}:
\begin{itemize}
    \item right ascension $\alpha$,
    \item declination $\delta$.
\end{itemize}

However, historically, sky coordinates have been defined as the ecliptic longitude $\lambda$ and ecliptic latitude $\beta$ attached to the ecliptic reference frame centered on the \gls{ssb}. We provide a definition of this frame in Sec.~\ref{sec:ecliptic_frame}, together with how to transform from the equatorial coordinate system to the ecliptic coordinate system.

\subsection{Parameters of a binary system}

Masses will be given in the source frame. In the detector frame, the masses (and other dimensionful quantities) will be redshifted as $m_{det} = (1+z) m_{source} $, where $z$ is the redshift. An analogous scaling also applies to other dimensionful quantities.

The 10 {\em intrinsic} parameters of the binary system, which do not depend on the location or orientation of the binary in the universe, are the following:

\begin{itemize}
\item Component masses:               $m_1 \geq m_2$.
\item Component dimensionless spin vectors:   $\chi_1$, $\chi_2$.
\item Eccentricity and mean anomaly for eccentric systems:  $e$ and $l$.
\end{itemize}


For quasi-circular systems the eccentricity $e$ vanishes. In the case of equatorial symmetry, the spins are orthogonal to the orbital plane, and the orbital plane and their directions are preserved.
In general, the eccentricity, mean anomaly and direction of the spin vectors show 
a significant time dependence, and for purposes of parameterization need to be specified at some reference time, see Sec.~\ref{sec:reference_time}. At least in the case of black holes in vacuum, the masses and spin magnitudes exhibit only a very weak time dependence due to the influx of a small amount of gravitational waves into the black holes, and this time dependence is often neglected.

We also define several derived quantities:
\begin{itemize}
    \item The total mass is defined as $M = m_1 + m_2$.
    \item The dimensionful angular momentum of the objects is $S_i = \frac{G}{c} \chi_i m_i^2 $.
    \item The mass-ratio will be denoted as $q=m_2/m_1 \leq 1$ or equivalently $Q=m_1/m_2 \geq 1$.
    \item The reduced mass is $\mu = m_1m_2/(m_1+m_2)$.
    \item The symmetric mass-ratio is $\eta= \mu/M = m_1m_2/(m_1+m_2)^2$.
    \item The chirp mass is ${\cal M} = (m_{1} m_{2})^{3/5}/(m_1+m_2)^{1/5}$.
\end{itemize}

Some notes are in order: 

The symmetric mass ratio is not to be confused with a quantity $\eta_{ij}$ that is used to label laser combinations on the optical bench, see Eq.~\ref{def:laser_eta}. 

The definitions of many standard quantities can be ambiguous in general relativity, which is often related to the lack of a natural coordinate gauge.
We do not resolve such ambiguities here, but note that further specifications may be required:
e.g. a precise definition of even the spin magnitude requires a  spin supplementary condition (see e.g.~\cite{Schafer:2018jfw}). In the absence of symmetry (e.g. equatorial symmetry), the angles of the spins are coordinate-dependent. The definition of 
eccentricity and mean anomaly in the strong field regime also suffers from ambiguities,  see e.g. the discussion in \cite{Shaikh2023}, where
definitions are provided 
in terms of the gravitational wave signal, which are free of gauge ambiguities and have the correct Newtonian limit. For specific definitions of intrinsic parameters, in particular, those related to eccentricity and spin precession, it is recommended to supply information on gauge conditions whenever relevant and to provide information on how definitions relate to the standard definitions, e.g. those of Newtonian physics or post-Newtonian expansions, when the separation of the binary components is sufficiently large.
For simple definitions of eccentricity based on the orbit see Secs.~
\ref{sec:L_frame} and \ref{sec:EMRI_frame} below.

We also define the {\em extrinsic} parameters of the binary system, which describe the  location and orientation of the binary system in the Universe. These parameters are: the luminosity distance $D_L$, see Sec.~\ref{sec:universe}, the sky coordinates, see Sec.~\ref{sec:sky_coordinates},
the inclination angle $\iota$ (note however that alternative definitions are possible, see Sec.~\ref{sec:reference_frames} for details) between the observers line of sight and the orbital angular momentum or angular velocity, the polarization angle $\psi$, see Sec.~\ref{sec:equatorial_frame}, and the reference time and phase. The phase and inclination angles are defined at some reference time or frequency, similar to the spins, eccentricity, and mean anomaly, see also Sec.~\ref{sec:reference_time}.

For a detailed discussion of different source frames and their relation to the observer's frame see Sec.~\ref{sec:source_frames}.


\section{Instrument response}

\subsection{Metric signature and metric fluctuation}

While the physics is fully independent of the metric signature, it is important to choose one common metric signature that needs to be consistently used everywhere in the data modelling and data analysis. Here, we will use the mostly + signature, i.e. the metric signature is
\begin{equation}\label{eq:metric}
	\left(-,+,+,+\right).
\end{equation}
Note that the metric signature can be generally written as $\varepsilon_g \left(-,+,+,+\right)$, where $|\varepsilon_g|=1$. Therefore, the \gls{sgs} convention adopts $\varepsilon_g = + 1$.

The metric deviation $h_{\mu\nu}$ is defined as
\begin{equation}\label{eq:h}
    g_{\mu\nu}=\eta_{\mu\nu}+h_{\mu\nu}\, ,
\end{equation}
where $\eta_{\mu\nu}=\mathrm{diag}\left(-1,1,1,1\right)$ is the Minkowski metric.

\subsection{Element indexing}

Spacecraft are indexed clockwise when looking down from the ecliptic North (i.e. when looking down at the solar panels). Spacecraft 1 is the \textit{reference spacecraft}. The reference spacecraft refers to the spacecraft on top of the stack, and the first to be separated from the upper stage(we do not distinguish individual spacecraft). At the time of writing, this definition requires confirmation. These definitions match \gls{esa} conventions. 

\begin{figure}
    \centering
    \includegraphics[width=0.6\textwidth]{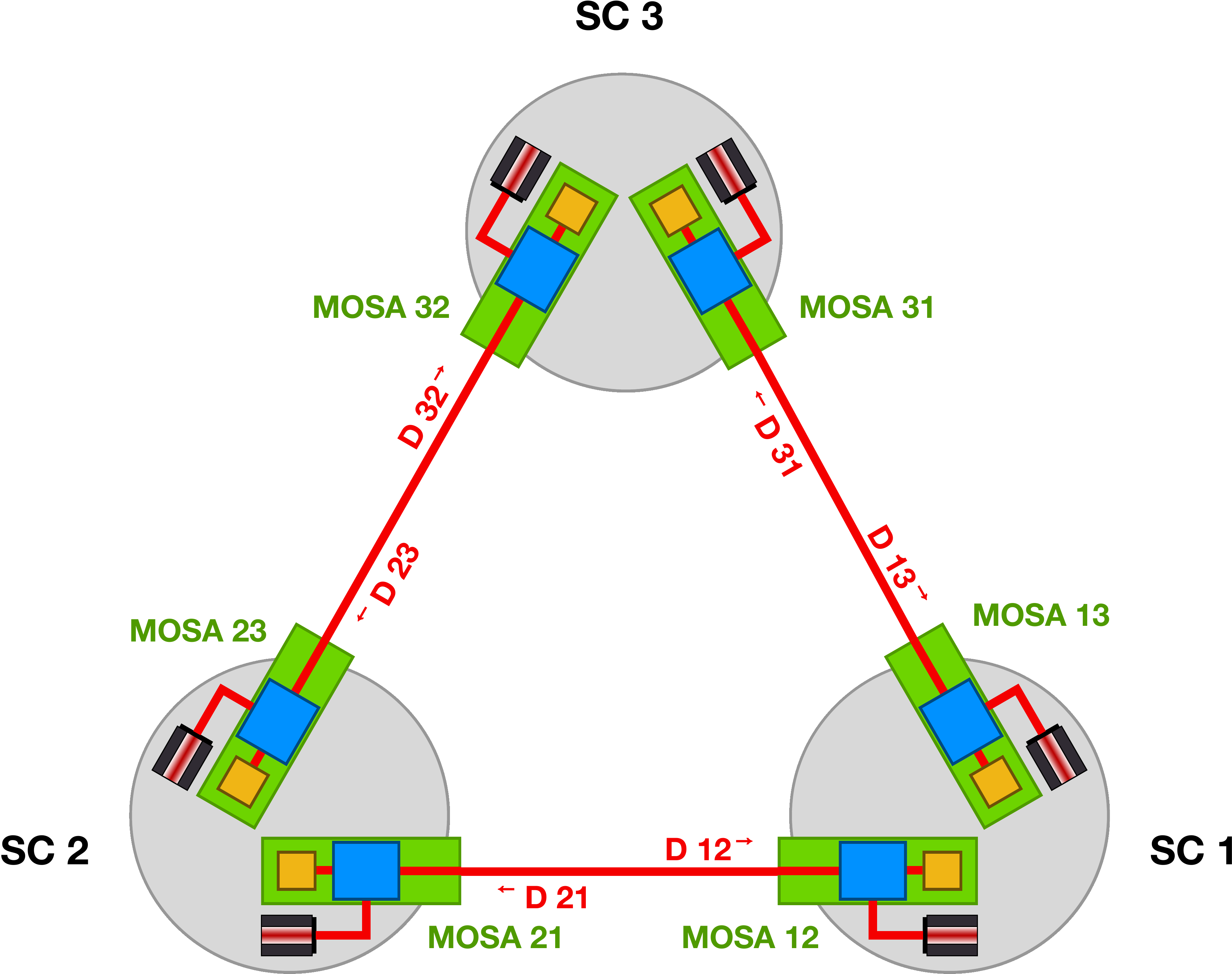}
    \caption{Indexing conventions. Figure extracted from~\cite{Bayle:2021mue}, based on~\protect\hyperlink{AD1}{AD1}.}
    \label{fig:indexing}
\end{figure}

The \glspl{mosa} are labelled with two indices~$ij$ as shown in Fig.~\ref{fig:indexing}. The former matches the index~$i$ of the spacecraft hosting the \gls{mosa} (the local spacecraft), while the second index is that of the spacecraft~$j$ exchanging light with the considered \gls{mosa} (also called distant spacecraft). Any subsystem or quantity uniquely attached to a spacecraft or a \gls{mosa} will be labelled according to the latter. For example, the reference interferometer on the optical bench~$ij$ will be indexed~$ij$.

The first index~$i$ denotes the \textit{receiving} spacecraft and the second index~$j$ denotes the \textit{emitting} spacecraft. For example,
\begin{equation}
    \vb{r}_{ij} = \vb{x}_i - \vb{x}_j.
\end{equation}
The same convention is used to denote \gls{ltt} $L_{ij}$ from spacecraft~$j$ to~$i$, which can only be measured on optical bench~$ij$.

\subsection{Interferometric measurements}

The optical bench $ij$ contains 3 interferometers,
\begin{itemize}
    \item The science interferometer $\textrm{SCI}_{ij}$ (also referred to as inter-spacecraft interferometer), monitoring the distance between two optical benches, which is the only interferometer containing gravitational-wave signals,
    \item The test-mass interferometer $\textrm{TMI}_{ij}$, monitoring the distance between the test mass and the optical bench, which is the only interferometer with a beam reflected on the test mass,
    \item The reference interferometer $\textrm{REF}_{ij}$, comparing two lasers hosted on the same spacecraft.
\end{itemize}

The heterodyne superposition of modulated beams produces multiple oscillating components in the \si{\mega\hertz} frequency range, which we call beatnotes. In particular, we often refer to the carrier or upper and lower sideband beatnotes, denoted $\textrm{SCI}_{\textrm{c}, ij}$, $\textrm{SCI}_{\textrm{usb}, ij}$, and $\textrm{SCI}_{\textrm{lsb}, ij}$ (and similar for other interferometers).

In the following sections, we describe how the gravitational-wave signals appear in the science interferometers; we leave out the noise terms. For a complete description of the content of these interferometric measurements, refer to \cite{Bayle:2022okx,Perf_Model_Description_Doc}.

Note that, because of laser locking (adjusting the laser frequencies so that certain beatnotes vanish), the gravitational-wave signals may not appear in all science interferometers, but instead are \textit{folded in} the non-locking beatnotes. Similarly, noises are folded in non-locking beatnotes.

\subsubsection{Sign of the beatnote}
\label{sec:instru_response_ifo_sign_beatnote}
The phasemeters measure the absolute value of the beatnote phase, i.e., the absolute value of the difference between the phase of the laser beam coming from the distant spacecraft and that of the laser beam from the local optical bench.

The exact frequencies of the laser beams vary by tens of MHz during mission time but are actively controlled to follow a pre-computed \textit{Offset Lock Planning}, also referred to as a frequency plan. As a consequence, we know \textit{a priori}, at any time, which beam has the higher frequency (the beatnote polarity) in each interferometer.

To simplify all subsequent analyses, the currently baseline is to correct for this beatnote polarity by adding the correct sign as part of the L0-L1 processing. As a consequence, all beatnote phase (or frequencies) can be expressed as the difference between two laser beam phases (or frequencies), following a single convention.

The choice of beatnote sign is conventional. We decide to define it such that the beatnote phase (in all interferometers) is the difference between the remote beam (corresponding to the distant beam for the science interferometers, and the adjacent beam for the test-mass and reference interferometers) and the local beam phases.

In particular, for the science interferometer,
\begin{equation}
    \Phi_{ij}(t) = \Phi_j\left(t-L_{ij}(t)\right)-\Phi_i(t),
    \label{eq:BN_phase}
\end{equation}
where $\Phi_j(t)$ is the phase (expressed in rad) of the emitting (distant) laser, $\Phi_i(t)$ is the phase of the receiving laser and $L_{ij}$ is the LTT between the two S/C. In this expression, $t$ is the time of the receiving S/C.

The current LISA Simulation tools uses this convention~\cite{Bayle:2022okx}. This convention is never specified in any LISA Data Analysis software. The ESA Convention document~\hyperlink{AD1}{AD1} aligns with this convention for the \gls{tmi}; nothing is specified for the \gls{sci} and \gls{ref}.

\Cref{eq:BN_phase} defines the SCI beatnote in terms of phase. Differentiating this equation leads to the definition of the beatnote in terms of frequency,
\begin{align}
    \nu_{ij}(t) &= \frac{1}{2\pi} \dv{\Phi_{ij}}{t}
    = \frac{1}{2\pi}\Bigg[\left. \dv{ \Phi_e}{t}\right|_ {t - L_{ij}(t)} \times  \left(1 - \dv{L_{ij}}{t} \right) -\left.\dv{ {\Phi}_i}{t}\right|_{t}\Bigg] \nonumber
    \\
    &= \frac{1}{2\pi}  \dot \Phi_j\left(t-L_{ij}(t)\right) \times  \left(1 - \dot{L}_{ij}(t) \right) - \frac{1}{2\pi} \dot \Phi_i(t)\nonumber\\
    &= \nu_j \left(t-L_{ij}(t)\right)\times  \left(1 - \dot{L}_{ij}(t) \right)- \nu_i(t)
\end{align}
where $\nu_j$ is the local frequency of the laser emitting the signal from the distant spacecraft at the time of emission, and $\nu_i$ is the local frequency of the laser on the receiving spacecraft. 

To identify the contribution from GW on this observable, let us decompose the LTT into a a slowly evolving part $L_{ij}^{(0)}$ and an in-band contribution due to GW: $\delta L_{ij}$. Introducing this decomposition into the last expression leads to
\begin{align}
	\nu_{ij}(t) &\approx \nu_j \left(t-L^{(0)}_{ij}(t)\right)\times  \left[1 - \dot{L}^{(0)}_{ij}(t) \right]- \nu_i(t) -{\delta \dot L}_{ij} \nu_j\left(t-L^{(0)}_{ij}(t)\right) \, ,
\end{align}
where terms proportional to the derivative of the laser frequency have safely been neglected. The in-band fluctuation due to the \gls{gw} is therefore given by $-{\delta \dot L}_{ij} \nu_j$, such that the relative frequency fluctuations are defined by
\begin{equation}
    y_{ij}(t)= -{\delta \dot L_{ij}}.
\end{equation}

Note that this corresponds to a decrease of the beatnote frequency for an increase of the optical path length. Consequently, the beatnote phase decreases if the optical pathlength increases. To obtain a calibrated length readout signal (sometimes called longitudinal pathlength signal, LPS which will be denoted also as $x$ in the equation below), it is, therefore, necessary to account for the sign and use
\begin{equation}
    \text{LPS}_{ij} (t) = x_{ij}(t) =  - \frac{\lambda}{2\pi} \Phi_{ij}(t)\,,
    \label{eq:signed-LPS}
\end{equation}
where $\lambda$ is the laser central wavelength.

\subsection{One-link response function}

The quantity  $\delta \dot L_{ij}$ which appears in the expression of the beatnote can be expressed, to first order, as a function of the metric perturbation (defined by Eq.~(\ref{eq:h})), here considered in the plane-wave approximation, $h_{ij}(t,\mathbf{x})=h_{ij}\left(t-\frac{1}{c} \mathbf{\hat{k}}\cdot\mathbf{x}\right)$ (where is the GW direction of propagation and $\mathbf{x}$ the position) as \cite{cornish:2009aa}: 
\begin{equation}
     \delta \dot L_{ij} =  \frac{1}{2}\frac{H_{ij}(\xi_i)-H_{ij}(\xi_j)}{1-\mathbf{\hat k} \cdot \mathbf{\hat x_{ij}}} \, ,
\end{equation}
which depends on the sign convention for the space-time metric. We defined the variable $\xi_{i}(t) \equiv t - \frac{1}{c} \mathbf{\hat{k}}\cdot\mathbf{x}_{i}(t)$ where  $\mathbf{x}_{i}$ is the position vector of the S/C $i$. In addition  $\xi_i \equiv \xi_i(t)$ while $\xi_j \equiv \xi_j(t_j) = \xi\left(t - L_{ij}(t)\right)$. Finally, in the last equation
\begin{equation}
	H_{ij}\left(\xi \right) = \left[ \mathbf{\hat x}_{ij} \otimes \mathbf{\hat x}_{ij} \right] : \mathbf{h}(\xi) = \hat x_{ij}^k \hat x_{ij}^l h_{kl}(\xi) \, ,
\end{equation}
where 
\begin{equation}
	\mathbf{x}_{ij} = \mathbf x_i(t) -\mathbf x_j\left(t-L_{ij}\right) \, \quad 	\mathbf{\hat x}_{ij} =\frac{\mathbf{x}_{ij} }{\left|\mathbf{x}_{ij} \right|}\, .
\end{equation}

As a consequence,\footnote{Note that the response depends on both the beatnote definition and the metric signature $\varepsilon_g$ although this is not explicited here.}
\begin{equation}\label{eq:y}
    y_{ij}(t) = -\frac{1}{2}\frac{H_{ij}(\xi_i)-H_{ij}(\xi_j)}{1-\mathbf{\hat k} \cdot \mathbf{\hat x_{ij}}}\, .
\end{equation}

\section{Time-delay interferometry}
\label{sec:tdi}

The following definitions are compatible with the ESA P\&O Performance Budget Technical Note and in~\hyperlink{AD2}{AD2}.

\subsection{Generalities}

\Gls{tdi} combinations are linear combinations of time-shifted measurements. They were initially introduced to reduce the otherwise-overwhelming laser noise in the raw interferometric measurements.

The current baseline is to first compute intermediary variables, which take advantage of the split interferometry optical design to suppress spacecraft jitter and reduce the problem from 6 to 3 lasers~\cite{Otto:2015erp,Hartwig:2021dlc}. Using these results as inputs, a variety of laser noise-reducing combinations can then be computed, including variables synthesizing Sagnac and Michelson-like interferometers.

The space of possible laser noise-suppressing combinations depends on orbital assumptions. In the case of a non-rotating rigid constellation, we use first-generation \gls{tdi} (and the space is algebraically described by a set of 4 generators~\cite{tintoTimedelayInterferometry2020}). We call second-generation combinations, those which can reduce laser noise according to requirements in a realistic orbital setup\footnote{Note that various definitions for these \gls{tdi} exist in the literature. We choose here a practical definition, based on the effective arm-length mismatch. Refer to~\cite{Muratore:2020mdf} for discussion.}.

Lastly, these laser noise-free combinations can be further combined into quasi-orthogonal channels~\cite{Prince:2002hp} (see Sec.~\ref{sec:tdi-aet}).

\subsection{Delay operators}
\label{sec:delay_operators}
We define the delay operator $\delay_{ij}$ by its action on a time series $x(t)$,
\begin{equation}
    \delay_{ij} x(t) = x(t - L_{ij}(t)),
\end{equation}
where $L_{ij}(t)$ is the light travel time\footnote{In the final L1 data, the delays represent the \glspl{ltt} in the \gls{bcrs}. Note that the delays actually applied inside the L0-L1 pipelines need to account for the desynchronization of the spacecraft clocks and might differ from these \glspl{ltt}~\cite{Hartwig:2022yqw}. This should be transparent to the end user of the data.} from spacecraft $j$ to spacecraft $i$, at reception time $t$.

One can chain delay operators, applying them from right to left. For two delays, we obtain,
\begin{equation}
    \delay_{ij} \delay_{kl} x(t) = \delay_{ij} x(t - L_{kl}(t)) = x(t - L_{ij}(t) - L_{kl}(t - L_{ij}(t))).
\end{equation}
From this result, we trivially deduce that delay operators do not commute in general. Expressions for an arbitrary number of chained delay operators can be found in the literature~\cite{Bayle:2018hnm}. Under the approximation that the relevant light travel times are constant, we can commute the operators.

For conciseness, we introduce the following shorthand notation for chained delay operators, when the indices also chain up,
\begin{equation}
    \delay_{i_1 i_2 \dots i_n} = \delay_{i_1 i_2} \delay_{i_2 i_3} \dots \delay_{i_{n-1} i_n}.
\end{equation}

Note that the \gls{tdi} combinations defined in this section can be computed from data expressed as a total phase, phase fluctuations, total frequency, or frequency fluctuations. In the two latter cases, one should make sure to replace the usual delay operator $\delay_{ij}$ by the Doppler-shifted delay operator $\dot \delay_{ij}$~\cite{Bayle:2021mue},
\begin{equation}
    \dot \delay_{ij} x(t) = (1 - \dot L_{ij}(t)) \times \delay_{ij} x(t),
\end{equation}
where $\dot L_{ij}(t)$ is the time derivative of the light travel time along arm $ij$.

\subsection{Spacecraft jitter reduction}

The $\xi_{ij}$ combinations are constructed to reduce spacecraft jitter. They combine the inter-spacecraft beatnote with the difference of reference and test-mass beatnotes to construct a virtual test-mass to test-mass measurement. They are written as

\begin{equation}
    \xi_{12} = \text{sci}_{12} + \frac{\text{ref}_{12} - \text{tmi}_{12}}{2} + \frac{\delay_{12}(\text{ref}_{21} - \text{tmi}_{21})}{2}.
\end{equation}

The expressions for all other 6 optical benches can be deduced by applying the rotation and reflection of the indices.

\subsection{Reduction to 3 lasers}
\label{sec:Reduction_to_3_lasers}

The $\eta_{ij}$ combinations remove the noise of half the lasers in the constellation.

\begin{equation}\label{def:laser_eta}
    \eta_{12} = \xi_{12} + \frac{\delay_{12}(\text{ref}_{21} - \text{ref}_{23})}{2}
    \qand
    \eta_{13} = \xi_{13} + \frac{\text{ref}_{12} - \text{ref}_{13}}{2}.
\end{equation}
The expressions for the other spacecraft can be deduced from cyclic index permutation.

\subsection{Sagnac combinations}

First and second-generation Sagnac combinations $\alpha_i$, $\beta_i$, $\gamma_i$ (with $i=1, 2$) synthesize the interference of photons circulating clockwise and counterclockwise the constellation.

The fully symmetric first-generation Sagnac combination $\zeta_1$ combines all measurements with exactly one delay.

Note that $\alpha_1$, $\beta_1$, $\gamma_1$, $\zeta_1$ are generators of the first-generation \gls{tdi} combination space~\cite{Tinto:2002de}. Therefore, all first-generation combinations can be written as a linear combination of the latter.

\subsubsection{First generation}

The first Sagnac combination is given by
\begin{equation}
    \alpha_1 = \eta_{13} + \delay_{13} \eta_{32} + \delay_{132} \eta_{21} - (\eta_{12} + \delay_{12} \eta_{23} + \delay_{123} \eta_{31}).
\end{equation}
with $\beta_1$ and $\gamma_1$ given by circular permutation of the indices. Note that we fix the sign convention somewhat differently than most papers in the literature, so that we are consistent with the definition of Michelson combinations (i.e., $\eta_{13}$ appears as the positive-signed unshifted measurement).

The fully-symmetric Sagnac combination reads
\begin{equation}
    \zeta_1 = \delay_{23} (\eta_{13} - \eta_{12})  + \delay_{12} (\eta_{32} - \eta_{31}) + \delay_{31}( \eta_{21} -  \eta_{23}).
\end{equation}
Note that this variable is strictly only defined for first-generation \gls{tdi}, i.e., a static non-rotating constellation with $\delay_{ij} = \delay_{ji}$. Under that assumption, $\zeta_1$ is fully symmetric, and the usual \gls{lisa} transformations (circular permutation of indices and reflections) only yield the same combination up to a sign. We fix here the sign convention.

\subsubsection{Second generation}

The first Sagnac combination is given by
\begin{equation}
    \alpha_2 = \alpha_1 + \delay_{1321} \eta_{12} + \delay_{13212} \eta_{23} + \delay_{132123} \eta_{31} - (\delay_{1231} \eta_{13} + \delay_{12313} \eta_{32} + \delay_{123132} \eta_{21}).
\end{equation}
with $\beta_2$ and $\gamma_2$ given by circular permutation of the indices. Again, we flipped the sign with respect to most of the literature to preserve consistency in our conventions.

The original second-generation version for the fully-symmetric Sagnac combination proposed in \cite{Tinto:2003vj} has been shown to not suppress laser noise to the same level as other second generation variables, but alternatives exist. We choose here the variable labelled $C_{27}^{16}$ in~\cite{Hartwig:2021mzw}, for which we further adjust the sign and overall time-shift to define 
\begin{equation}
\begin{split}
    \zeta_2 ={}& \delay_{23} [
    (1-\delay_{13123}\adv_{31})(\eta_{13} - \eta_{12})
    \\ 
    &+(\delay_{131} -\delay_{12}\adv_{23}\delay_{31}\adv_{12}\delay_{231})\eta_{12}
    \\
    &+ (\delay_{12}\adv_{23}-\delay_{13123}\adv_{31}\delay_{12}\adv_{23})(\eta_{32} - \eta_{31})
    \\
    &+ (\delay_{13} -\delay_{12}\adv_{23}\delay_{31}\adv_{12}\delay_{23})\eta_{31}
    \\
    &+(\delay_{12}\adv_{23}\delay_{31}\adv_{12} - \delay_{12}\adv_{23}\delay_{31}\adv_{12}\delay_{2312})\eta_{21}
    \\
    & - (\delay_{12}\adv_{23}\delay_{31}\adv_{12} - \delay_{1312})\eta_{23}]
\end{split}
\end{equation}
Note that this variable uses both delays and their inverse advancement operators, defined via 
\begin{equation}
    \delay_{ij} \adv_{ji} = \adv_{ji} \delay_{ij} = 1.
\end{equation}
$\zeta_2$ as defined above remains approximately fully symmetric, in that it simplifies to
\begin{equation}
    \zeta_2 \approx (1 - \delay_{1231}) \zeta_1
\end{equation}
when assuming $\delay_{ij} = \delay_{ji}$. Similarly, we have
\begin{equation}
    \alpha_2 \approx (1 - \delay_{1231}) \alpha_1
    \qcomma
    \beta_2 \approx (1 - \delay_{1231}) \beta_1
    \qcomma
    \gamma_2 \approx (1 - \delay_{1231}) \gamma_1.
\end{equation}

\subsection{Michelson combinations}

Michelson combinations synthesize a virtual Michelson interferometer to reduce laser noise in a 3-laser configuration. Note that the combinations' signs are arbitrary; we fix the conventions here.

\subsubsection{First generation}

The first generation Michelson combination $X_1$ is given by
\begin{subequations}
\begin{align}
    X_1 &= (1 - \delay_{121})(\eta_{13} + \delay_{13} \eta_{31}) - (1 - \delay_{131}) (\eta_{12} + \delay_{12} \eta_{21}),
    \\
    X_1 &= \eta_{13} + \delay_{13} \eta_{31} + \delay_{131} \eta_{12} + \delay_{1312} \eta_{21} - (\eta_{12} + \delay_{12} \eta_{21} + \delay_{121} \eta_{13} + \delay_{1213} \eta_{31}).
\end{align}
\end{subequations}
with $Y_1$ and $Z_1$ given by circular permutation of the indices.

We can decompose $X_1$ in terms of Sagnac combinations~\cite{Hartwig:2021mzw},
\begin{equation}
    X_1 = \alpha_1 - \delay_{12} \beta_1 - \delay_{13} \gamma_1 + \delay_{13} \delay_{12} \zeta_1.
\end{equation}

\subsubsection{Second generation}

The second generation Michelson combination $X_2$ is given by
\begin{subequations}
\begin{align}
    \begin{split}
        X_2 ={}& (1 - \delay_{121} - \delay_{12131} + \delay_{1312121})(\eta_{13} + \delay_{13} \eta_{31})
        \\
        &- (1 - \delay_{131} - \delay_{13121} + \delay_{1213131}) (\eta_{12} + \delay_{12} \eta_{21}),
    \end{split}
    \\
    \begin{split}
        X_2 ={}& X_1 + \delay_{13121} \eta_{12} + \delay_{131212} \eta_{21} + \delay_{1312121} \eta_{13} + \delay_{13121213} \eta_{31}
        \\
        &- (\delay_{12131} \eta_{13} + \delay_{121313} \eta_{31} + \delay_{1213131} \eta_{12} + \delay_{12131312} \eta_{21}).
    \end{split}
\end{align}
\end{subequations}
with $Y_2$ and $Z_2$ given by circular permutation of the indices.

Note that, under the assumption of equal light travel times, we can write
\begin{equation}
    X_2 = (1 - \delay^4) X_1.
\end{equation}

\subsection{\label{sec:tdi-aet}Orthogonal combinations}

The orthogonal combinations are linear combination of the first or second-generation combinations, in which noises are uncorrelated (under simplifying assumptions). Their definition is not unique, and we fix here the conventions.

\begin{equation}
    A_i = \frac{Z_i - X_i}{\sqrt{2}}
    \qcomma
    E_i = \frac{X_i - 2Y_i + Z_i}{\sqrt{6}}
    \qcomma
    T_i = \frac{X_i + Y_i + Z_i}{\sqrt{3}}.
\end{equation}
Here, $T_i$ is sometimes called the \textit{null channel} due to its suppressed sensitivity to gravitational-wave signals at low frequencies. Note that this property has been shown to be highly sensitive to small mismatches in the light travel times, and does not hold in a realistic instrumental setup with the above definitions~\cite{Adams:2010vc,Hartwig:2023pft}.

Note that one can define the same kind of orthogonal combinations starting from most sets of three base variables, including the aforementioned Sagnac combinations~\cite{Vallisneri:2007xa}.

\section{Reference frame definitions}\label{sec:reference_frames}

In this section, we describe the commonly used reference frames and how to transform from one to the other. We start with the closest frame to the detector, called LISA frame, and continue with the \gls{ssb} frame, then the wave frame, and finally the source frame.
We define a preferred source frame for all \gls{gw} sources. However, we also describe other frames that can be used to describe specific source types, along with transformations to translate from these source-specific frames to the preferred source frame.

\subsection{Constellation reference frame}


\begin{figure}[tbh]
	\centering
	\includegraphics[width=0.6\linewidth]{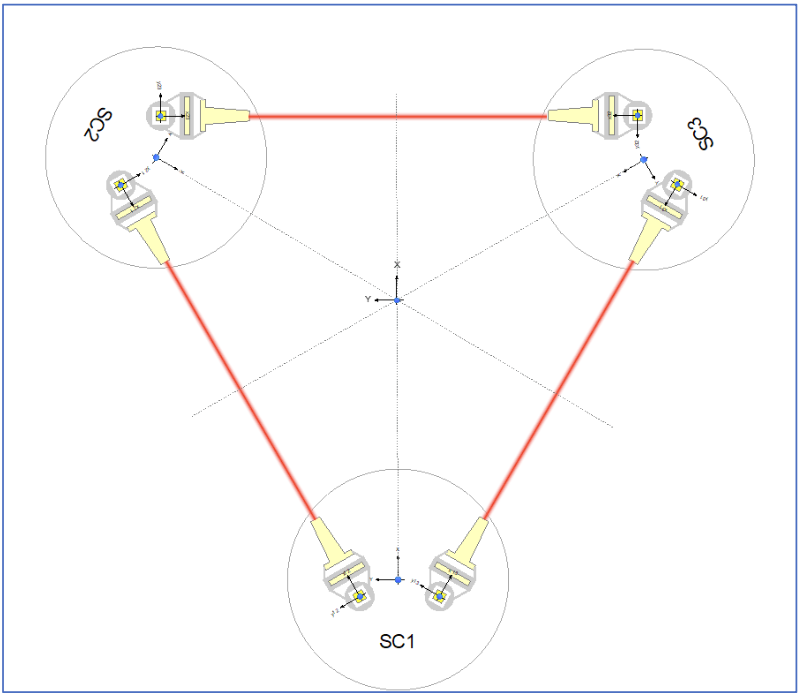}
	\caption{Representation of the constellatoin reference frame (source: \protect\hyperlink{AD1}{AD1}).}
	\label{fig:lisa-frame}
\end{figure}

The constellation reference frame, also called LISA frame, is useful for data analysis purposes. Here we follow the ESA Conventions document~\hyperlink{AD1}{AD1}.

The origin of the frame is located at the geometrical center of the constellation as depicted in Fig.~\ref{fig:lisa-frame}. Hence, if $\mathbf{x}_i$ is the location vector of \gls{sc} $i$ \gls{com} in the constellation frame, then $\mathbf{x}_1 + \mathbf{x}_2 + \mathbf{x}_3 = \mathbf{0}$.

The $X$ axis of the constellation frame is oriented from spacecraft \gls{sc} 1 along the bisector between the true instantaneous
directions of the distant spacecraft and towards the inside of the triangle formed by the 3 spacecraft.

The $Z$ axis of the constellation frame is perpendicular to the plane containing the 3 spacecraft \gls{com} and oriented such that the Sun has a positive component along the $Z$ axis.

The $Y$ axis of the constellation frame completes the right-handed set.

\subsection{Equatorial frame}\label{sec:equatorial_frame}
\glsreset{icrf}

We define the equatorial frame, also named LISA Default Inertial Reference Frame in~\hyperlink{AD1}{AD1}, as the \gls{icrf}. We label its axes $(\bm{x}_{0}, \bm{y}_{0}, \bm{z}_{0})$. We follow exactly the ESA Conventions document~\hyperlink{AD1}{AD1} for its definition.

The \gls{icrs} has its origin at the Solar system barycenter and has ``fixed'' axis directions. It is meant to represent the most appropriate coordinate system for expressing reference data on the positions and motions of celestial objects.

The \gls{icrf} is a realization of the \gls{icrs} using reference extragalactic radio sources observed with \gls{vlbi}. The \gls{icrf} is the inertial reference frame used for LISA's orbit propagations and mission analysis and is defined as follows.

The $\bm{z}_{0}$ axis is aligned close to the north celestial pole of J2000 but fixed to the radio sources
based reference frame. The celestial pole of J2000 is normal to the mean (precession model only, no
nutation) celestial equator of date at epoch 1 January 2000 at 12:00:00 TDB (Julian Day JD 2451545).

The $\bm{x}_{0}$ axis points in the direction that best aligns with the mean equinox of J2000 but is defined by
radio sources positions rather than Earth's motion. The J2000 (vernal) equinox, or Line of Aries direction, is the intersection of the equatorial and the ecliptic planes at epoch 1 January 2000 at 12:00:00 TDB (Julian Day JD 2451545).

The $\bm{y}_0$ axies completes the right-handed trihedron.

The \gls{icrf} is associated with a time metric, the \gls{tcb}.

The corresponding class definining the equatorial frame in \texttt{LISA Constants} is \textsf{lisaconstants.coords.LISAEquatorialFrame}. In Astropy v7.1.1, it is \textsf{astropy.coordinates.ICRS}\footnote{\url{https://docs.astropy.org/en/stable/api/astropy.coordinates.ICRS.html}}.

\begin{figure}[tbh]
  \centering
  \begin{tikzpicture}[scale=4,tdplot_main_coords]
    \coordinate (O) at (0,0,0);
    \draw[thick,->] (0,0,0) -- (1,0,0) node[anchor=north east]{$\bm{x}_0$};
    \draw[thick,->] (0,0,0) -- (0,1,0) node[anchor=north west]{$\bm{y}_0$};
\draw[thick,->] (0,0,0) -- (0,0,1) node[anchor=south]{$\bm{z}_0$};
    \tdplotsetcoord{P}{\rvec}{\thetavec}{\phivec}
    \tdplotsetcoord{r}{1.0}{\thetavec}{\phivec};
    \draw (r) node {.};
    \draw[thick,color=red] (P) -- (O) ;
    \draw[dashed, color=red] (O) -- (rxy);
    \draw[dashed, color=red] (r) -- (rxy);
    \tdplotdrawarc{(O)}{0.2}{0}{\phivec}{anchor=north}{$\alpha = \phi$}
    \tdplotsetthetaplanecoords{\phivec}
    \tdplotdrawarc[tdplot_rotated_coords]{(0,0,0)}{0.3}%
        {\thetavec}{90}{anchor=south west}{$\delta$};
    \tdplotdrawarc[tdplot_rotated_coords]{(0,0,0)}{0.2}{0}%
        {\thetavec}{anchor=south west}{$\theta$};
    \tdplotsetcoord{rth}{1.0}{\thetavec-12.0}{\phivec};
    \tdplotsetcoord{rph}{1.0}{\thetavec}{\phivec-20};
    \tdplotsetcoord{rr}{0.7}{\thetavec}{\phivec};
    \draw[-stealth, color=blue] (r)--(rth) node[above] {$ \bf{v}_0 = -\bm{e}_{\theta}$};
    \draw[-stealth, color=blue] (r)--(rph) node[below left] {$\bf{u}_0 = -\bm{e}_{\phi}$};
    \draw[-stealth,color=ForestGreen,line width=1pt] (r)--(rr) node[below right] {$\bf{k} = -\bm{e}_{r}$};
    \fill (r) circle[radius=0.3pt] node[anchor=south west]{};
    
    \draw[dashed,tdplot_rotated_coords] (1.0,0,0) arc (0:90:1.0);
    \draw[dashed] (1.0,0,0) arc (0:90:1.0);
    \draw[dashed] (1,0,1.19) arc (0:60:1.0);
     
\end{tikzpicture}
  \caption{Representation of the equatorial reference frame and its reference polarization vectors. The equatorial reference frame basis vectors $\bm{x}_0, \bm{y}_0, \bm{z}_0$ are in black. The \gls{gw} propagation vector $\bm{k}$ is in green. }
  \label{F:equatorial_reference_frame}
\end{figure}
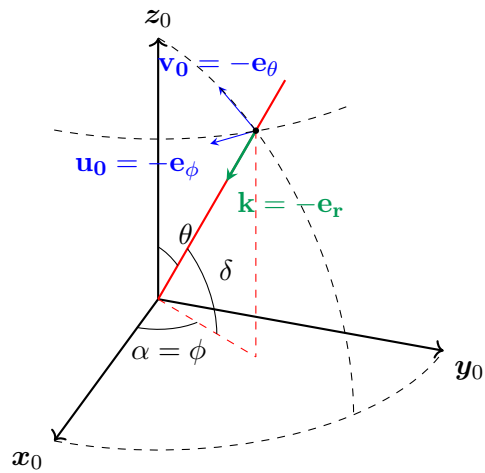

We associate an equatorial coordinate system based on right ascension $\alpha$ and declination $\delta$, as illustrated in Fig.~\ref{F:equatorial_reference_frame}.
We also introduce standard spherical coordinates in the equatorial frame $(r,\theta, \phi)$, and the associated spherical orthonormal basis vectors $(\bm{e}_{r}, \bm{e}_{\theta}, \bm{e}_{\phi})$. The position of the source in the sky will be parametrized by the declination $\delta = \pi/2-\theta$ and the right ascension $\alpha = \phi$ (see Fig.~\ref{F:equatorial_reference_frame}). Note that the ranges of these parameters are $\delta \in [-\pi/2, +\pi/2]$ and $\alpha \in [0, 2\pi]$.

 The GW propagation vector $\bm{k}$ in Cartesian $(\bm{x}_0, \bm{y}_0, \bm{z}_0)$ components is given by
\begin{equation}\label{eq:k}
	\bm{k} = - \bm{e}_{r} = \left(-\cos{\delta}\cos{\alpha}, -\cos{\delta}\sin{\alpha}, -\sin{\delta}\right) \,.
\end{equation}
The explicit expressions for the other vectors of the spherical basis are
\begin{subequations}
\begin{align}
	\bm{e}_{\theta} &= \left(\sin{\delta}\cos{\alpha}, \sin{\delta}\sin{\alpha},-\cos{\delta}\right) \,,\\
	\bm{e}_{\phi} &= \left(-\sin{\alpha}, \cos{\alpha}, 0\right) \,.
\end{align}
\end{subequations}
Introduce reference polarization vectors for the equatorial frame as
\begin{subequations}\label{eq:uv}
\begin{align}
	\bm{u}_0 = -\bm{e}_{\phi} & = \left(\sin{\alpha}, -\cos{\alpha}, 0\right) \,,\\
	\bm{v}_0 = -\bm{e}_{\theta} & = \left(-\sin{\delta}\cos{\alpha}, -\sin{\delta}\sin{\alpha}, \cos{\delta}\right) \,,
\end{align}
\end{subequations}
so that $(\bm{u}_0, \bm{v}_0, \bm{k})$ form a direct orthonormal triad. Equivalent expressions directly in terms of $\bm{k}$ and $\bm{z}_0$ are
\begin{equation}
	\bm{u}_0 = \frac{\bm{z}_0 \times \bm{k}}{|\bm{z}_0 \times \bm{k}|} \,, \quad \bm{v}_0 = \bm{k} \times \bm{u}_0.
\end{equation}
The equatorial frame and its reference polarization vectors are given schematically in Fig.~\ref{F:equatorial_reference_frame}.

Defining the polarization tensors
\begin{equation}\label{eq:defeplusecross}
	\bm{\epsilon}^{+}_{ij} = (\bm{u}_0 \otimes \bm{u}_0 - \bm{v}_0 \otimes \bm{v}_0)_{ij} \,, \quad
	\bm{\epsilon}^{\times}_{ij} = (\bm{u}_0 \otimes \bm{v}_0 + \bm{v}_0 \otimes\bm{u}_0)_{ij} \,,
\end{equation}
the \gls{gw} strain in transverse-traceless gauge propagating in the direction $\bm{k}$ takes the form
\begin{equation}
	h_{ij}^{\rm TT} = \bm{\epsilon}^{+}_{ij} h^{\rm SSB}_{+} + \bm{\epsilon}^{\times}_{ij} h^{\rm SSB}_{\times} \,.
\end{equation}
This defines the polarizations $h^{\rm SSB}_{+}$ and $h^{\rm SSB}_{\times}$, functions of time. They are also given by the inverse relations
\begin{equation}
	h^{\rm SSB}_{+} = \frac{1}{2} h_{ij}^{\rm TT}\bm{\epsilon}^{+}_{ij} \,, \quad h^{\rm SSB}_{\times} = \frac{1}{2} h_{ij}^{\rm TT} \bm{\epsilon}^{\times}_{ij} \,.
\end{equation}

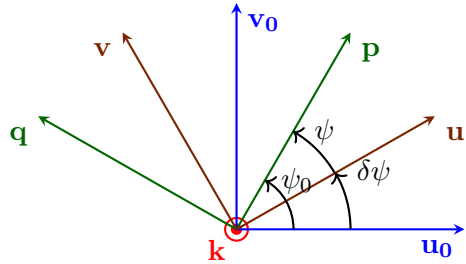
\begin{figure}[htbp]
\centering
\begin{tikzpicture}[scale=3]
    \coordinate (O) at (0,0);
    \draw[thick, color=red] (0,0) circle (0.05);
    \draw[thick, color=red, fill=red] (0,0) circle (0.02) node[anchor=north east]{$\bf{k}$};
    
    \draw[thick,blue,-stealth] (0,0) -- (1,0) node[anchor=north east]{$\bf{u}_0$};
    \draw[thick,blue,-stealth] (0,0) -- (0,1) node[anchor=north west]{$\bf{v}_0$};
    
    \draw[thick,color=Darkgreen,-stealth] (0,0) -- (0.5, 0.87) node[anchor=north west]{$\bf{p}$};
    \draw[thick,color=Darkgreen,-stealth] (0,0) -- (-0.87, 0.5)  node[anchor=north east]{$\bf{q}$};

    \draw[thick,color=Brown,-stealth] (0,0) -- (0.87, 0.5) node[anchor=north west]{$\bf{u}$};
    \draw[thick,color=Brown,-stealth] (0,0) -- (-0.5, 0.87)  node[anchor=north east]{$\bf{v}$};
    
    \draw[->, thick] (0.25,0) arc (0:60:0.25) node[right=2pt]{$\psi_0$};
    \draw[->, thick] (0.5,0) arc (0:30:0.5) node[right=4pt]{$\delta\psi$};
    \draw[->, thick] (0.43,0.25) arc (30:60:0.5) node[right=4pt]{$\psi$};
    
     
\end{tikzpicture}

\caption{Polarization angle definitions in the equatorial and ecliptic frame. The source frame polarization vectors (green) differ from the equatorial reference polarization vectors (blue) by a rotation $\psi_0$ around the GW propagation vector $\bm{k}$. Similarly, a rotation by $\psi$ relates the ecliptic reference polarization vectors to the source frame polarization vectors.}
\label{fig:PolarisationAngle1}
\end{figure}

\gls{gw} sources will generally be described in a source frame with polarization vectors $\bm{p}, \bm{q}$ (see Section~\ref{sec:source-frame-general}) that differ from the \gls{ssb}-polarization vectors $\bm{u}_0, \bm{v}_0$ by a rotation in the plane orthogonal to $\bm{k}$.
%
%
We define the polarization angle $\psi$ to be the angle of the rotation around $\bm{k}$ that maps $\bm{u}_0$ to $\bm{p}_0$ (see figure~\ref{fig:PolarisationAngle1}):
\begin{subequations}
\begin{align}
    \bm{p} &= \bm{u}_0 \cos{\psi} + \bm{v}_0 \sin{\psi}, \\
    \bm{q} &= -\bm{u}_0 \sin{\psi} + \bm{v}_0 \cos{\psi}.
\end{align}
\end{subequations}
The polarization angle can be computed as\footnote{With the convention that $\arctan_{2}[y, x]$ is the polar angle of the point of coordinates $(x,y)$.}
\begin{equation}
	\psi = \arctan_{2} \left[ \bm{p} \cdot\bm{v}_0, \bm{p} \cdot\bm{u}_0  \right] \,.
\end{equation}

\subsection{Ecliptic reference frame}\label{sec:ecliptic_frame}

\subsubsection{Definition}

Another commonly used reference frame is the ecliptic frame, with Cartesian basis vectors $(\bm{x}, \bm{y}, \bm{z})$. We choose to use the \gls{bme} reference frame at J2000. Its origin is also the solar system barycenter, with the $\bm{x}$ axis coinciding with the $\bm{x}_0$ axis of the ICRS (mean equinox at J2000). The $\bm{z}$ axis is taken as the vector normal to the mean ecliptic plane at Julian Day 2000.0, pointing toward the Northern Hemisphere. The $\bm{y}$ axis completes the triad.

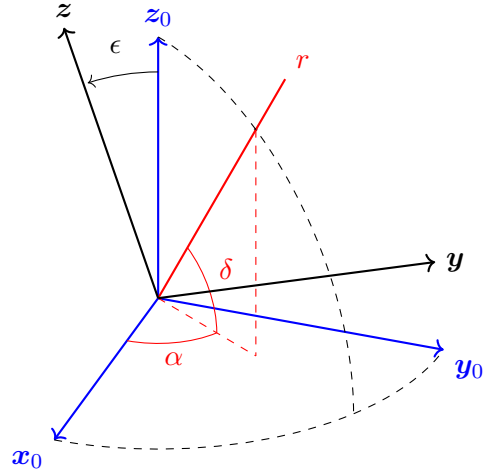
\begin{figure}[tbh]
	\centering
	\begin{tikzpicture}[scale=4,tdplot_main_coords]
		\def\sunangle{23.4}
		\coordinate (O) at (0,0,0);
		\draw[thick,->,color=blue] (0,0,0) -- (1,0,0) node[anchor=north east](x0){$\bm{x}_{0}$};
		\draw[thick,->,color=blue] (0,0,0) -- (0,1,0) node[anchor=north west](y0){$\bm{y}_{0}$};
		\draw[thick,->,color=blue] (0,0,0) -- (0,0,1) node[anchor=south](z0){$\bm{z}_{0}$};
		\tdplotsetcoord{P}{\rvec}{\thetavec}{\phivec}
		\tdplotsetcoord{r}{1.0}{\thetavec}{\phivec};
		\draw (r) node {.};
		\draw[thick, color=red]  (P) node[anchor=south west,color=red](rvect){$r$} -- (O) ;
		\draw[dashed, color=red] (O) -- (rxy);
		\draw[dashed, color=red] (r) -- (rxy);
		\tdplotdrawarc[color=red]{(O)}{0.3}{0}{\phivec}{anchor=north,color=red}{$\alpha$}
		\tdplotsetthetaplanecoords{\phivec}
		\tdplotdrawarc[tdplot_rotated_coords,color=red]{(0,0,0)}{0.3}{\thetavec}{90}{anchor=south west,color=red}{$\delta$};

		\draw[dashed,tdplot_rotated_coords] (1.0,0,0) arc (0:90:1.0);
		\draw[dashed] (1.0,0,0) arc (0:90:1.0);
		
		
		\draw[thick,->, color=black] (0,0,0) -- (0, 0.97, 0.33) node[anchor=west] (y) {$\bm{y}$};
		\draw[thick,->, color=black] (0,0,0) -- (0, -0.33, 0.97) node[anchor=south] (z) {$\bm{z}$};
		\draw[color=black] pic[draw,->,"$\epsilon$",angle eccentricity=1.2,angle radius=3cm,anchor=north] {angle=z0--O--z};
		
		\tdplotsetcoord{Y}{1.0}{90}{90};
		\tdplotsetcoord{Z}{1.0}{0}{90};
		
		
		
	\end{tikzpicture}
	\caption{Representation of the equatorial frame (blue basis vectors) and of the ecliptic frame (black basis vectors). The $\bm{x}_0$ vector of the equatorial frame coincides with the $\bm{x}$ vector of the ecliptic plane, not represented here. One transforms from the equatorial (blue) to the ecliptic frame (black) by applying a rotation of $\epsilon \approx 23.4$ degrees about $\bm{x}_0$.}
	\label{F:equatorial_ecliptic_frames}
\end{figure}

The corresponding class definining the ecliptic frame in \texttt{LISA Constants} is \textsf{lisaconstants.coords.LISAEclipticFrame}. In \texttt{Astropy} v7.1.1, it is defined as  \textsf{astropy.coordinates.BarycentricMeanEcliptic}\footnote{See \url{https://docs.astropy.org/en/stable/api/astropy.coordinates.BarycentricMeanEcliptic.html}. It must be instantiated with the attribute \textsf{equinox = "J2000"}.}.
The transformation of axes from the equatorial to ecliptic basis is illustrated in Fig.~\ref{F:equatorial_ecliptic_frames} and can be written as:
\begin{eqnarray*}
    \bm{x} &=& \bm{x}_0 \\
    \bm{y} &=& \cos(\epsilon) \bm{y}_{0} + \sin(\epsilon) \bm{z}_0 \\
    \bm{z} &=& - \sin(\epsilon) \bm{y}_{0} + \cos(\epsilon) \bm{z}_0
\end{eqnarray*}
where $\epsilon$ is the mean obliquity of the ecliptic at J2000, or approximately $23.43927945$ degrees. The value is stored in OBLIQUITY variable in the \texttt{LISA Constants} software~\cite{bayle_2022_6627346}. Note that since the chosen date is the same as IRCF, there is no need to apply a precession rotation to convert ICRF coordinates to \gls{bme} coordinates.

The sky position in the ecliptic frame is described by the ecliptic longitude and latitude $(\lambda, \beta)$, defined in the same way as $(\alpha, \delta)$ for the equatorial frame. One can define reference polarization vectors $(\bm{u}, \bm{v})$ and a polarization angle $\psi$ for the ecliptic frame by repeating the definitions for the equatorial frame shown in Fig.~\ref{F:equatorial_reference_frame}, with the replacements
\begin{equation}
    (\bm{x}_0, \bm{y}_0, \bm{z}_0) \rightarrow (\bm{x}, \bm{y}, \bm{z}), \quad (\bm{u}_0, \bm{v}_0) \rightarrow (\bm{u}, \bm{v}), \quad (\alpha, \delta, \psi_0) \rightarrow (\lambda, \beta, \psi) \,.
\end{equation}

\subsubsection{From equatorial to ecliptic coordinate system}

We can transform the equatorial coordinate system (radius $r$, right ascension $\alpha$ and declination $\delta$) to ecliptic coordinates (radius $r$, longitude $\lambda$ and latitude $\beta$) as
\begin{eqnarray}
r & = & r_0 \nonumber \\
\sin \beta & = & \sin \delta \cos \epsilon-\cos \delta \sin \epsilon \sin \alpha \nonumber \\
\cos \lambda & = & \cos \alpha \cos \delta / \cos \beta \nonumber \\
\sin \lambda & = & [\sin \delta \sin \epsilon+\cos \delta \cos \epsilon \sin \alpha] / \cos \beta 
\end{eqnarray}
The transformation can be performed with the method \textsf{lisaconstants.coords.equatorial\_to\_ecliptic} in \texttt{LISA Constants}.
Conversely, we transform ecliptic to equatorial coordinates as
\begin{eqnarray}
r_0 & = &  r \nonumber \\
\sin \delta & = & \sin \beta \cos \epsilon+\cos \beta \sin \epsilon \sin \lambda \nonumber \\
\cos \alpha & = & \cos \lambda \cos \beta / \cos \delta \nonumber \\
\sin \alpha & = & [-\sin \beta \sin \epsilon+\cos \beta \cos \epsilon \sin \lambda] / \cos \delta
\end{eqnarray}
The transformation can be performed with the method \textsf{lisaconstants.coords.ecliptic\_to\_equatorial} in \texttt{LISA Constants}.

Since the $z$-axis of the equatorial and ecliptic frames differ, as well as their conventional polarization vectors $(\bm{u},\bm{v})$ and $(\bm{u}_0,\bm{v}_0)$, the polarization angles $\psi$ and $\psi_0$ are different. We can convert between the two following
\begin{subequations}
\begin{align}
    \psi_0 &= \psi + \delta \psi \,,\\
    \cos\delta\psi &= \frac{1}{\cos\beta} \left( \sin\epsilon \sin\delta \sin\alpha + \cos\epsilon \cos\delta \right) \,,\\
    \sin\delta\psi &= -\frac{1}{\cos\beta} \sin\epsilon \cos\alpha \,,\\
    \cos\delta\psi &= \frac{1}{\cos\delta} \left( -\sin\epsilon \sin\beta \sin\lambda + \cos\epsilon \cos\beta \right) \,,\\
    \sin\delta\psi &= -\frac{1}{\cos\delta} \sin\epsilon \cos\lambda \,.
\end{align}
\end{subequations}




%

\subsection{Galactocentric reference frame}\label{sec:galactocentric_frame}

We introduce Galactocentric coordinates for the purpose of astrophysical populations. In particular, we provide the transformation to and from the ICRS coordinate system (equatorial frame). 

We adopt the definition of Astropy~\cite{price2018astropy} to define the Galactocentric coordinate system. The transformation from ICRS cartesian coordinates to Galactocentric cartesian coordinates can be summarized by the following equations~\footnote{For more details, see \url{https://docs.astropy.org/en/stable/coordinates/galactocentric.html\#coordinates-galactocentric}}
\begin{align}
	\label{eq:galactocentric-to-icrs}
	\mathbf{r}_{\rm GC} & = \mathbf{H}\,\big(\mathbf{R}\,\mathbf{r}_{0} - d_{\rm GC}\,\hat{\mathbf{x}}_{\rm GC}\big),
\end{align}
where $\mathbf{r}_{0}$ and $\mathbf{r}_{\rm GC}$ are the position vectors in the \gls{icrs} and Galactocentric frames, respectively. We denote $d_{\rm GC}$ the distance between the solar system barycenter and the Galactic center. The matrix $\mathbf{R}$ represents a succession of 2 rotations. The first one rotates the frame about $\mathbf{z}_0$ by an angle $\alpha_{\rm GC}$. The second one rorates the frame about $\mathbf{y}_0$ by an angle $\delta_{\rm GC}$. The angles $(\alpha_{\rm GC}, \delta_{\rm GC})$ are the right ascension and declination of the Galactic center in the \gls{icrs} coordinate system. This results in
\begin{equation}
	\mathbf{R} = \left[\begin{array}{ccc}
		\cos \delta_{\mathrm{GC}} & 0 & \sin \delta_{\mathrm{GC}} \\
		0 & 1 & 0 \\
		-\sin \delta_{\mathrm{GC}} & 0 & \cos \delta_{\mathrm{GC}}
	\end{array}\right] \left[\begin{array}{ccc}
	\cos \alpha_{\mathrm{GC}} & \sin \alpha_{\mathrm{GC}} & 0 \\
	-\sin \alpha_{\mathrm{GC}} & \cos \alpha_{\mathrm{GC}} & 0 \\
	0 & 0 & 1
	\end{array}\right] .
\end{equation}
Note that in some conventions a third rotation of a roll angle $\eta$ about the $x$ axis can be applied to orient the galactic plane, but we do not include it in this definition.

We use $\hat{\mathbf{x}}_{\rm GC} = \left(1, 0, 0\right)^\intercal$ to shift the frame origin at the Galactic center. 

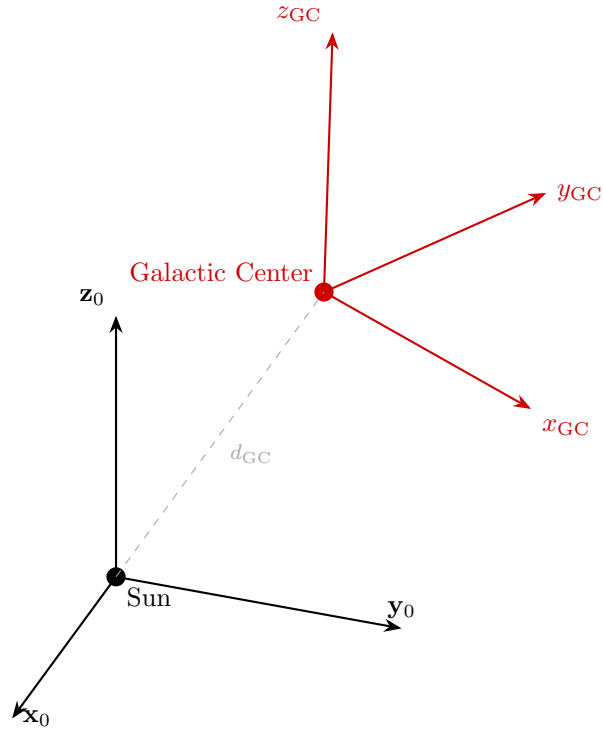
\begin{figure}[tbh]
	\centering
\begin{tikzpicture}[tdplot_main_coords, scale=4.0,
	axis/.style={-Stealth, thick},
	icrs/.style={black, thick},
	galcen/.style={red!80!black, thick},
	every node/.style={font=\small}
	]
	
	\coordinate (Sun) at (0,0,0);
	\coordinate (GC) at (-2, 0,  0.005); 
	
	\draw[axis,icrs] (Sun) -- ++(1,0,0) node[right]{$\mathbf{x}_{0}$};
	\draw[axis,icrs] (Sun) -- ++(0,1,0) node[above]{$\mathbf{y}_{0}$};
	\draw[axis,icrs] (Sun) -- ++(0,0,1) node[above left]{$\mathbf{z}_{0}$};
	
	\tdplotsetrotatedcoords{93}{1.7}{-30} 
	\draw[axis,galcen,tdplot_rotated_coords] (GC) -- ++(1,0,0) node[below right]{$x_{\rm GC}$};
	\draw[axis,galcen,tdplot_rotated_coords] (GC) -- ++(0,1,0) node[right]{$y_{\rm GC}$};
	\draw[axis,galcen,tdplot_rotated_coords] (GC) -- ++(0,0,1) node[above left]{$z_{\rm GC}$};
	
	\filldraw[icrs] (Sun) circle (0.8pt) node[below right]{Sun};
	\filldraw[galcen] (GC) circle (0.8pt) node[above left]{Galactic Center};
	
	\draw[dashed,gray!70] (Sun) -- (GC)
	node[midway,below right,font=\scriptsize]{$d_{\rm GC}$};
	

	
\end{tikzpicture}
\caption{\label{fig:galactocentric}Schematic representation of the Galactocentric frame (red) and the equatorial \gls{icrf} (black), whose origins are separated by a distance $d_{\rm GC}$.}
\end{figure}

The overall transformation $\mathbf{H}$ serves to rorate the frame by an angle $\theta = \arcsin (z_{\odot}/d_{\rm GC})$ about its $y$ axis to account for the height $z_{\odot}$ of the Sun above the Galactic midplane~:
\begin{equation}
	\boldsymbol{H}=\left[\begin{array}{ccc}
		\cos \theta & 0 & \sin \theta \\
		0 & 1 & 0 \\
		-\sin \theta & 0 & \cos \theta
	\end{array}\right].
\end{equation}

We obtain the transformation from Galactocentric coordinates to \gls{icrs} coordinates by inverting the relation \eqref{eq:galactocentric-to-icrs}~:
\begin{align}
	\label{eq:icrs-to-galactocentric}
\mathbf{r}_{0} & =	\mathbf{R}^\intercal \,\big( \mathbf{H}^\intercal \mathbf{r}_{\rm GC} + d_{\rm GC}\,\hat{\mathbf{x}}_{\rm GC} \big).
\end{align}

In \texttt{LISA Constants}, the class defining the galactocentric frame is \textsf{lisaconstants.coords.LISAGalactocentricFrame}. In \texttt{Astropy} v7.1.1, the counterpart is  \textsf{astropy.coordinates.Galactocentric}~\footnote{\url{https://docs.astropy.org/en/stable/api/astropy.coordinates.Galactocentric.html}}, with the default set of parameters labelled as ``v4.0''.

\begin{table}[tbh]
	\centering
	\begin{tabular}{c c c c }
		$\alpha_{\rm GC}$ [deg] & $\delta_{\rm GC}$ [deg]& $d_{\rm GC}$ [kpc] & $z_{\odot}$ [pc] \\
				\hline
		266.4051 & -28.936175 & 8.122 & 20.8
	\end{tabular}
	\caption{\label{tab:galactocentric-parameters}Galactocentric coordinate system parameters adopted in the conventions, corresponding to ``v4.0'' settings in Astropy v7.1.1.}
\end{table}

\subsection{\label{sec:source-frame-general}Source frame: general conventions}

We continue by generically introducing a source frame defined by unit vectors $(\bm{x}_{S}, \bm{y}_{S}, \bm{z}_{S})$. 
We will define each source class's preferred source frame (orientation and time) below. For now, we introduce a general convention to label its axes, angles, and how it relates to the wave frame.

In this source frame, we introduce standard spherical coordinates $(r_{S}, \theta_{S}, \phi_{S})$, and the associated spherical orthonormal basis vectors $(\bm{e}^S_{r}, \bm{e}^S_{\theta}, \bm{e}^S_{\phi})$. The unit vector $\bm{k}$ defines the direction of propagation of the gravitational waves, from the source towards the observer. Its Cartesian $(\bm{x}_{S}, \bm{y}_{S}, \bm{z}_{S})$ components are
\begin{equation}
	\bm{k} = \bm{e}_{r}^{S} = \left(\sin{\theta_{S}}\cos{\phi_{S}}, \sin{\theta_{S}}\sin{\phi_{S}}, \cos{\theta_{S}}\right) \,.
\end{equation}
The explicit expressions for the other vectors of the spherical basis are
\begin{subequations}
\begin{align}
	\bm{e}^S_{\theta} = { \partial  \bm{e}_{r}^{S} \over \partial \theta_{S}} &= \left(\cos{\theta_{S}}\cos{\phi_{S}}, \cos{\theta_{S}}\sin{\phi_{S}},-\sin{\theta_{S}}\right) \,,\\
	\bm{e}^S_{\phi} = {1 \over \sin \theta_{S}} { \partial  \bm{e}_{r}^{S} \over \partial \phi_{S}}  &= \left(-\sin{\phi_{S}}, \cos{\phi_{S}}, 0\right) \,.
\end{align}
\end{subequations}

\begin{figure}[tbh]
  \centering
\begin{tikzpicture}[scale=4, tdplot_main_coords]
    \coordinate (O) at (0,0,0);
    \draw[thick,->] (0,0,0) -- (1,0,0) node[anchor=north east]{$\bm{x}_{S}$};
    \draw[thick,->] (0,0,0) -- (0,1,0) node[anchor=north west]{$\bm{y}_{S}$};
\draw[thick,->] (0,0,0) -- (0,0,1) node[anchor=south]{$\bm{z}_{S}$};
    \tdplotsetcoord{P}{\rvec}{\thetavec}{\phivec}
    \draw[-stealth,color=red] (O) -- (P) node[above right] {$\bm{k}$};
    \draw[dashed, color=red] (O) -- (Pxy);
    \draw[dashed, color=red] (P) -- (Pxy);
    \tdplotdrawarc{(O)}{0.2}{0}{\phivec}{anchor=north}{$\;\;\;\phi_{S} = \varphi$}
    \tdplotsetthetaplanecoords{\phivec}
    \tdplotdrawarc[tdplot_rotated_coords]{(0,0,0)}{0.2}{0}%
        {\thetavec}{anchor=south west}{$\theta_{S} = \iota$};
    \tdplotsetcoord{r}{1.0}{\thetavec}{\phivec};
    \draw (r) node {.};
    \tdplotsetcoord{rth}{1.0}{\thetavec+12.0}{\phivec};
    \tdplotsetcoord{rph}{1.0}{\thetavec}{\phivec+25};
    \draw[-stealth, color=Darkgreen] (r)--(rth) node[right] {$\bm{e}_{\theta}^S = \bf{p}$};
    \draw[-stealth, color=Darkgreen] (r)--(rph) node[right] {$\bm{e}_{\phi}^S = \bf{q}$};
    \fill (r) circle[radius=0.3pt] node[anchor=south west]{};
    
    \draw[dashed,tdplot_rotated_coords] (1.0,0,0) arc (0:90:1.0);
    \draw[dashed] (1.0,0,0) arc (0:90:1.0);    
\end{tikzpicture}
\caption{General representation of the source frame. }
\label{F:source_frame} 
\end{figure}
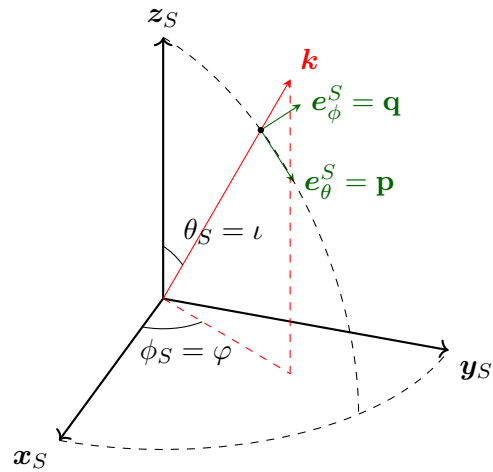

As a function of the source frame basis vectors, the polarization basis vectors are: 
\begin{equation}
\label{eq:polarization-basis-vectors-pq}
	\bm{p} = \bm{e}^S_{\theta}, \;\; \bm{q} = \bm{e}^S_{\phi}
\end{equation}
which form together with $\bm{k}$ the radiation frame or wave-frame: $(\bm{p}, \bm{q}, \bm{k})$. They can be defined from $\bm{z}_{S}$ and $\bm{k}$ alone as
\begin{equation}\label{eq:defpq}
	\bm{q} = \frac{\bm{z}_{S} \times \bm{k}}{|\bm{z}_{S} \times \bm{k}|} \,, \quad \bm{p} = \bm{q} \times \bm{k}.
\end{equation}
The source frame and the polarization vectors are shown in Fig.~\ref{F:source_frame}. Defining the polarization tensors
\begin{equation}\label{eq:defeplusecross}
	e^{+}_{ij} = (\bm{p} \otimes \bm{p} - \bm{q} \otimes \bm{q})_{ij} \,, \quad
	e^{\times}_{ij} = (\bm{p} \otimes \bm{q} + \bm{q} \otimes\bm{p})_{ij} \,,
\end{equation}
the \gls{gw} strain in transverse-traceless gauge takes the form
\begin{equation}
\label{eq:httdecomposition}
	h_{ij}^{\rm TT} = e^{+}_{ij} h_{+} + e^{\times}_{ij} h_{\times} \,.
\end{equation}
This defines the polarizations $h_{+}$ and $h_{\times}$, functions of time. They are also given by the inverse relations
\begin{equation}\label{eq:hplushcrossPols}
	h_{+} = \frac{1}{2} h_{ij}^{\rm TT}e^{+}_{ij} \,, \quad h_{\times} = \frac{1}{2} h_{ij}^{\rm TT} e^{\times}_{ij} \,.
\end{equation}

Note that we often call $\theta_S$ the inclination, and $\phi_S$ the observer phase, and we note them:
\begin{subequations}
\begin{align}
	\text{Inclination :} & \quad \iota \equiv \theta_{S} \,, \\
	\text{Observer phase :} & \quad \varphi \equiv \phi_{S} \,.
\end{align}
\end{subequations}

If we label $(1, \theta_{zS}, \phi_{zS})$ the spherical coordinates of the vector $\bm{z}_S$ in the ecliptic reference frame such that $z_S = \left(\cos{\phi_{zS}}\sin{\theta_{zS}}, 
\sin{\phi_{zS}}\sin{\theta_{zS}}, \cos{\theta_{zS}}\right)$, we can compute the source inclination and polarization as
\begin{eqnarray}
\iota &=& \arccos\left[ 
-\cos{\theta_{zS}}\sin{\beta} - \cos{\beta}\sin{\theta_{zS}}\cos{(\lambda-\phi_{zS})}
\right]\\
\tan \psi & = &    \frac{- \sin \beta \sin \theta_{zS} \cos \left( \lambda - \phi_{zS} \right) + \cos \theta_{zS} \cos \beta } 
{ \sin \theta_{zS} \sin \left( \lambda  - \phi_{zS} \right)}
\end{eqnarray}

The relation between the polarization tensors $\bm{\epsilon}^+, \bm{\epsilon}^\times$ associated to $(\bm{u}, \bm{v})$ (as defined in Eq.~\eqref{eq:defeplusecross}) and the polarization tensors $(\bm{e}^+, \bm{e}^\times)$ associated to $(\bm{p}, \bm{q})$ is
\begin{subequations}
\begin{align}
	\bm{e}^+ &= \bm{\epsilon}^+ \cos{2\psi} + \bm{\epsilon}^{\times} \sin{2\psi},\\
	\bm{e}^{\times} &= -\bm{\epsilon}^+ \sin{2\psi} + \bm{\epsilon}^{\times} \cos{2\psi}.
\end{align}
\end{subequations}
The corresponding representation of the strain in the ecliptic SSB frame is 
\begin{equation}
 h_{ij}^{\mathrm{SSB}}  =  \left( h_{+} \cos 2 \psi  -  h_{\times} \sin 2 \psi  \right) \bm{\epsilon}^{+}_{ij} 
  + \left( h_{+} \sin 2 \psi  + h_{\times} \cos 2 \psi  \right) \bm{\epsilon}^{\times}_{ij}.  
\end{equation}

\subsection{Source frames: variations and transformation}\label{sec:source_frames}
Above we have considered the observer's orientation in the source frame without specifying the orientation of the frame itself. 
In this section, we introduce several choices of source frame $(\bm{x}_{S}, \bm{y}_{S}, \bm{z}_{S})$ relevant to different types of binaries. In this picture, the source frame is to be defined from the properties of the binary system, namely its trajectory, and its precise definition might vary depending on the class of sources considered (e.g., EMRIs vs. MBHBs). The aim is to describe the main source frames and provide the transformations from one to another.

The source frames described below are constructed from kinematic and/or conserved quantities evaluated at a reference time along the binary's trajectory. To fully define the source frame, one must describe how this reference time or reference point in the orbit is chosen (see Sec.~\ref{sec:reference_time}).

The four main frames that we propose are:
\begin{itemize}
    \item the $\bm{L}_N$\textbf{-frame}, or \textbf{kinematic frame} where the $z$-axis is the normal to the orbital plane (equivalently, the direction of the Newtonian orbital angular momentum) $\hat{\bm{L}}_N$;
    \item the $\bm{L}$\textbf{-frame}, where the $z$-axis is the direction of the total orbital angular momentum $\bm{L}$, which differs slightly in direction from $\hat{\bm{L}}_N$;
    \item the $\bm{J}$\textbf{-frame}, where the $z$-axis is the direction of the total angular momentum $\bm{J}$;
    \item the $\bm{S}_1$\textbf{-frame}, better adapted to the EMRI setting, where the $z$-axis is chosen to be the direction of the primary spin, that is to say the spin of the central Kerr black hole.
\end{itemize}

We will also describe two other choices of source frame used in current software packages:
\begin{itemize}
    \item the \textbf{LAL-frame} used as a convention in \texttt{LALsuite} as of 2024, as well as for \texttt{SXS} waveforms. It is closely related to the $\bm{L}_N$\textbf{-frame} but with an important change in the chosen convention for the separation unit vector;
    \item the \textbf{FEW-frame} used by the software suite \texttt{FastEMRIWaveforms} (\texttt{FEW}) in the EMRI setting. The frame is defined here in the 0th post-adiabatic order.
\end{itemize}

Note that all these frames make reference to the binary's trajectory and are therefore gauge-dependent. Efforts are underway to define reference frames from the waveform alone, relying only on gauge-independent observables.

\subsubsection{Preliminaries: Kinematic quantities}


Let us first introduce the kinematic frame, defined from the trajectory of the separation vector $\bm{x} = r \bm{n} = \bm{y}_2 - \bm{y}_1$, with $r$ the separation, $\bm{n}$ the separation unit vector, and $\bm{y}_{1,2}$ the positions of the two bodies in the convention $m_1 > m_2$. This means that the unit separation vector $\bm{n}$ points \textbf{from the heavier object towards the lighter object}. Note that the PN literature and the \texttt{LALsuite} conventions use the opposite sign convention for $\bm{n}$. The reason for our choice is a better continuity between comparable-mass systems such as MBHBs and EMRIs: in an EMRI setting, it is natural to think of the trajectory as that of the lighter object around a central heavy black hole that defines the origin of coordinates.

We can then write:
\begin{subequations}
\begin{align}
    \bm{x} &= r \bm{n} \,,\\
    \bm{v} &= \dot{\bm{x}} = \dot{r}\bm{n} + r \omega \bm{\lambda} \,,\\
    \bm{\ell} &= \frac{\bm{x} \times \bm{v}}{|\bm{x} \times \bm{v}|} \,,
\end{align}
\end{subequations}
where $\bm{\lambda} = \frac{\bm{\ell} \times \bm{n}}{|\bm{\ell} \times \bm{n}|}$. We obtain a unit triad $(\bm{n}, \bm{\lambda}, \bm{\ell})$ defined at any point along the trajectory (in a given coordinate frame -- the trajectory is gauge-dependent). The unit vector $\bm{\ell}$ is the direction of the instantaneous normal to the orbital plane, and corresponds to that of the Newtonian orbital angular momentum: $\bm{\ell} = \hat{\bm{L}}_N$.

The total angular momentum of the system is given as $\bm{J} = \bm{L} + \bm{S}/c$, where $\bm{L}$ is the orbital angular momentum and $\bm{S} = \bm{S}_1 + \bm{S}_2$ is the total spin of the binary (the factor $c$ is conventional in PN theory to be able to treat spins as Newtonian-order quantities). We can write the individual spins in terms of dimensionless spins as $\bm{S}_1 = m_1^2 \chi_1 \hat{\bm{S}}_1$ (similarly for 2). At low PN order, we have the approximation
\bea
\bm{J} \simeq |\bm{L}| \bm{\ell} + \bm{S}/c, 
\label{Eq:J}
\ena
where
\bea
|\bm{L}| = \frac{M\eta}{v} \left[ 
1 + v^2\left(\frac3{2} + \frac{\eta}{6}\right) + \mathcal{O}(v^3)
\right], \;\;\;\; v = (M \omega)^{1/3}
\ena
is the 1-PN expression for the norm of the orbital angular momentum. Note that at higher PN order, $\bm{L}$ acquires more PN contributions, including terms starting at 1.5PN that depend on the spins, so that the direction $\hat{\bm{L}} \neq \bm{\ell}$ in general. One advantage of using $\bm{L}$ over $\bm{L}_N$ is that precession can induce nutations of $\bm{L}_N$ on the orbital timescale, while $\bm{L}$ has a smoother evolution; however, for e.g. an NR simulation, $\bm{J}$, $\bm{L}$ might not be readily available, being built from conserved quantities, while the definition of $\hat{\bm{L}}_N = \bm{\ell}$ is purely kinematic. In the following, we will keep the distinction between $\hat{\bm{L}}$ and $\hat{\bm{L}}_N$.

\subsubsection{$L_N$-frame or Kinematic frame}

To build the $\bm{L}_N$\textbf{-frame} or \textbf{kinematic frame}, one orients the frame with respect to the binary at some fiducial moment of time (or at some orbital frequency if it can be defined unambiguously). The $z$-axis is chosen to be the instantaneous normal to the orbital plane. The source frame is then completed using the unit separation vector defined above (from the heavier to the lighter object), at that reference time. Using the axes labeling introduced in Fig.~\ref{F:source_frame}, this definition sets
\begin{alignat}{2}
  & \begin{dcases}
    \bm{x}_{S} = \bm{x}_{L_N}\\
    \bm{y}_{S} = \bm{y}_{L_N}\\
    \bm{z}_{S} = \bm{z}_{L_N}
  \end{dcases}
    & \quad \text{such that} \quad & 
    \begin{dcases}
    \bm{x}_{L_N} = \bm{n}\,,\\
    \bm{y}_{L_N} = \bm{\lambda} = \frac{\bm{\ell} \times \bm{n}}{|\bm{\ell} \times \bm{n}|}\,,\\
    \bm{z}_{L_N} = \bm{\ell} = \hat{\bm{L}}_N\,.
  \end{dcases}
\label{eq:lnframe}
\end{alignat}
The $\bm{L}_N$\textbf{-frame} axes as $\left(\bm{x}_{L_N}, \bm{y}_{L_N}, \bm{z}_{L_N}\right)$ are depicted in blue in~\autoref{F:Jsource_frame}. 

\subsubsection{$\bf L$-frame}
\label{sec:L_frame}

To define the $\bm{L}$\textbf{-frame}, we take the $z$-axis in Fig.~\ref{F:source_frame} to be $\bm{z}_{S} = \bh{L}$, the unit vector along the total orbital angular momentum $\bm{L} = \bm{J} - \bm{S}/c$, evaluated at some reference time. This direction slightly differs from $\hat{\bm{L}}_N$. As a result, $\bm{n}$ is not orthogonal to $\bh{L}$ and we need another way to set the $x$-axis. We can choose $\bm{x}_S$ so that the unit vector $\bm{n}$ lies in the $(\bh{L},\bm{x}_S)$ plane, with the extra condition $\bm{x}_S \cdot \bm{n} \geq 0$. This definition sets
\begin{alignat}{2}
  & \begin{dcases}
    \bm{x}_{S} = \bm{x}_{L}\\
    \bm{y}_{S} = \bm{y}_{L}\\
    \bm{z}_{S} = \bm{z}_{L}
  \end{dcases}
    & \quad \text{such that} \quad & 
    \begin{dcases}
    \bm{x}_{L} = \frac{\bm{n} - (\bh{L} \cdot \bm{n}) \bh{L}}{|\bm{n} - (\bh{L} \cdot \bm{n}) \bh{L}|} \,,\\
    \bm{y}_{L} = \bm{z}_{L} \times \bm{x}_{L}\,,\\
    \bm{z}_{L} = \hat{\bm{L}}\,.
  \end{dcases}
\label{eq:lnframe}
\end{alignat}

Let us emphasize again that this frame is defined at some instance of time because the angular momentum (in general) is a function of time, as well as the position of the secondary body.  This frame is very convenient for setting the initial conditions and specifying the spins orientation $\bh{S}_{i} = \left(\sin{\theta_{S_{i}}}\cos{\phi_{S_{i}}}, \sin{\theta_{S_{i}}}\sin{\phi_{S_{i}}}, \cos{\theta_{S_{i}}} \right)$.

In general, we expect the motion to be eccentric and we define the instantaneous ellipse by specifying the position of the periapse $\vec{a}$, such that the angle between the periapse to $\bm{x}_L$ is $\xi$ and the distance to the secondary object is defined as 
\bea
r = \frac{pM}{1+e\cos{\xi}},
\ena
where $p$ is dimensionless semi-latus rectum and $e$ is orbital eccentricity. 

\subsubsection{$\bf J$-frame}

Another possible choice of source frame is the \textbf{J-frame}, where the principal axis is oriented along the total momentum of the system: $\bm{z}_{S} = \bh{J}$. We can choose $\bm{x}_S$ so that the unit vector $\bm{x}_{L_N} = \bm{n}$ from the $L_N$-frame (or kinematic frame) lies in the $(\bh{J},\bm{x}_S)$ plane, with the extra condition $\bm{x}_S \cdot \bm{n} \geq 0$. This definition sets
\begin{alignat}{2}
  & \begin{dcases}
    \bm{x}_{S} = \bm{x}_{J}\\
    \bm{y}_{S} = \bm{y}_{J}\\
    \bm{z}_{S} = \bm{z}_{J}
  \end{dcases}
    & \quad \text{such that} \quad & 
    \begin{dcases}
    \bm{x}_{J} = \frac{\bm{n} - (\bh{J} \cdot \bm{n}) \bh{J}}{|\bm{n} - (\bh{J} \cdot \bm{n}) \bh{J}|} \,,\\
    \bm{y}_{J} = \bm{z}_{J} \times \bm{x}_{J}\,,\\
    \bm{z}_{J} = \hat{\bm{J}}\,.
  \end{dcases}
\label{eq:Jframe}
\end{alignat}

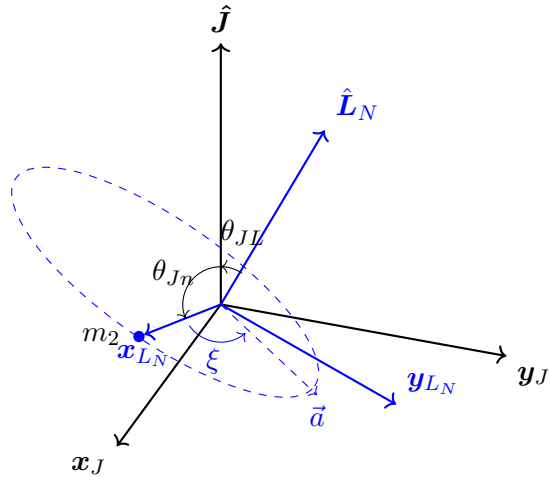
\begin{figure}[tbh]
  \centering
\begin{tikzpicture}[scale=4, tdplot_main_coords]
    \coordinate (orig) at (0,0,0);
    \draw[thick,->] (0,0,0) -- (1,0,0) node[anchor=north east]{$\bm{x}_{J}$};
    \draw[thick,->] (0,0,0) -- (0,1,0) node[anchor=north west]{$\bm{y}_{J}$};
    \draw[thick,->] (0,0,0) -- (0,0,1) node[anchor=south] (Jax){$\bm{\hat{J}}$};
   \draw[thick,->, color=blue] (0,0,0) -- (0.1, 0.4, 0.8) 
   node[anchor=south west] (Lvec) {$\hat{\bm{L}}_N$};
   \draw[thick,->, color=blue] (0,0,0) -- (0.6, -0.05,0.2) 
   node[anchor=north](m2) {$\bm{x}_{L_N}$};
   \draw[thick,->, color=blue] (0,0,0) -- (0.1, 0.65, -0.2) 
   node[anchor=south west](yL) {$\bm{y}_{L_N}$};
   \draw[dashed, ->, color=blue] (0,0,0) -- (0.31, 0.45, -0.09) node[anchor=north](periaps) {$\vec{a}$};
   \draw[dashed, color=blue, rotate around={-35:(0.0,0.0)}] (0.15,-0.15) ellipse (0.6cm and 0.2cm);
   \coordinate[label = left:$m_2\;$]  (pm2) at (0.65, -0.05, 0.22);
   \node at (pm2)[circle,fill, color=blue, inner sep=1.5pt]{};
   \draw pic[draw, ->, "$\theta_{Jn}$", angle eccentricity=1.5] {angle=Jax--orig--pm2};
   \draw pic[draw, ->, "$\theta_{JL}$", angle eccentricity=2.] {angle=Lvec--orig--Jax};
   
   \draw pic[draw, ->, color=blue, "$\xi$", angle eccentricity=1.5] {angle=m2--orig--periaps}; 
\end{tikzpicture}
\caption{Relationship between the $L_N$-frame and the $J$-frame.}
\label{F:Jsource_frame} 
\end{figure}

The \textbf{J-frame} axes $\left(\bm{x}_{J}, \bm{y}_{J}, \bm{z}_{J}\right)$ are depicted in black in \autoref{F:Jsource_frame}. The convenience of this frame is twofold. First, the direction $\hat{\bm{J}}$ is time independent to a high PN degree and can be considered the same at instances of time during the inspiral. Second, it is conveniently close to what is used to describe EMRIs, as the dominant contribution to $\bm{J}$ comes from $\bm{S}_1$ in the extreme mass ratio limit, so that its orientation closely follows the spin of the MBH.

\subsubsection{$S_1$-frame}

Another possible choice of source frame is the $\bm{S}_1$\textbf{-frame}.  Here the principal axis is oriented along the spin of the most massive object: $\bm{z}_{S} = \bh{S}_1$. We can choose $\bm{x}_{S}$ so that the unit vector $\bm{x}_{L_N} = \bm{n}$ from the $L_N$-frame (or kinematic frame) lies in the $(\bh{S}_1,\bm{x}_{S})$ plane, with the extra condition $\bm{x}_{S} \cdot \bm{n} \geq 0$. This definition sets
\begin{alignat}{2}
  & \begin{dcases}
    \bm{x}_{S} = \bm{x}_{S_1}\\
    \bm{y}_{S} = \bm{y}_{S_1}\\
    \bm{z}_{S} = \bm{z}_{S_1}
  \end{dcases}
    & \quad \text{such that} \quad & 
    \begin{dcases}
    \bm{x}_{S_1} = \frac{\bm{n} - (\bh{S}_1 \cdot \bm{n}) \bh{S}_1}{|\bm{n} - (\bh{S}_1 \cdot \bm{n}) \bh{S}_1|} \,,\\
    \bm{y}_{S_1} = \bm{z}_{S_1} \times \bm{x}_{S_1}\,,\\
    \bm{z}_{S_1} = \hat{\bm{S}}_1\,.
  \end{dcases}
\label{eq:S1frame}
\end{alignat}

By using as a $z$-axis the spin of the most massive object, this frame offers a natural connection to test-mass orbits, where a test particle orbits in a Kerr spacetime. 

\subsubsection{LAL-frame}

The \textbf{LAL-frame}, used in \texttt{SXS} catalogs and in \texttt{LALsuite}~\cite{Schmidt:2017btt,lal} is closely related to the $\bm{L}_N$-frame, but uses the opposite sign convention for the unit separation vector. We keep $\bm{z}_{S} = \bh{L}$, the unit vector along the orbital angular momentum (at some fiducial instance). However, we define $\bm{x}_{S} = -\bh{n}$ pointing from the less heavy object (secondary $m_2$) to the more heavy (primary $m_1$)\footnote{In the case of equal mass binaries, the labeling is arbitrary.}. This definition sets
\begin{alignat}{2}
  & \begin{dcases}
    \bm{x}_{S} = \bm{x}_{\rm LAL}\\
    \bm{y}_{S} = \bm{y}_{\rm LAL}\\
    \bm{z}_{S} = \bm{z}_{\rm LAL}
  \end{dcases}
    & \quad \text{such that} \quad & 
    \begin{dcases}
    \bm{x}_{\rm LAL} = -\bm{x}_{L_N} = -\bm{n} \,,\\
    \bm{y}_{\rm LAL} = -\bm{y}_{L_N} = -\bm{\lambda} \,,\\
    \bm{z}_{\rm LAL} = \bm{z}_{L_N} = \bm{\ell} = \hat{\bm{L}}_N\,.
  \end{dcases}
\label{eq:LALframe}
\end{alignat}

\subsubsection{FEW-frame}
\label{sec:EMRI_frame}

In the \texttt{FEW} package, the $S_1$-frame convention $\bm{z}_S = \bh{S}_1$ is adopted. This is natural because the spin of the primary black hole provides a $z$-axis for the Kerr spacetime, and both $\bh{J}$ and $\bh{S}$ reduce to $\bh{S}_1$ in the limit of an infinitesimal mass ratio. However, the convention used in \texttt{FEW} differs in its logic from that of the $S_1$-frame, as the \texttt{FEW} ``source frame'' definition makes use of the direction towards the observer. In contrast, in all the other conventions considered above, the source frame is completely specified only from the orbit of the binary (at some reference time).

\begin{figure}
    \centering
    \includegraphics[width=0.32\textwidth]{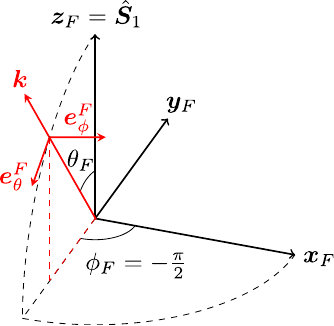}
    \caption{\texttt{FEW} frame as defined in the code \texttt{FastEMRIWaveforms} (\texttt{FEW}). Figure adapted from~\cite{katz2021fast}. The $\bm{z}_F$ axis is aligned with the spin of the more massive object, and $\bm{x}_F$ is orthogonal to both the wave propagation vector $\bm{k}$ and $\bm{z}_F$.}
    \label{fig:few_conventions}
\end{figure}

Denoting by $(\bm{x}_F,\bm{y}_F,\bm{z}_F)$ the \texttt{FEW} source frame, and by $(\bm{e}_\theta^F,\bm{e}_\phi^F)$ the spherical vectors forming a triad with $\bm{k}$ (and similarly $(\bm{e}_\theta^J,\bm{e}_\phi^J)$ for the J-frame), if $(\theta_F,\phi_F)$ are the spherical angles of $\bm{k}$ in the \texttt{FEW}-frame, the frame is chosen so that $\bm{z}_F = \bh{S}_1$ and $\phi_F=-\pi/2$ (meaning $\bm{k}$, $\bh{S}_1$, and $\bm{y}_F$ are coplanar). The \texttt{FEW} convention consists therefore in choosing
\begin{alignat}{2}
  & \begin{dcases}
    \bm{x}_{S} = \bm{x}_{F}\\
    \bm{y}_{S} = \bm{y}_{F}\\
    \bm{z}_{S} = \bm{z}_{F}
  \end{dcases}
    & \quad \text{such that} \quad & 
    \begin{dcases}
    \bm{z}_F = \bh{S}_1 \,,\\
    \bm{x}_F = \frac{\bh{S}_1 \times \bm{k}}{|\bh{S}_1 \times \bm{k}|} \,,\\
    \bm{y}_F = \bm{z}_F \times \bm{x}_F \,,
  \end{dcases}
\label{eq:def_xyzF}
\end{alignat}
together with a definition of the polarization vectors that differs from what we provided in Eq.~\eqref{eq:polarization-basis-vectors-pq}:
\begin{subequations}
\begin{align}
    \bm{p}_F &= \bm{e}_\phi^F \,,\\
    \bm{q}_F &= -\bm{e}_\theta^F \,.
\end{align}
\end{subequations}
Note that $(\bm{p}_F,\bm{q}_F)$ differ from $(\bm{p}_J,\bm{q}_J)=(\bm{e}_\theta^J,\bm{e}_\phi^J)$ in two ways: first, one chooses $(\bm{e}_\phi,-\bm{e}_\theta)$ instead of $(\bm{e}_\theta,\bm{e}_\phi)$; second, the  vectors $(\bm{e}_\theta^F,\bm{e}_\phi^F)$ differ from $(\bm{e}_\theta^J,\bm{e}_\phi^J)$ because they are spherical vectors associated with different frames. To convert between the \texttt{FEW}-frame and the other frames described above, we must have access to the unit separation vector $\bm{n}$ between the two bodies. This vector $\bm{n}$ is not natively computed in \texttt{FEW}. We outline our routine for constructing it below.

\begin{figure}
		\centering
		\includegraphics[width = \textwidth]{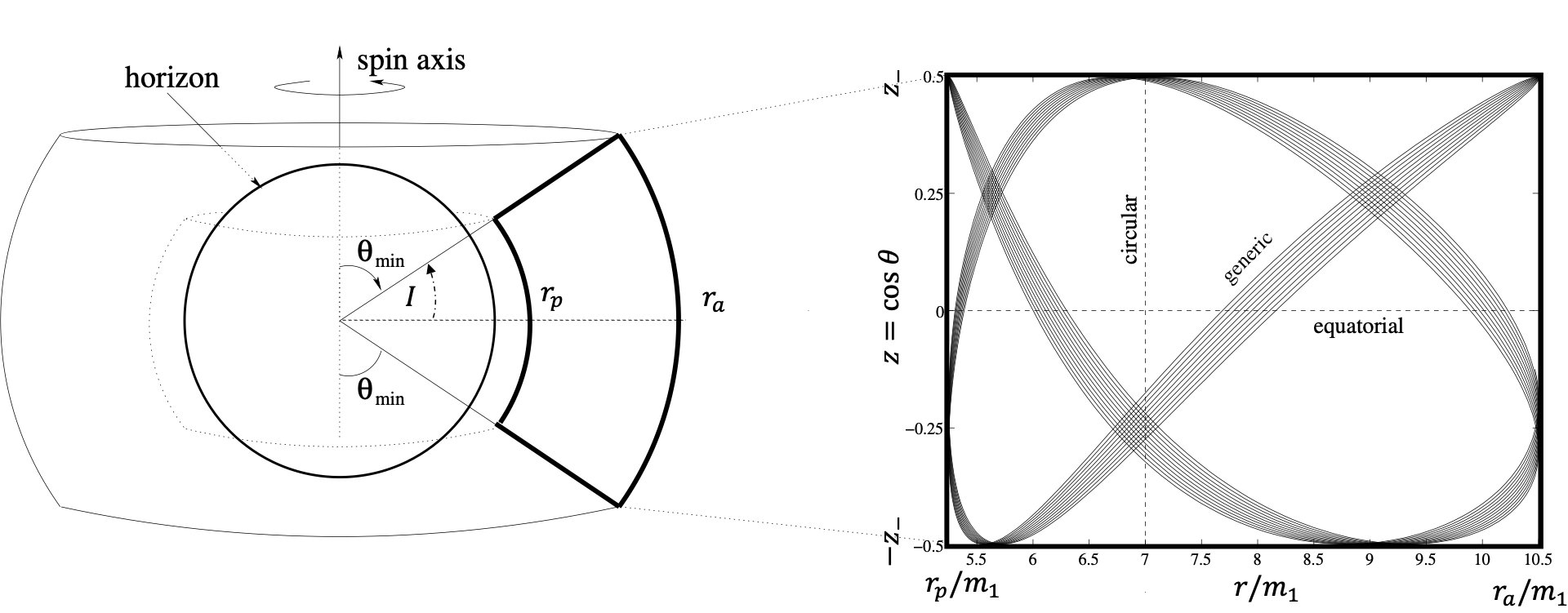}
		\caption{Illustration of the radial and polar motion of generic orbits around a Kerr primary. The graph shows how the motion fills a torus which is bounded by the radial roots $r_a$ and $r_p$ and the value of the polar root $z_-$. This illustration was taken from \cite{Drasco:2005kz} with minor alterations. }
		\label{fig:Kerr_Geodesics}
	\end{figure}

The separation  vector $\bm{x}$ between the primary and the secondary can be described using a quasi-Keplerian parametrization of the secondary's Boyer-Lindquist coordinate trajectory. The secondary's orbital radius $r$ around the primary is written in terms of a radial phase $\psi\in[0,2\pi)$ as
\begin{equation}
r(\psi) = \frac{p M}{1 + e\cos\psi}\,,
\end{equation}
with dimensionless semi-latus rectum $p$ and eccentricity $e$ given by 
\begin{subequations}\label{eq:primary_roots_inverse}
		\begin{gather}
			p = \frac{2 r_a r_p}{M(r_a + r_p)} \quad \text{and}   \quad e = \frac{r_a-r_p}{r_a+r_p}.    \tag{\theequation a-b}
		\end{gather}
	\end{subequations}
Here $\psi$ is the relativistic anomaly, not to be confused with the polarization, defined such that $r_{p} = r(\psi = 0)$ and $r_{a} = r(\psi = \pi)$ respectively represent the periapsis and apoapsis, the Boyer-Lindquist orbital radii at closest and furthest approach. Similarly, the secondary's Boyer-Lindquist polar angle $\theta$ is parametrized in terms of a phase $\chi\in[0,2\pi)$ as
\begin{equation}
\cos \theta(\chi) = z(\chi) = z_{-}\cos(\chi),     
\end{equation}
with $\theta=\pi/2$ corresponding to the equatorial plane normal to the spin vector $\hat{\boldsymbol{S}}_{1}$. 
Here $z_{-}$ is the smallest root of the polar potential in the Kerr geodesic equations~\cite{schmidt2002celestial}, and $r_a z_-$ is the orbit's maximum elevation above the equatorial plane. We also define the inclination angle $I$ as the orbit's maximum Boyer-Lindquist polar angle relative to the equator (i.e., $I=\pi/2-\theta_{\rm min}$). Concretely, we define\footnote{Semi-relativistic ``kludge" waveforms~\cite{babak2007kludge,chua2017augmented} use a slightly different definition of inclination given by $Y = L_{z} / \sqrt{L_{z}^2 + Q}$. Conversion from $x_{I}$ to $Y$ is trivial, and a function to numerically convert from $Y$ to $x_I$ is provided in the \texttt{FEW} package~\cite{katz2021fast}.} 
\begin{equation}\label{eq:FEW_inclination_reparam}
    x_{I} = \cos I = \pm\sqrt{1 - z_{-}^2}.
\end{equation}
If $x_{I} > 0$ $(< 0)$, the orbit is prograde (retrograde). A useful schematic describing the orbit is given in Fig.~\ref{fig:Kerr_Geodesics}. Under radiation reaction, $(p,e,x_I)$ all become time-varying quantities, $(p(t),e(t),x_I(t))$. 

Rather than using the quasi-Keplerian phases $(\psi, \chi, \phi)$ to describe the orbit, \texttt{FEW} uses Boyer-Lindquist-time action angles $(\Phi_{r},\Phi_{\theta},\Phi_{\phi})$; the angle $\Phi_r$ (for example) is the relativistic mean anomaly. A trajectory $(p(t),e(t),x_{I}(t), \Phi_{r}(t), \Phi_{\theta}(t),\Phi_{\phi}(t))$ as a function of coordinate time $t$ is built by solving differential equations describing the orbital evolution, given a set of initial input parameters $\{m_{1},m_{2},\chi_1,p_{0},e_{0},x_{I,0},\Phi_{r_,0}, \Phi_{\theta,0}, \Phi_{\phi,0}\}$; additional details are provided in Appendix~\ref{appendix:emris}. Extracting the Boyer-Lindquist coordinate separation at a given time $t$ then requires the transformation from the action angles $(\Phi_{r}(t), \Phi_{\theta}(t), \Phi_{\phi}(t))$ to the quasi-Keplerian angles $(\psi(t), \chi(t), \phi(t))$.  Reference~\cite{Lynch:2024hco} provides an efficient numerical root-finding algorithm, with C and Python implementations, to perform this transformation at leading, adiabatic (0PA) order. Access to $(\psi(t), \chi(t), \phi(t))$ then allows us to describe the position of the secondary in Cartesian coordinates defined from the Boyer-Lindquist coordinates in the natural way:
\begin{subequations} \label{eq:Cartesian_Coordinates_FEW}
    \begin{gather}
			x_C(t) = r(t) \sin \theta(t) \cos \phi(t),   \quad  y_C(t)  = r(t) \sin \theta(t) \sin \phi(t), \quad \quad  z_C(t) = r(t) \cos \theta(t).  \tag{\theequation a-c}
		\end{gather}
\end{subequations}
The unit separation vector at an instant of time $t$ is then given by 
\begin{equation}
\bm{n}(t) = \left(\frac{x_{C}(t)}{r(t)}, \frac{y_{C}(t)}{r(t)}, \frac{z_{C}(t)}{r(t)}\right).    
\end{equation}

Beyond 0PA order, an EMRI waveform will not naturally be in the \texttt{FEW} frame presented here because the primary will no longer be stationary relative to the Boyer-Lindquist coordinate origin, and its spin might not be aligned with the Boyer-Lindquist $z$ axis. The relationship between the waveform and the orbital trajectory $(x_C(t), y_C(t),z_C(t))$ will also depend on the spacetime foliation along with a number of other gauge choices. Controlling the frame at 1PA order and transforming it to the frames presented here remains an open problem.

\subsubsection{Transformation from the J-frame to the FEW-frame}
Given $J$-frame components of the spins, and spherical angles $(\theta_J,\phi_J)$ of the direction to the observer, we can use
\begin{equation}
    \bm{k} = \sin\theta_J \cos\phi_J\bm{x}_J + \sin\theta_J \sin\phi_J\bm{y}_J + \cos\theta_J \bm{z}_J \,,
\end{equation}
and compute $\bm{x}_F,\bm{y_F}$ using~\eqref{eq:def_xyzF}.

Regardless of their definitions, we can also use directly vector data to determine the polarization angle change (rotation around $\bm{k}$) induced by the change of definition of $(\bm{p},\bm{q})$: if $(\bm{p}_F,\bm{q}_F)$ are obtained from $(\bm{p}_J,\bm{q}_J)$ by a direct rotation of angle $\delta \psi_F$ around $\bm{k}$, then we can compute that angle as (in the convention $\mathrm{arctan}[y/x] = \mathrm{arctan}_2[y,x]$)
\begin{equation}
    \delta \psi_F = \mathrm{arctan}_2  \left[ \bm{p}_F \cdot \bm{q}_J , \bm{p}_F \cdot \bm{p}_J \right] \,.
\end{equation}


\subsubsection{Transformation from the L-frame to the J-frame}

We start with the $L$-frame. 
We presume that the orientation of spins is defined in this frame by angles $\theta_{S_{1,2}}, \phi_{S_{1,2}}$, or by the spin components on the basis vectors $(\bm{x}_L,\bm{y}_L,\bm{z}_L)$ (as the \texttt{LAL} interface does for the LAL-frame). The total momentum is already given by Eq.~\eqref{Eq:J}. The fact that $(\bh{J},\bm{x}_L,\bm{x}_J)$ are in the same plane can be expressed as 
\bea
 \bm{x}_L = \bh{J} \cos{\theta_{Jn}} + \bm{x}_J\sin{\theta_{Jn}},
 \label{Eq:xL}
\ena
where $\cos{\theta_{Jn}} = (\bm{x}_L.\bh{J}) = \sin{\theta_{JL} \cos{\phi_{J}}}$, which gives $(\bm{x}_L.\bm{x}_J) = \sin{\theta_{Jn}}$. From the same equation, we can deduce
$$
(\bm{x}_J.\bh{L}) = -\frac{ \cos{\theta_{Jn}} }{ \sin{\theta_{Jn}} }  \cos{\theta_{JL}},
$$
and writing explicitly the rotation we also get
$$
(\bm{x}_J.\bm{y}_L) = -\frac{ \cos{\theta_{Jn}} }{ \sin{\theta_{Jn}} }  \bm{x}_L \cdot \left( \hat{\bm{J}} \times \hat{\bm{L}}\right) \,.
$$
We arrive at the expression of $\bm{x}_J$ decomposed on the L-frame:
\bea
\bm{x}_J = \sin{\theta_{Jn}} \bm{x}_L - 
\left(\frac{ \cos{\theta_{Jn}} }{ \sin{\theta_{Jn}} }  \bm{x}_L \cdot \left( \hat{\bm{J}} \times \hat{\bm{L}}\right) \right) \bm{y}_L -
\left(\frac{ \cos{\theta_{Jn}} }{ \sin{\theta_{Jn}} }  \cos{\theta_{JL}}\right) \bh{L}.
\ena
The third axis can be obtained from the vector product $\bm{y}_J = \bh{J}\times \bm{x}_J$. Finally, the direction of the periapse is given as $\bh{a} = \cos{\xi} \bm{x}_L + \sin{\xi} \bm{y}_L.$

\subsubsection{Transformation from the J-frame to the L-frame}

Now we start from the $J$-frame. The spins are assumed to be given in that frame $(\bm{x}_J,\bm{y}_J,\bm{z}_J)$, and again, we are using~\autoref{Eq:J}.
We express the unit vector $\bh{L}$ in $J$-frame: 
\begin{equation}
    \bh{L} = \left(\sin{\theta_{JL}} \cos{\phi_L}, \sin{\theta_{JL}} \sin{\phi_L}, \cos{\theta_{JL}}\right)
\end{equation},
\begin{subequations}
\begin{align}
    \bm{L} \cdot \bm{x}_J &= -\frac{1}{L_N}\left(\bm{S} \cdot \bm{x}_J \right) \,,\\
    \bm{L} \cdot \bm{y}_J &= -\frac{1}{L_N}\left(\bm{S} \cdot \bm{y}_J\right) \,,\\
    \bm{L} \cdot \bh{J} &= \frac{1}{L_N}\left(|\bm{J}| - \bm{S}\cdot \bh{J}\right) \,,
\end{align}
\end{subequations}
where $\bm{S}$ is the total spin $\bm{S} = \bm{S}_1 + \bm{S}_2$, of which we know the components in the $J$-frame.
The $\bm{x}_L$-axis can be found using \autoref{Eq:xL} with the angle $\theta_{Jn} \in [0,\pi]$ given by: 
\bea
    \tan{\theta_{Jn}} = -\frac{(\bh{L}.\bh{J})}{(\bh{L}.\bm{x}_J)} = -\frac{\cos{\theta_{JL}}}{\sin{\theta_{JL}}\cos{\phi_L}}
\ena
Finally, we can find $\bm{y}_L = \bh{L} \times \bm{x}_L$ and  $\bh{a} = \cos{\xi} \bm{x}_L + \sin{\xi} \bm{y}_L$. Sometimes parameters could be specified in the mixed coordinates, the popular choice is we use $\bh{J}$-frame where we specify the orientation of $\bh{L}$ but the spins components are given in $\bh{L}$-frame. The transformation can be obtained using \autoref{Eq:J} and \autoref{Eq:xL}, we will give it explicitly if needed later.

\subsection{Rotation of the waveforms}

Often the two waveform polarizations are decomposed in the spin-weighted (-2) spherical harmonics:
\bea
\label{eq:spherical-harmonics-decomposition}
  h_{+}(t) - i h_{\times}(t) = \sum_{\ell=2}^{+\infty} \sum_{m=-\ell}^{\ell} h_{\ell m}(t) \sYlm(\iota, \varphi) \,,
\ena
where $\iota$ and $\varphi$ are the polar and azimuthal angles in the source coordinate system, as depicted in Fig.~\ref{F:source_frame}.
Our convention for the spin-weighted spherical harmonics is the same as the one used in \texttt{LAL}~\cite{lal}, in~\cite{ajith2011dataformats} and~\cite{blanchet_post-newtonian_2024}:
\bea\label{eq:defsYlm}
	&\sYlm (\iota_{0}, \varphi_{0}) = \sqrt{\frac{2\ell +1}{4\pi}} d^{\ell}_{m,2}(\iota_{0}) e^{i m \varphi_{0}} \,,\\
	&d^{\ell}_{m,2}(\iota_{0}) = \sum_{k = k_{1}}^{k_{2}} \frac{(-1)^{k}}{k!} \frac{\sqrt{(\ell+m)! (\ell - m)! (\ell+2)! (\ell-2)!}}{(k-m+2)! (\ell+m-k)! (\ell-k-2)!} \left( \cos \frac{\iota_{0}}{2} \right)^{2\ell + m - 2k -2} \left( \sin \frac{\iota_{0}}{2} \right)^{2k - m +2} \,,
\ena
with $k_{1} = \max (0,m-2)$ and $k_{2} = \min(\ell +m, \ell -2)$. The polarizations can be expressed as
\bsub
\begin{align}\label{eq:hphcmodesgeneric}
	h_{+} = \frac{1}{2} \sum_{\ell, m} \left( \sYlm h_{\ell, m} + \sYlmstar h_{\ell, m}^{*} \right) \,,\\
	h_{\times} = \frac{i}{2} \sum_{\ell, m} \left( \sYlm h_{\ell, m} - \sYlmstar h_{\ell, m}^{*} \right) \,,
\end{align}
\esub
Note that our polarization vectors~\eqref{eq:defpq} differ from the PN convention of~\cite{Blanchet:2013haa} by a rotation of $\pi/2$, which translates into an overall factor $(-1)$ in the polarizations $h_{+,\times}$ and in the modes $h_{\ell m}$.

For non-precessing binary systems, with a fixed equatorial plane of orbit, an exact symmetry relation between modes holds:
\be\label{eq:nonprecsymmetry}
	h_{\ell, -m} = (-1)^{\ell} h_{\ell m}^{*} \,.
\en
When this symmetry is verified, we can write
\be\label{eq:hpcKpc}
	h_{+,\times} = \sum_{\ell, m} K_{\ell m}^{+, \times} h_{\ell m} \,, 
\en
with
\bsub\label{eq:defKpc}
\begin{align}
	K_{\ell m}^{+} =\frac{1}{2} \left( \sYlm + (-1)^{\ell} \sYlminusmstar \right) \,,\\
	K_{\ell m}^{\times} = \frac{i}{2} \left( \sYlm - (-1)^{\ell} \sYlminusmstar \right) \,.
\end{align}
\esub

\subsection{Frequency domain}\label{sec:frequency}

For the conventions regarding Fourier transforms see Sec.~\ref{sec:FT}.

A useful relation is
\begin{equation}
	\widetilde{F^*} (f) = \widetilde{F}(-f)^*.
\end{equation}
For a real-valued time-domain signal one therefore obtains that
\begin{equation}
	F(t) \in \mathbb{R} \implies \widetilde{F}(-f) = \widetilde{F}(f)^{*}.
    \label{eq:FTreal}
\end{equation}
For non-precessing systems, \eqref{eq:nonprecsymmetry} implies
\be
	\tilde{h}_{\ell, -m} (f) = (-1)^{\ell} \tilde{h}_{\ell m} (-f)^{*} \,.
\en

Since a given mode has a phase dependency $h_{\ell m} \propto \exp[- i m \phi_{\rm orb}]$, with the  orbital phase verifying $\dot{\phi}_{\rm orb} > 0$, an approximation often used for non-precessing systems (or in the precessing frame for a binary with misaligned spins) is
\bsub\label{eq:neglectposmposf}
\begin{align}
	\tilde{h}_{\ell m} (f) &\simeq 0 \; \text{for} \; m > 0 \,, f > 0 \,,\\
	\tilde{h}_{\ell m} (f) &\simeq 0 \; \text{for} \; m < 0 \,, f < 0 \,,\\
	\tilde{h}_{\ell 0} (f) &\simeq 0 \,.
\end{align}
\esub
Note that in the Fourier convention~\eqref{eq:defFourier}, this approximation means that for positive frequencies $f>0$ the mode $\tilde{h}_{2,-2}(f)$ has support while $\tilde{h}_{22}(f)$ is negligible\footnote{Note that because of this, some parts of the \gls{ldc} code internally use the opposite sign convention in the exponentials of~\eqref{eq:defFourier}, so as to have support for the modes $m>0$ for $f>0$. This amounts to a mapping $f\leftrightarrow -f$, which can be undone at the end of the computation by conjugating the FT of the observables, which are real signals (see~\eqref{eq:FTreal}).}. When using the approximation~\eqref{eq:neglectposmposf}, \eqref{eq:hpcKpc} becomes for $f>0$
\be\label{eq:hpcKpcFT}
	\tilde{h}_{+,\times} (f) = \sum_{\ell} \sum_{m<0} K_{\ell m}^{+, \times} \tilde{h}_{\ell m} \,.
\en

\subsection{\label{sec:eccentricity}Definition of eccentricity}

We adopt the gauge-independent definition of eccentricity presented in Refs.~\cite{Ramos-Buades:2022lgf,Shaikh2023}, which is based on the waveform in the source frame.

We consider the decomposition of the waveform in spin-weighted spherical harmonics introduced in Eq.~\eqref{eq:spherical-harmonics-decomposition}. We can write the $(2, 2)$ mode in terms of complex amplitude and phase:
\begin{equation}
    h_{22}(t) = A_{22}(t) e^{-\Phi_{22}(t)}\,.
\end{equation}
We can then define the instantaneous angular frequency of the $(2, 2)$ mode as
\begin{equation}
    \omega_{22}(t) = \frac{d \Phi_{22}(t)}{dt}\,.
\end{equation}
We start adopting the definition of Ref. \cite{Mora:2002gf}, which is based on the orbital frequency,
\begin{equation}
    e_{\omega_{\rm orb}}(t)=\frac{\sqrt{\omega_{\rm orb}^{\mathrm{p}}(t)}-\sqrt{\omega_{\rm orb}^{\mathrm{a}}(t)}}{\sqrt{\omega_{\rm orb}^{\mathrm{p}}(t)}+\sqrt{\omega_{\rm orb}^{\mathrm{a}}(t)}},
    \label{eq:eqEccOmOrb}
\end{equation}
where the orbital frequency is defined as the time derivative of the orbital phase $\omega_{\rm orb} = d\phi_{\rm orb}/dt$, and the superscripts $\rm{a,p}$ refer to the apastron and periastron, respectively. The definition in Eq. \eqref{eq:eqEccOmOrb} reduces to the Newtonian definition of eccentricity \cite{Mora:2002gf}.

To avoid the coordinate dependence of Eq. \eqref{eq:eqEccOmOrb} one can apply Eq. \eqref{eq:eqEccOmOrb} to the $(2,2)$ mode angular frequency \cite{Ramos-Buades:2019uvh}
\begin{equation}
    e_{\omega_{22}}(t)=\frac{\sqrt{\omega_{22}^{\mathrm{p}}(t)}-\sqrt{\omega_{22}^{\mathrm{a}}(t)}}{\sqrt{\omega_{22}^{\mathrm{p}}(t)}+\sqrt{\omega_{22}^{\mathrm{a}}(t)}},
\label{eq:eqEccOmega2}
\end{equation}    
where $\omega_{22}^{\mathrm{p}}(t)$ and $\omega_{22}^{\mathrm{a}}(t)$ are interpolants through $\omega_{22}(t)$ evaluated at pericenters and apocenters, respectively. More precisely, at a given time $t$, we take $\omega_{22}^{\mathrm{p}}(t)$ to be the local maximum of $\omega_{22}(t)$, and $\omega_{22}^{\mathrm{a}}(t)$ to be the local minimum.

However, the definition in Eq. \eqref{eq:eqEccOmega2} does not reduce to the Newtonian definition of eccentricity in the Newtonian limit \cite{Ramos-Buades:2022lgf}. To obtain the correct Newtonian limit at large binary separations, one can define the eccentricity as \cite{Ramos-Buades:2022lgf} 
\begin{eqnarray}
    e_{\mathrm{gw}} & = & \cos (\Psi / 3)-\sqrt{3} \sin (\Psi / 3)\,, \\
   \Psi & = & \arctan \left(\frac{1-e_{\omega_{22}}^2}{2 e_{\omega_{22}}}\right) \, .
\label{eq:eqEgw}
\end{eqnarray}

This definition is applicable to any binary independent of the mass ratio as long as there are enough orbits to define periastra and apastra. For large mass ratios $Q$, $e_{\mathrm{gw}}$ remains close (within a few $10^{-3}$, see \cite{Ramos-Buades:2022lgf,Shaikh2023} for details) to the geodesic eccentricity $e$ of a test particle around a Kerr black hole used in EMRI computations, which we introduced in Eq.~\eqref{eq:primary_roots_inverse}.

Regarding the second parameter describing the ellipse, different sources use distinct quantities to generate the waveforms. For instance, in the EMRI case the waveform is generated specifying the semilatus rectum (see Eq. \eqref{eq:primary_roots_inverse}), while waveform models for massive black hole binaries typically use an anomaly angle \cite{Nagar:2021gss,Ramos-Buades:2021adz,Gamboa:2024hli}. Similarly as in the case of the eccentricity parameter one can use the mean anomaly as the second parameter characterizing the ellipse based on the waveform modes. In particular, the mean anomaly describes the fraction of the orbital period that has elapsed since the last pericenter passage ~\cite{Schmidt:2017btt,Shaikh2023}
\begin{equation}
\label{eq:mean_anomaly_definition}
  l_{\rm gw} (t) = 2\pi \, \frac{t - t^P_{i}}{t^P_{i+1} - t^P_{i}}.
\end{equation}
The interval $t^P_{i} \leq t < t^P_{i+1}$ defines any two consecutive pericenter passages $t^P_i$ and $t^P_{i+1}$.  Similarly as in Eq. \eqref{eq:eqEccOmega2}, the pericenter passages entering the calculation of Eq. \eqref{eq:mean_anomaly_definition} can also be computed from the waveform \cite{Shaikh2023}. In Newtonian gravity the period of the orbit $T = t^P_{i+1} - t^P_{i}$ remains constant, while in General Relativity, due to gravitational radiation reaction $T$ decreases over time, making  $l_{\rm gw}(t)$ a stepwise linear function whose slope increases as the binary evolves to merger. 

Both $e_{\rm gw}$ and $l_{\rm gw}$ can be a (2,2)-mode from any source with enough orbital cycles (at least one orbital period is needed) \cite{Ramos-Buades:2022lgf,Shaikh2023}, and they have been used for parameter estimation analyses in the context of the LVK \cite{Bonino:2022hkj,Ramos-Buades:2023yhy,Gupte:2024jfe}.

\subsection{Reference time}\label{sec:reference_time}

This section provides a standard method to choose the reference time in the binary system's orbit to fully define the source frames described in Section~\ref{sec:source_frames}. This method depends on the \gls{gw} source type.

\subsubsection{EMRI reference time}

EMRI waveforms generated in \texttt{FEW} are specified by parameters $(m_1,m_2,\chi_1,p_0,e_0,x_{I,0},\Phi_{r,0},\Phi_{\theta,0}$, $\Phi_{\phi,0})$ defined at some time $t_0$. Current infrastructure in \texttt{FEW} allows generation of waveforms by integrating equations of motion forward in time from $t_0=0$ or backward in time from $t_0 = t_{\text{final}}$. $t_0=0$ could be interpreted as the initial time the instrument is in ``science mode" and begins taking data, for example. Future developments are expected to allow users to specify any $t_0 = t_{\rm ref}$ with the freedom to either (1) integrate forwards until near plunge, (2) integrate backwards to a stopping point determined by the user, or (3) a mixture of the two. A choice of reference time that is useful for analysis purposes might not be appopriate for final summary statistics for cataloguing, and for that reason the choice can be left to each user. 

	\begin{figure}
		\centering
		\includegraphics[width = .75\textwidth]{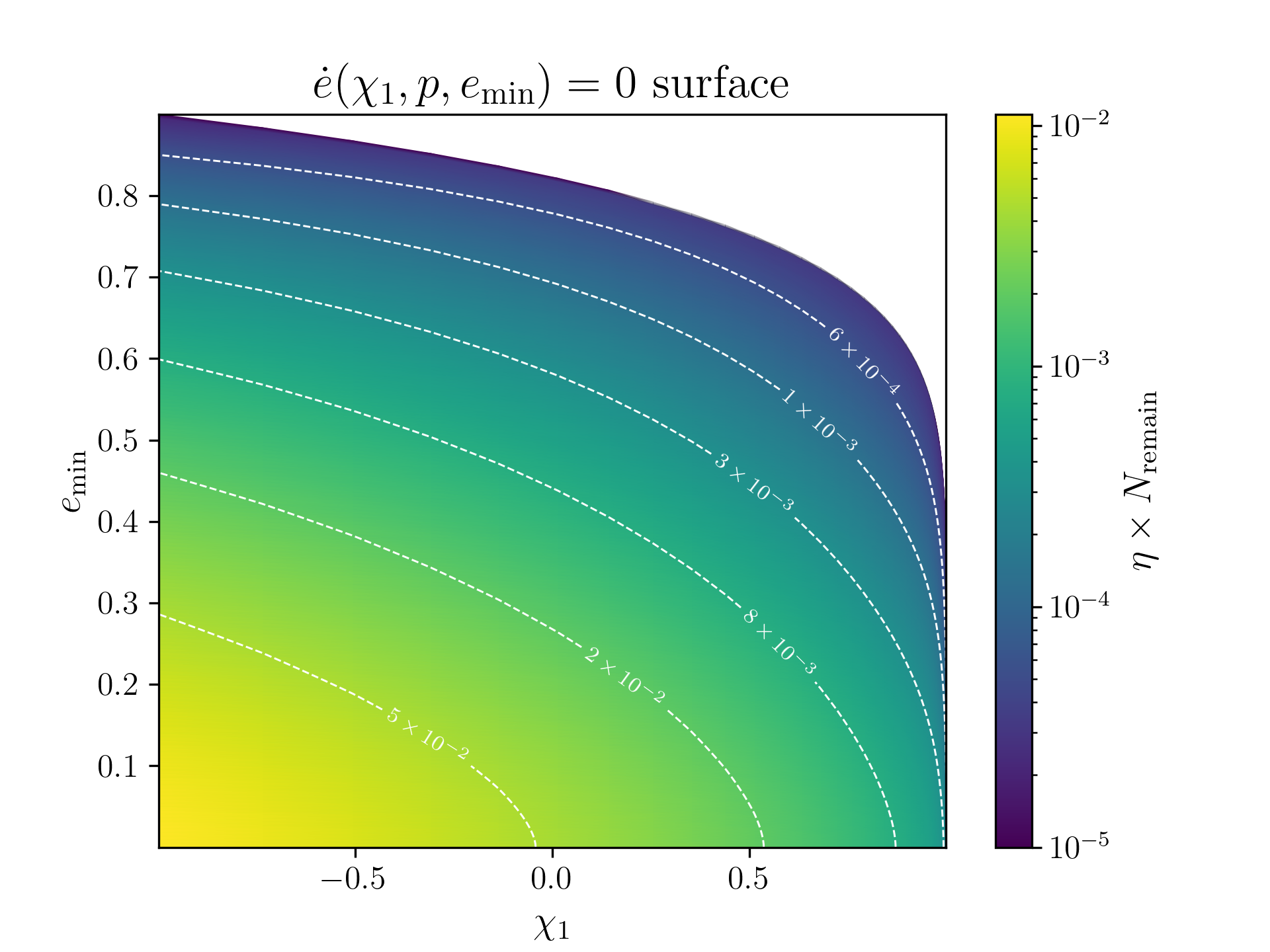}
		\caption{Number of radial cycles remaining in an equatorial EMRI when the EMRI reaches its minimum eccentricity. We take this minimum eccentricity as the definition of ``eccentricity at plunge''. Here negative spin values correspond to retrograde orbits and positive ones to prograde orbits. The number of remaining cycles displayed is scaled by the symmetric mass ratio; for an EMRI with mass ratio $\eta=10^{-5}$, the number ranges from $\approx 1$ cycle for highly eccentric, prograde orbits to $\approx 10^{3}$ cycles for low-eccentricity, retrograde orbits.}
		\label{fig:e_at_plunge}
	\end{figure}

However, since the system's final eccentricity (``at plunge'') is an important parameter for catalogues, one desirable type of reference time is the time at which the orbit satisfies some identifiable criterion near the end of the inspiral. There is no unique definition of the time at which the transition from inspiral to plunge occurs, and the separatrix between inspiral and plunge is not currently within the parameter range of \texttt{FEW}. But a robust notion that can be used and precisely measured is the time at which the secondary's orbit reaches its minimum eccentricity, where we make use of the quasi-Keplerian definition of eccentricty in \eqref{eq:primary_roots_inverse}. This occurs somewhere slightly outside the separatrix (at a dimensionless semi-latus-rectum distance $p-p_{\rm sep}\lesssim 1$), after which time the eccentricity increases until the final plunge.

Figure~\ref{fig:e_at_plunge} displays the number of remaining radial cycles that occur after the orbit reaches its minimum eccentricity, in the special case of equatorial orbits (i.e., orbits with inclination angle $I=0$). Although the number of remaining cycles can be high, it typically falls in the range $1$--$100$ for expected mass ratios. Moreover, the minimum eccentricity has several advantages: it is uniquely identifiable and gauge invariant at 0PA order; it lies in a regime where \texttt{FEW} models are accurate (as opposed to the transition to plunge); and it always lies within the \texttt{FEW}v2.0.0 grid (unlike the separatrix). We therefore take the minimum eccentricity in an EMRI as our definition of its ``eccentricity at plunge''. To extend this to non-eccentric orbits, we introduce a reference semilatus rectum $p_\mathrm{ref}(e_\mathrm{min}, x_I)$ corresponding to the value at which this minimum occurs. This value remains well-defined as $e_\mathrm{min} \rightarrow 0$. The time at which an EMRI reaches $p_\mathrm{ref}(e_\mathrm{min}, x_I)$ then provides a candidate reference time for both eccentric and non-eccentric systems. At 1PA, this definition is no longer gauge invariant; it will be the focus of future work to extend this definition to higher PA orders, potentially making use of $e_{\rm gw}$ as defined in Eq. \eqref{eq:eqEgw}, rather than the quasi-Keplerian eccentricity.

\subsubsection{MBHB reference time}

For massive black hole binaries (MBHBs), we follow the strategy used by the LVK collaboration, where parameters are defined at a chosen reference frequency $f_{\rm ref}$ in SI units. This frequency is selected by the user, typically near the starting frequency of the (2,2) gravitational wave (GW) mode and located in the inspiral phase. This choice allows one to approximate the Fourier variable $f$ as the instantaneous GW frequency of the (2,2) mode (expressed in the co-precessing frame for precessing systems), making it easier to associate a corresponding time.

The advantage of using the frequency in SI units is that it can be related to the observable frequencies in the detector. However, depending on the total mass of the system, it migth significantly change the time to merger between different waveforms. Very high mass systems would locate the reference frequency close to merger or in the ringdown, making the assumption that it locates in the inspiral invalid. The user will need to be aware and adapt the frequency if needed.

The associated time is known as the reference time $t_{\rm ref}$, which is particularly relevant for time-domain models that evolve the system in a discrete time array rather than in the Fourier domain. When necessary, time-domain models compute $t_{\rm ref}$ from $f_{\rm ref}$ using their internal description of the binary’s dynamics, assuming that the reference frequency corresponds to the (2,2) GW frequency, which is related to the orbital frequency as $f_{22}\approx 2f_{\rm orbital}$. For example,  the IMRPhenomT model provides an analytical description of the frequency as a function of time, and $t_{\rm ref}$ is computed through a root finding of this expression for the input value of $f_{\rm ref}$. 

In the case of precessing systems, the reference frequency is associated to the (2,2) GW frequency in a co-precessing frame (which might not be exactly the same for different models), since the inertial frame mixes both the orbital and precessing frequencies. In the co-precessing frame, the precessing motion is minimized and the GW frequency is closer to the orbital one. In summary, the orbital frequency in the co-precessing frame equals $f_{\rm ref}/2$ at $t=t_{\rm ref}$.

Whether using $t_{\rm ref}$ or $f_{\rm ref}$, this instant defines the {\it reference frame} in which the binary’s spin components are specified, as well as its orientation relative to the observer. Specifically, it sets the inclination angle $\iota$ and the reference phase $\phi_{\rm ref}$ (or azimuthal angle), both of which enter the spin-weighted spherical harmonics that describe the GW emission.

The origin of time $t=0$, is typically defined at the \textit{peak time} of the merger. However, the definition of the waveform peak is not uniform across models. A widely used approach defines the peak as the maximum of the squared sum of all harmonic amplitudes, see Eq.~4 in~\cite{schmidt2017}. In other cases, the peak is determined by the maximum of the squared sum of the two polarization components. In the \textsc{IMRPhenomT} model, the peak time corresponds to the maximum amplitude of the (2,2) mode in the co-precessing frame.

\subsubsection{Stellar-mass BHB reference time}

SBHBs are long-lived sources in the LISA data stream. Some of these systems will merge during the lifetime of LISA, and will exit the LISA band towards higher frequencies at some point during LISA observations. Others will only merge later, and LISA will only observe a snapshot of their inspiralling phase, during which the system will only chirp by a small amount. This motivates a difference in treatment with respect to MBHBs, which are primarily transient and merger-dominated (except at intermediate masses). Using a reference point at or near merger is not relevant for SBHBs, as that point is completely out-of-band. The definition of the orbital frequency for eccentric and precessing systems should follow the one chosen for MBHBs.

If the reference epoch (time or frequency) is used to define the source frame, which in turns defines the orbital phase, we presumably need to chose a reference that is relevant for all systems. The LISA observation window will select different snapshots of these system's inspiral tracks. For SBHBs, the start of the signal will always be in-band, with the track starting somewhere in LISA's sensitivity bucket.

For this reason, the preferred reference point could be taken at the starting time of LISA observations $t_{\rm ref} = t_{\rm start}$: this reference point is guaranteed to be relevant for all SBHBs as the starting point of their track. The waveform would then be parametrized by $f_{\rm start} = f(t_{\rm tstart})$ (or alternatively by time-to-merger). 

Contrasting this choice with different source classes, we note that picking a reference point as a given geometric frequency $M \omega_{22}^{\rm ref}$ could land during LISA observations but also before the signal starts or after it ends; it should also be checked as landing inside LISA's sensitivity for all masses (considering the lowest possible mass). Similarly, a reference point at a given Fourier frequency, for instance $f_{\rm ref} = 20 \mathrm{mHz}$, would present the advantage that we can choose this frequency to be always within LISA's sensitivity, but suffers from the same caveat that some systems might stop before reaching it or start after it.

It is probably important to remain flexible and allow for a different choice, if necessary, for each system. For the purpose of sampling, the optimal choice for reducing correlations may not be obvious beforehand, and $t_{\rm start}$ might be better replaced by a later point in the signal. For example, the start of LISA observations might correspond to a low point in the LISA response to a specific SBHB system.

\subsubsection{Galactic binaries reference time}

White dwarf binaries are mildly relativistic, most of them are expected to be on the circular orbits emitting GW at twice the orbital frequency. They are present in the LISA band throughout the mission duration. The frequency evolution (orbital dynamics) is driven by gravitational radiation for the detached systems. For the interacting systems (when one component overflows the Roche lobe) with accretion, the dynamics is defined by the interplay of mass transfer and GW dissipation. When the donor is the primary companion, we expect the outspiral (negative orbital frequency derivative). The most natural way to characterize the system is to set the orbital frequency and its derivative at some fiducial moment of time, which is usually chosen as \emph{start of observations}. The constant frequency derivative assumes a steady dynamic (does not cover on-off accretion). 

Eccentric binaries are expected in some cases (like hierarchical triplets), in which case the dynamics is described by a change of the (azimuthal) orbital frequency and eccentricity. In any case, that change in orbit is linear over the observation span, and \emph{start of observations} also seems to be a good choice. 

Triplet system: a fair amount (10\% or so) of WD binaries could be in the hierarchical triplet system. For those, we might measure the acceleration of the binary barycentre which adds 4 more parameters describing outer perturbed defined in \cite{Katz:2022izt}.

\subsection{Link to the LAL conventions (2018)}\label{sec:Link_to_LAL}

In conclusion of this section we explain how our conventions relate to those used in \texttt{LAL} (software library used by LIGO/Virgo collaborations~\cite{lal}), which are standardized in the document~\cite{Schmidt:2017btt}. First, in \texttt{LAL} the source frame is constructed by the convention of~\cite{Schmidt:2017btt}, setting the unit separation vector $\bm{n} = \vx_{S}$ and the normal to the orbital plane $\hat{\bm{L}}_{N} = \vz_{S}$, at the start of the waveform. In our case we have not specified how this source-frame is built, as this construction might differ for different physical systems (SMBHs, EMRIs, \dots).

Translated in the notations we used above, the choice of polarization vectors for LAL is
\bsub\label{eq:pqLAL}
\begin{align}
	\vp_{\rm LAL} &= - \vp \,,\\
	\vq_{\rm LAL} &= - \vq \,,
\end{align}
\esub
which amounts to a rotation of $\pi$ of the polarization basis. Because the polarizations $h_{+,\times}$ transform with $\cos 2\psi,\sin 2\psi$ under a change of polarization, this means that there is no difference here, provided the source-frame is identical:
\bsub
\begin{align}
	h_{+}^{\rm LAL} &= h_{+} \,,\\
	h_{\times}^{\rm LAL} &= h_{\times} \,.
\end{align}
\esub
The phase quantity used in \texttt{LAL} differs from our definition by
\be
	\Phi_{\rm LAL} = \frac{\pi}{2} - \varphi_{0} \,.
\en
Although the geometric definition of the reference polarization vectors $\vu$, $\vv$ is the same, the \gls{ssb} frame $(\vx, \vy, \vz)$ is different. In the case of \texttt{LAL}, it is a geocentric frame based on the celestial equator, while in our case it is the SSB frame based on the ecliptic plane. For the polarization, a difference of $\pi$ in the definition also comes from~\eqref{eq:pqLAL}:
\be
	\psi_{\rm LAL} \leftrightarrow \psi \,: \text{diff. of } \pi \text{ and different frame} \,.
\en
For the sky position, \texttt{LAL} uses $\mathrm{ra}, \mathrm{dec}$ instead of our ecliptic $\lambda, \beta$:
\be
	(\alpha_{\rm LAL}, \beta_{\rm LAL}) \leftrightarrow (\lambda, \beta) \,: \text{ different frame} \,.
\en
We leave for future work this exact map, relating the SSB frame to the geocentric frame.

\subsection{Example: quadrupole formula for a binary system}

In order to complete our conventions, we present a self-contained derivation of gravitational polarizations at leading order for a binary system on elliptical orbits, hopefully ironing out sign issues. Einstein's quadrupole formula for the gravitational radiation produced, in radiative coordinates and in the transverse-traceless gauge, reads
\begin{equation}
    h_{ij}^{\rm TT} = \frac{2G}{D_L c^4} \mathcal{P}_{ijab} \ddot{Q}_{ab}(t_{ret}) \,,
    \label{eq:hijTT}
\end{equation}
where $D_L$ is the luminosity distance to the source, 
$t_{ret}$ is the retarded time ($t_{ret}=T - R/c$ in flat space, where $R$ is the Euclidean distance, equal in this case to the luminosity distance),
$Q_{ij}$ is the mass quadrupole of the system $Q_{ij} = \int d^3x \, \rho x_i x_j$, and we introduced the projector
\begin{align}
    \mathcal{P}_{ijab} &= \mathcal{P}_{ia}\mathcal{P}_{jb} - \frac{1}{2} \mathcal{P}_{ij}\mathcal{P}_{ab}\,, \\
    \mathcal{P}_{ab} &= \delta_{ij} - k_i k_j \,,
\end{align}
that ensure that the resulting $h_{ij}^{\rm TT}$ is transverse to the direction of propagation $\bm{k}$ and traceless\footnote{For more details on this, see Sec. \ref{sec:source-frame-general}.}.

For a binary system represented as two point masses $m_1$, $m_2$, introducing the separation vector $\bm{x} = \bm{x}_1 - \bm{x}_2 = r \bm{n}$ with $\bm{x}_1 = m_2/M \bm{x}$, $\bm{x}_2 = -m_1/M \bm{x}$ for the total mass $M=m_1 + m_2$, the mass quadrupole is
\begin{equation}
    Q_{ij} = M \eta x_i x_j
    \label{eq:Qdef}
\end{equation}
with $\eta = m_1 m_2 / M^2$ the symmetric mass ratio. 

\begin{figure}[!]
    \centering
    \includegraphics[width=0.8\textwidth]{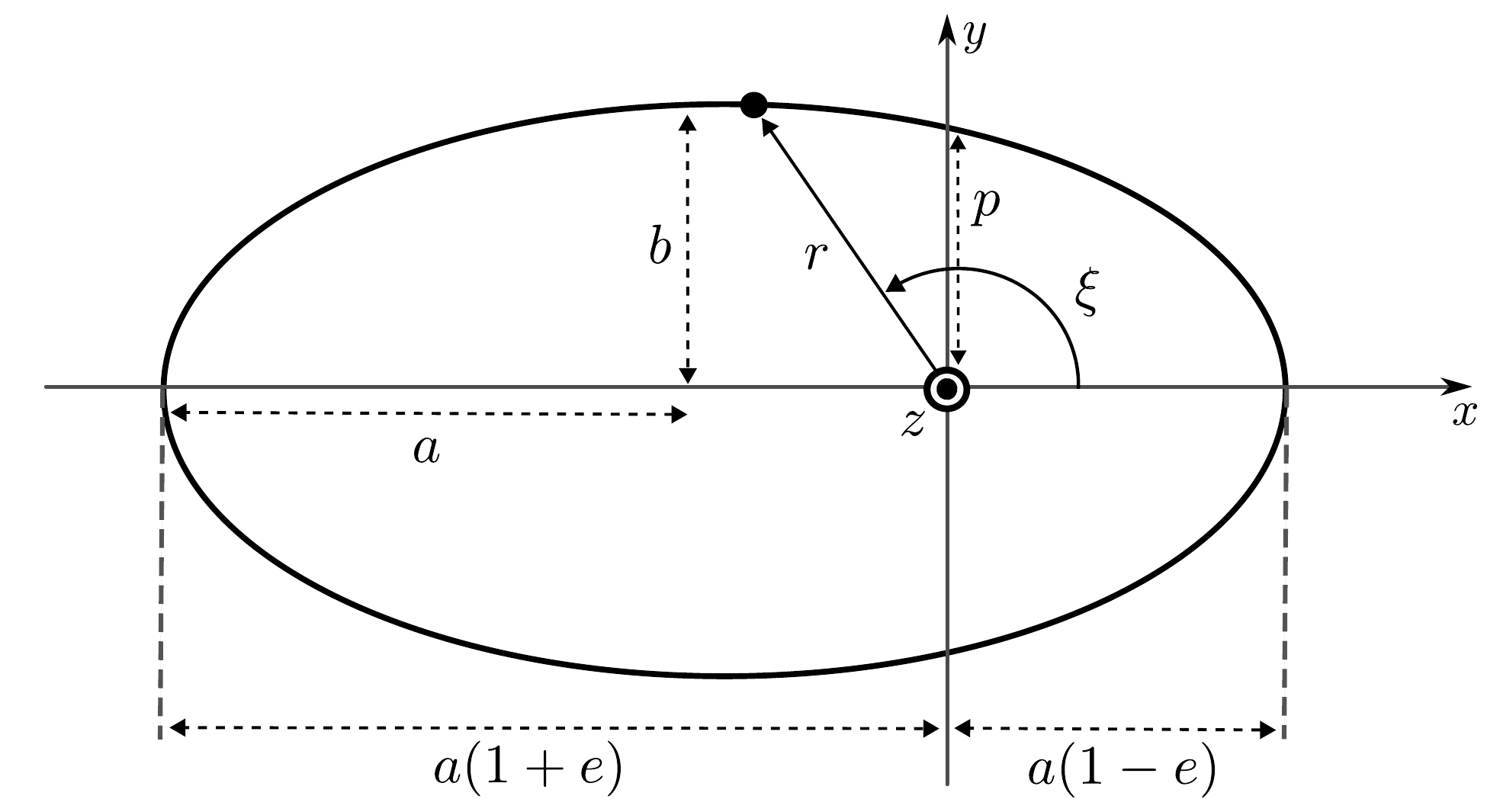}
    \caption{Definitions used for an elliptical orbit. The Cartesian coordinates $x,y$ and polar coordinates $r,\xi$, are centered on the focus of the ellipse. The angle $\xi$ is measured counterclock-wise, from the y-axis. The semiaxes $a,b$ are indicated. The focus splits the major axis into two segments of length $a(1\pm e)$, where $e$ is the orbital eccentricity.}
    \label{fig:eccOrbit}
\end{figure}

In the case of elliptical orbits the motion can be can be described in the center-of-mass frame by a one-body problem with a particle of mass $\mu = m_1 m_2/M$. For Newtonian orbits the conservation of energy, $E$, and angular momentum, $L$, confines the motion to a plane, i.e., in the notation of Sec. \ref{sec:source-frame-general} one can choose  $\bm{z}_S=0$. Then, one can introduce polar coordinates $(r,\xi)$ centered on the focus of the ellipse, see Fig. \ref{fig:eccOrbit}. The solution of the Kepler problem leads to the following parametrization of the orbit \cite{Maggiore:2007ulw,AIHPA_1986__44_3_263_0}
\begin{equation}
    r=  \frac{p}{1+e \cos \xi},
    \label{eq:rEcc}
\end{equation}
where $p$ is the semi-latus rectum and $e$ is the orbital eccentricity of the binary. Both are constant of motion and are related to the energy and angular momentum, 
\begin{equation}
    p=  \frac{L^2}{G M \mu^2}, \quad e^2= 1+ \frac{2E L^2}{G^2 M^2 \mu^3}.
    \label{eq:pEcc}
\end{equation}

For a bounded system $(E<0)$ the eccentricity of the binary satisfies $0\leq e < 1$. The two semiaxes of the ellipse can be expressed as 
\begin{equation}
    a  =  \frac{p}{1-e^2}, \quad  b  = \frac{p}{\sqrt{1-e^2}}, 
    \label{eq:semiAxes}
\end{equation}
which are consistent with the geometric definition of eccentricity $e^2 = 1+ b^2/a^2$. Using Eqs. \eqref{eq:pEcc} and \eqref{eq:semiAxes} the semi-major axis can also be written in terms of the energy as 
\begin{equation}
    a  =  \frac{G M \mu}{2 |E|}.
    \label{eq:semiMajorAxis}
\end{equation}

Note that the angle $\xi$, also called  \textit{true anomaly} \cite{Maggiore:2007ulw,AIHPA_1986__44_3_263_0} only corresponds to the orbital phase $\Phi$ at Newtonian order. At high post-Newtonian orders the true anomaly is shifted by the periastron advance as well as higher order post-Newtonian corrections \cite{Boetzel:2017zza}. The conservation of energy and orbital angular conservation imply,
\begin{eqnarray}
    E =& \frac{1}{2}\mu (\dot{r}^2+r^2\dot{\xi}^2)-\frac{G M \mu}{r},\\
    L =& \mu r^2 \dot{\xi}.
    \label{eq:EnLconservation}
\end{eqnarray}
whose solution can be can be obtained as
\begin{eqnarray}
    r = a (1- e \cos u), \\
    \cos \psi = \frac{\cos u - e }{1- e \cos u},
    \label{eq:eqsMotion}
\end{eqnarray}
where $u$ is the eccentric anomaly, and is related to time via the Kepler equation
\begin{equation}
    l\equiv n(t-t_0) = u-e \sin u ,
\end{equation}
where $l$ is the mean anomaly, $t_0$ is the initial time, and $n= \frac{2\pi}{P}$ is the mean motion defined in terms of the orbital period $P$.  The true anomaly $\psi$  and the eccentric anomaly $u$ can be related through Eq. \eqref{eq:eqsMotion} as 
\begin{equation}
    \psi = \psi_0+2 \arctan \left[ \left(\frac{1+e}{1-e}\right)^{1/2} \tan \frac{u}{2}\right],
\end{equation}
where $\psi_0$ is the value of $\psi$ at $t_0$. 

With such a parametrization of the orbit, $\bm{x}=r(\cos\xi, \sin \xi,0)$, the mass quadrupole tensor in Eq. \eqref{eq:Qdef} can be expressed as 
\begin{equation}
    Q_{ij} = M \eta r^2 \begin{pmatrix}
\cos^2 \xi & \cos \xi \sin \xi  & 0\\
 \cos \xi \sin \xi  & \sin^2 \xi  & 0\\
 0  & 0  & 0   
\end{pmatrix}.
\label{eq:Qmass}
\end{equation}
In order to compute Eq. \eqref{eq:hijTT} one needs to compute two time derivatives of the mass quadrupole tensor. This calculation can be simplified by the use of the conservation of energy and orbital angular conservation, Eqs. \eqref{eq:EnLconservation}, and Eq. \eqref{eq:rEcc}, so that the components of the mass quadrupole tensor can be written in terms of the semi-major axis, $a$, and eccentricity, $e$, which are constants at Newtonian order, and the time-dependent variable $\xi$, 
\begin{equation}
    Q_{ij} = M \eta \frac{a^2(1-e^2)^2}{(1+e \cos \xi)^2} \begin{pmatrix}
\cos^2 \xi & \cos \xi \sin \xi  & 0 \\
 \cos \xi \sin \xi  & \sin^2 \xi   & 0  \\ 
 0  & 0   & 0   
 \label{eq:Qmassphi}
\end{pmatrix}
\end{equation}
Combining Eqs. \eqref{eq:EnLconservation}, \eqref{eq:rEcc} and \eqref{eq:pEcc} the time derivatives of the angular coordinate can be expressed as 
\begin{equation}
     \dot{\xi} = \frac{\sqrt{GM}}{a^{3/2} (1-e^2)^{3/2}}(1+e \cos \xi)^2.
    \label{eq:phidot}
\end{equation}
Then, taking two time derivatives of the quadrupole mass tensor, Eq. \eqref{eq:Qmassphi}, and using Eq. \eqref{eq:phidot} one obtains, 
\begin{equation}
    \ddot{Q}_{ij} = -\frac{G M^2 \eta}{2 a (1-e^2)}
    \begin{pmatrix} 
     \ddot{Q}_{00}  &  \ddot{Q}_{01} & 0\\
 \ddot{Q}_{10} & \ddot{Q}_{11} & 0 \\
 0 & 0 & 0
 \label{eq:Qdotdotmass}
\end{pmatrix} ,\\
\end{equation}
where the components in Eq. \eqref{eq:Qdotdotmass} are given by 
\begin{eqnarray}
\ddot{Q}_{00} =&  3 e \cos(\xi)+ 4 \cos(2 \xi) + e \cos (3 \xi) \\
\ddot{Q}_{10} = \ddot{Q}_{01} =& -7 e \cos(\xi) - 4 \cos (2\xi) -e (4 e + \cos (3 \xi)) \\
\ddot{Q}_{11} =& 2 [ 4 \cos (\xi)+ e (3+ \cos (2\xi))] \sin (\xi).
 \label{eq:QdotdotmassComponents}
\end{eqnarray}
As shown in Eq. \eqref{eq:hplushcrossPols} the calculation of the polarizations requires the contraction of $h_{ij}^{\rm TT}$ with the polarization tensors $e^{+,\times}_{ij}$. In Sec. \ref{sec:source-frame-general} the radiation frame is introduced by defining the triad $(\bm{p},\bm{q},\bm{k})$ with the choice of the basis vectors in terms of a spherical basis $\bm{k}=\bm{e}^S_r$, $\bm{p}=\bm{e}^S_\theta$ and $\bm{q}=\bm{e}^S_\phi$. Using these definitions and Eq. \eqref{eq:defeplusecross}, the polarization tensors can be expressed as 
\begin{eqnarray}
    \bm{e}^{+}_{ij} = \bm{p} \otimes \bm{p} - \bm{q}\otimes \bm{q} =&
\begin{pmatrix}
 \cos ^2\theta \cos ^2\varphi-\sin ^2\varphi & \left(1+\cos ^2\theta\right) \sin \varphi \cos
   \varphi & -\sin\theta \cos\theta \cos\varphi \\
 \left(1+\cos ^2\theta\right) \sin\varphi \cos \varphi & \cos ^2\theta \sin ^2\varphi-\cos
   ^2\varphi & -\sin \theta\cos\theta \sin\varphi \\
 -\sin \theta \cos \theta  \cos \varphi  & -\sin \theta\cos \theta \sin \varphi & \sin
   ^2\theta 
\end{pmatrix}, \nonumber \\
 \label{eq:eplusCoords}\\
    \bm{e}^{\times}_{ij} = \bm{p} \otimes \bm{q} + \bm{q}\otimes \bm{p} =&
\begin{pmatrix} 
 -2 \cos \theta  \sin \varphi  \cos \varphi & \cos \theta  \cos 2 \varphi  & \sin \theta  \sin \varphi  \\
 \cos \theta  \cos 2 \varphi  & \cos \theta  \sin 2 \varphi  & -\sin \theta  \cos \varphi \\
 \sin \theta  \sin \varphi  & -\sin \theta \cos \varphi  & 0 \\ 
\end{pmatrix}.
 \label{eq:ecrossCoords}
\end{eqnarray}
Once we have the expressions for the polarizations tensors we would need to compute the geometric projections $\bm{e}^{+,\times}_{ij} \mathcal{P}_{ijab}$. However, since the polarization tensors $\bm{e}^{+,\times}_{ij}$ are already transverse to $\bm{k}$ and traceless, the projection with $\mathcal{P}_{ijab}$ is redundant and we can check that $\bm{e}^{+,\times}_{ij} \mathcal{P}_{ijab} = \bm{e}^{+,\times}_{ab}$. As a consequence the calculation of the GW polarizations reduces to 
\begin{align}
    h_{+,\times} &= \frac{1}{2} e^{+,\times}_{ij} h_{ij}^{\rm TT} = \frac{1}{2} \frac{2 G}{D_L c^4} P_{ijab} \ddot{Q}_{ab} e^{+,\times}_{ij} = \frac{G}{D_L c^4}  \ddot{Q}_{ij} e^{+,\times}_{ij}.
    \label{eq:hpcePols}
\end{align}
Inserting Eqs. \eqref{eq:Qdotdotmass}, \eqref{eq:eplusCoords} and \eqref{eq:ecrossCoords} into Eq. \eqref{eq:hpcePols} one obtains,
\begin{eqnarray}
    h_+ =& - \frac{(G M)^2 \eta}{a c^4 D_L (1-e^2)}  \left\{ 
    \frac{1}{2} \left(1 + \cos ^2 \theta \right) \cos 2 \varphi (5 e \cos\xi+e (2 e+\cos 3 \xi)+4 \cos 2   \xi )+\right. \nonumber \\
    & \left. \left( 1 + \cos ^2\theta \right) \sin \xi \sin 2 \varphi (e (\cos 2 \xi+3)+4 \cos \xi)+e
   \sin ^2\theta (e+\cos \xi )   
    \right\} ,\\
    h_\times =&  - \frac{(G M)^2 \eta}{a c^4 D_L  (1-e^2)} \cos \theta  \left\{ 
     e (-2 e \sin 2 \varphi +5 \sin (\xi -2 \varphi )+\sin (3 \xi -2 \varphi ))+4 \sin (2 (\xi   -\varphi ))
    \right\}.
    \label{eq:hphc_ecc}
\end{eqnarray}

In the quasicircular limit, $e\rightarrow 0$ the polarizations reduce to 
\begin{eqnarray}
    h_+ =&  - \frac{(G M)^2 \eta}{a c^4 D_L} 2(1+\cos^2 \theta) \cos 2(\xi-\varphi),\\
    h_\times =&   - \frac{(G M)^2 \eta}{a c^4 D_L}  
   4 \sin 2 (\xi -\varphi ).
    \label{eq:hphc_qc}
\end{eqnarray}
If we use Kepler's law to relate the semimajor axis with the orbital frequency $a^3 \omega^2=GM$, and we introduce the dimensionless parameter $x=(GM \omega/c^3)^{2/3}$, then, the polarizations in Eq. \eqref{eq:hphc_qc} can be written as 
\begin{eqnarray}
    h_+ =&  - \mathcal{A} \frac{1}{2} (1+\cos^2 \theta) \cos 2(\xi-\varphi),\\
    h_\times =&   - \mathcal{A} \cos\theta  \sin 2 (\xi -\varphi ),
    \label{eq:hphc_qc}
\end{eqnarray}
where the overall amplitude factor is given by $ \mathcal{A} = \frac{4 G M \eta}{D_L c^2}x$.

Note the presence of an overall minus sign. This sign is tied to the choice of polarization vectors $\bm{p}$, $\bm{q}$, and there are essentially two conventions in the literature. In our conventions, we use the spherical coordinates vectors as $\bm{p} = \bm{e}_\theta^S$, $\bm{q} = \bm{e}_\phi^S$. In the conventions of Ref. \cite{Blanchet:2013haa}, the choice is $\bm{p} = -\bm{e}_\phi^S$, $\bm{q} = \bm{e}_\theta^S$ ($\bm{p}$ points toward the ascending node of the orbit). This other convention amounts to a rotation of these polarization vectors by $\pi/2$, which gives an overall minus sign.

The properties of the binary system presented in this section can be specified at a particular reference time or frequency. Following the definitions of reference time, $t_{\rm{ref}}$,  and frequency, $f_{\rm{ref}}$, for massive black-hole binaries introduced in Sec. \ref{sec:reference_time}, one needs to define the $(2,2)$-mode GW frequency, $f_{22}$, and relate it to the specified frequency or time. Thus, using Eq. \eqref{eq:spherical-harmonics-decomposition} the polarizations $h_{+,\times}$ in Eq. \eqref{eq:hphc_qc} can be decomposed in terms of waveform modes, $h_{lm}$. Specifically, the $(l,m) =\{(2,\pm 2)\}$ multipoles can be expressed as \cite{Blanchet:2013haa} 
\begin{eqnarray}
    h_{22} =&  \frac{- 2 GM \eta x}{D_L c^2}\sqrt{\frac{16 \pi}{5}} e^{- 2 i \xi},\quad h_{2-2} =& h^*_{22},
    \label{eq:h22_qc}
\end{eqnarray}
where $h^*_{22}$ is the complex conjugate of the (2,2)-mode. The GW frequency can be computed as the time-derivative of the phase of the (2,2)-mode $\omega_{22} = d \phi_{22}/dt$, with the phase of the (2,2)-mode given by $\phi_{22} = \arg[h_{22}]$, 
where $\arg$ is the complex argument. In the case of a quasicircular binary the GW frequency is a monotonic function, 
while for eccentric binaries one can use an orbit-averaged frequency, $\bar{f}_{22}$, 
to map the reference frequency or time to the binary evolution 
by the relation  $\bar{f}_{22} \sim  2  \bar{f}_{\rm{orb}}$, 
where $\bar{f}_{\rm{orb}}$ is the orbit-averaged orbital frequency. 

\section{Stochastic gravitational waves}

\subsection{Energy density}

Defining the GW energy-momentum tensor requires defining perturbations of a curved background. In cosmology, this would be the Friedmann-Lemaître-Robertson-Walker (FLRW) metric. To do so, we assume that the GW wavelengths $\lambda$ are much smaller than the length scale $L_B$ over which the background metric varies. Practically, we consider averaged physical quantities over a length scale $\ell$ such that
\begin{equation}
\frac{\lambda}{2\pi} \ll \ell \ll L_B
\end{equation}
We denote the perturbation of the background metric $h_{\alpha \beta} = \delta g_{\mu \nu}$ and the associated trace-reverse metric perturbation as $\bar{h}_{\mu \nu}=h_{\mu \nu}-\frac{1}{2} \bar{g}_{\mu \nu} \bar{g}^{\alpha \beta} h_{\alpha \beta}$. In the Lorentz gauge ($\nabla^{\mu}\bar{h}_{\mu\nu} = 0$), we define the GW energy-momentum tensor as the spatial average (over lengths $\ell$) of second-order Ricci tensor (yielding the covariance derivative of $h_{\mu \nu}$ relative to the background metric~\cite{caprini_cosmological_2018}: 
\begin{equation}
T^{\mathrm{GW}}{ }_{\mu \nu}= \frac{c^2}{32 \pi G}\left\langle\nabla_\mu h_{\alpha \beta} \nabla_\nu h^{\alpha \beta}\right\rangle
\end{equation}
Using the $00$ element of the tensor, in the transverse traceless gauge ($\partial^i h_{ij} = 0$) we get the GW energy density
\begin{equation}
\rho_{\mathrm{GW}} =\frac{c^2}{32 \pi G}\left\langle\partial_t h_{ij}(\mathbf{x}, t) \partial_t h^{i j}(\mathbf{x}, t)\right\rangle
\end{equation}
where $ij$ denote the spatial indices and $t$ the physical time. If we consider the FLRW metric, we can write the energy density as a function of conformal time $\eta$ which verifies $\frac{ d\eta}{dt} = a(t)$, $a(t)$ begin the cosmological scale factor from the FLRW metric:
\begin{equation}
\label{eq:energy-density}
\rho_{\mathrm{GW}} =\frac{c^2}{32 \pi G a^{2}(\eta)}\left\langle\partial_\eta h_{ij}(\mathbf{x}, \eta) \partial_\eta h^{i j}(\mathbf{x}, \eta)\right\rangle,
\end{equation}

We can also define the normalized GW energy density per logarithmic frequency interval
\begin{equation}
\label{eq:normalized-energy-density}
    \Omega_{\mathrm{GW}}(f) = \frac{1}{\rho_c} \frac{d \rho_{\mathrm{GW}}}{d \log f},
\end{equation}
where $\rho_c(t) = 3 c^2 H^2(t) / (8 \pi G)$ is the critical energy density at physical time $t$.

\subsection{Isotropic  GW backgrounds}

SGWB can be written as decomposing the transverse traceless perturbation $h_{ij}$ on stochastic Fourier amplitudes. As in Eq.~\eqref{eq:httdecomposition}, we can decompose the metric tensor on its two polarization states. We assume the wave is a superposition of incoherent stochastic plane waves propagating in directions $\mathbf{k}$. We adopt the same convention for the (inverse) Fourier transform as the Cosmology White Paper~\cite{auclair_cosmology_2023}
\begin{equation}
\label{eq:strain-fourier-decomposition}
h_{i j}(\mathbf{x}, t)=\int_{{\mathbb{R}}^{3}} \frac{d^3 \mathbf{k}}{(2 \pi)^3} \left (h_{+}(\mathbf{k}, t) e^{+}_{i j}(\hat{\mathbf{k}}) + h_{\times}(\mathbf{k}, t) e^{\times}_{i j}(\hat{\mathbf{k}}) \right)e^{-i \mathbf{k} \cdot \mathbf{x}} 
\end{equation}
where $\hat{\mathbf{k}} = \mathbf{k}/k$ and $k = \lVert \mathbf{k} \rVert$.
We have the properties $e^{r}_{i j}(\hat{\mathbf{k}}) = e^{r}_{i j}(-\hat{\mathbf{k}})$ for all $r=+,\times$ and the orthonormal relation $e^{r}_{i j}(\hat{\mathbf{k}}) e^{r'}_{i j}(\hat{\mathbf{k}}) = 2 \delta_{rr'}$. 

We start with defining the power spectrum for statistically homogenous, isotropic, unpolarized and Gaussian backgrounds:
\begin{equation}
	\label{eq:power-spectrum-def}
    \left\langle h_r (\mathbf{k}, \eta) h_p^{\ast}(\mathbf{q}, \eta)\right\rangle=\frac{8 \pi^5}{k^3} \delta^{(3)}(\mathbf{k}-\mathbf{q}) \delta_{r p}{h_c^2}(k, \eta) \quad \forall r, p \in \{+, \times\}
\end{equation}
where $h_c$ is a real and dimensionless quantity depending on the comoving wavenumber $k$ and conformal time $\eta$. It represents a characteristic GW amplitude per logarithmic wave-number interval and per polarization state, at a conformal time $\eta$.
There is a specific convention underlying this definition. The factor $8 \pi^5$ is chosen so that if we plug Eq.~\eqref{eq:strain-fourier-decomposition} into $\left\langle h_{i j}(\mathbf{x}, \eta) h_{i j}(\mathbf{x}, \eta)\right\rangle$, we get
\begin{equation}
\label{eq:time-average}
    \left\langle h_{i j}(\mathbf{x}, \eta) h_{i j}(\mathbf{x}, \eta)\right\rangle=2 \int_0^{+\infty} \frac{d k}{k} h_c^2(k, \eta)
\end{equation}
where the factor of 2 is also chosen conventionally to account for the energy density coming from two independent polarizations. Note that sometimes the power spectrum is defined as $P_h(k, \eta) \equiv \frac{1}{2} h_c(k, \eta)^2$.

Now we can draw the relation between the energy density $\rho_{\mathrm{GW}}$ and the characteristic GW amplitude $h_c$ by using Eq.~\eqref{eq:time-average} and Eq.~\eqref{eq:energy-density}, and using the approximation $(\partial_{\eta} h_c)^{2}(k, \eta) \simeq k^2 h_c^2(k, \eta)$ valid for $k \gg \mathcal{H}$ (small scales compared to Hubble's radius):
\begin{equation}
\label{eq:energy-density-vs-hc}
\rho_{\mathrm{GW}} =\frac{c^2}{32 \pi G a^{2}(\eta)} 2 \int_0^{+\infty} k^2 h_c^2(k, \eta) \frac{d k}{k} 
\end{equation}
This can be compared to the relation
\begin{equation}
    \rho_{\mathrm{GW}} = \int_{0}^{+\infty} \frac{d \rho_{\mathrm{GW}}}{dk} dk = \int_{0}^{+\infty} \frac{d \rho_{\mathrm{GW}}}{d \log k} \frac{d \log k}{d k} dk = \int_{0}^{+\infty} \frac{d \rho_{\mathrm{GW}}}{d \log k} \frac{d k}{k}
\end{equation}
from which we can identify
\begin{equation}
\label{eq:derivative-log-energy-density}
    \frac{d \rho_{\mathrm{GW}}}{d \log k} =  \frac{c^2 k^2 h_c^2(k, \eta)}{16 \pi G a^{2}(\eta)} 
\end{equation}

Up to now, we have expressed the energy density as a function of conformal time. 
We need to write quantities at the present day. We define the characteristic GW amplitude per logarithmic wave-number interval and polarization state today as 
\begin{equation}
    h_c(f) \equiv h_c(k, \eta_0),
\end{equation}
where $\eta_0$ is today's conformal time.

By definition of the wave number $k$,
\begin{equation}
    f(\eta) = \frac{c}{2\pi} k(\eta) =  \frac{c}{2\pi} \frac{k(t)}{a(t)}
\end{equation}
Taking $t=0$ in the above equation, and labelling $f(0) = f$, $k(0) = k$ and $a(0) = a_0$, we get
\begin{equation}
    f = \frac{k c }{2 \pi a_0}
\end{equation}
Consequently, writing Eq.~\eqref{eq:time-average} at present-day gives us
\begin{equation}
\label{eq:time-average-present}
    \left\langle h_{i j}(\mathbf{x}, \eta_0) h_{i j}(\mathbf{x}, \eta_0)\right\rangle=2 \int_0^{+\infty} \frac{d f}{f} h_c^2(f) 
\end{equation}
By defining the associated one-sided, single-polarization power spectral density of the background as
\begin{equation}
    S_{h}(f) \equiv \frac{h^2_{c}(f)}{2f},
\end{equation}
we have
\begin{equation}
\label{eq:time-average-present-psd}
    \left\langle h_{i j}(\mathbf{x}, \eta_0) h_{i j}(\mathbf{x}, \eta_0)\right\rangle=4 \int_0^{+\infty} S_{h}(f) df.
\end{equation}
Here, one factor of 2 comes from the fact that we integrate over positive frequency only (one-sided), and another factor of 2 comes from the equal power distribution between the two polarizations. 

From the definition of the normalized energy density in Eq.~\eqref{eq:normalized-energy-density}, and using Eq.~\eqref{eq:derivative-log-energy-density}, we get at $t=0$
\begin{equation}
    \Omega_{\mathrm{GW}}(f) = \frac{f^2 2 \pi^2 h_c^2(f)}{3 H_0^2} 
\end{equation}
which yields 
\begin{equation}
\label{eq:normalized-energy-density-today}
    \Omega_{\mathrm{GW}}(f) = \frac{4 \pi^2 f^3 }{3 H_0^2} S_h(f)
\end{equation}



\subsection{Anisotropic GW backgrounds}

We now consider anisotropic stochastic backgrounds, and start back from the power sepctrum definition in Eq.~\eqref{eq:power-spectrum-def}, which we update as
\begin{equation}
	\label{eq:power-spectrum-anisotropic}
	\left\langle h_r (\mathbf{k}, \eta) h_p^{\ast}(\mathbf{q}, \eta)\right\rangle=\frac{8 \pi^5}{k^3} \delta^{(3)}(\mathbf{k}-\mathbf{q}) \delta_{r p}{h_c^2}(\mathbf{k}, \eta) \quad \forall r, p \in \{+, \times\}.
\end{equation}
Plugging Eq.~\eqref{eq:strain-fourier-decomposition} into $\left\langle h_{i j}(\mathbf{x}, \eta) h_{i j}(\mathbf{x}, \eta)\right\rangle$ using the above definition yields 
\begin{equation}
	\label{eq:time-average-anisotropic}
	\left\langle h_{i j}(\mathbf{x}, \eta) h_{i j}(\mathbf{x}, \eta)\right\rangle=2 \int_0^{+\infty} \frac{d k}{k} \int_{\mathbb{R}^2} \frac{d^2\hat{\mathbf{k}}}{4 \pi} h_c^2(\mathbf{k}, \eta).
\end{equation}
Again, we define the  characteristic GW amplitude per logarithmic wave-number interval and polarization state today in the anisotropic case as
\begin{equation}
	h_c(f, \hat{\mathbf{k}}) \equiv h_c(\mathbf{k}, \eta_0),
\end{equation}
which now depends on frequency and angular wave propagation direction $\hat{\mathbf{k}}$. Similarly,  the one-sided, single-polarization power spectral density of the anisotropic background is
\begin{equation}
	S_{h}(f, \hat{\mathbf{k}}) \equiv \frac{h^2_{c}(f, \hat{\mathbf{k}})}{2f}.
\end{equation}
The normalized energy density then relates to the power spectral density as
\begin{equation}
	\label{eq:normalized-energy-density-today}
	\Omega_{\mathrm{GW}}(f) = \frac{4 \pi^2 f^3 }{3 H_0^2} \int_{\mathbb{R}^2} \frac{d^2\hat{\mathbf{k}}}{4 \pi}  S_{h}(f, \hat{\mathbf{k}}).
\end{equation}

\section{\label{sec:timestamping}Time stamping}

\subsection{Initial time}

The reference for time grid initialization is chosen to be January, 1st 2035 at midnight in \gls{tcb} (see next section). It is a parameter of the \gls{pdb} and is noted LISA\_EPOCH\_TCB in the \texttt{LISA Constants} software~\cite{bayle_2022_6627346}. For example, the time grid where data series are provided and waveforms are evaluated should be given as the time elapsed since LISA\_EPOCH\_TCB.

This choice allows to handle relatively small numbers when providing time stamps in seconds. This value will be aligned to the official ESA LISA epoch when decided.

\subsection{Time frame}
\glsreset{tcb}
The standard time frame used for time stamping is  \gls{tcb}. The motivation behind this choice is to align with \gls{esa} conventions as it is more appropriate for a space-based observatory operating far from Earth’s gravitational potential. It is also consistent with references in the \gls{lisa-mrd-0}, \gls{scird} and \gls{sird}.

\begin{appendices}
	\section{\label{appendix:emris}Specific EMRI waveform conventions}
	
	In \texttt{FEW}, the gravitational waveform, expanded in spin-weighted spherical harmonics as in Eq.~\eqref{eq:spherical-harmonics-decomposition}, is given by a multi-voice decomposition in discrete Fourier harmonic modes $(m,n,k)$:
	\begin{equation}\label{eq:EMRI_wave_general}
		h_{+} - ih_{\times} = \frac{\mu}{D_{L}}\frac{G}{c^2}\sum_{\ell=2}^{+\infty} \sum_{m=-\ell}^{\ell}\sum_{n=-\infty}^{+\infty}\sum_{k=-\infty}^{+\infty}{\cal A}_{lmnk}(t){}_{-2}Y_{lm}(\iota,\varphi)\exp(-i\Phi_{mnk}(t))\,,
	\end{equation}
with Fourier coefficients that are stored as functions of orbital parameters, meaning ${\cal A}_{lmnk}(t)={\cal A}_{lmnk}(p(t),e(t),x_I(t))$. Here we follow the conventions of \texttt{FEW}v2.0.0 by factoring out the reduced mass $\mu$~\cite{Chapman-Bird:2025xtd}. The quantity 
	\begin{equation}
		\Phi_{mnk} = m\Phi_{\phi} + n\Phi_{r} + k\Phi_{\theta}
	\end{equation}
is a linear combination of the orbit's Boyer-Lindquist-time action angles $\Phi_A=(\Phi_r,\Phi_\theta,\Phi_\phi)$, discussed in Sec.~\ref{sec:EMRI_frame}. The action angles are related to the binary's three (gauge-invariant) fundamental frequencies $\Omega_A=(\Omega_r,\Omega_\theta,\Omega_\phi)$ via 
	\begin{equation}
		\Phi_{A}(t) = \int^{t}_{t_{0}} \Omega_{A}(t')dt' + \Phi_{A,0}\,.
	\end{equation}
Given the multivoice structure of the waveform, its frequencies at harmonic $(m,n,k)$ are then given by a linear combination of the three fundamental orbital frequencies,
	\begin{equation}\label{eq:FEW_geod_ang_freq_mnk}
		\omega_{mnk} = m\Omega_{\phi} + n\Omega_{r} + k\Omega_{\theta}\,.
	\end{equation}
	
Transforming between the \texttt{FEW} frame and other frames requires the secondary body's orbital position in Boyer-Lindquist coordinates at a chosen reference time, as described in Sec.~\ref{sec:EMRI_frame}. Analytical expressions for the angular frequencies $\Omega_A$ as functions of the quasi-Keplerian parameters $(p,e,x_I)$ at leading, 0PA order can be found in~\cite{schmidt2002celestial, Fujita:2009bp,Lynch:2024hco}. An efficient numerical transformation from $\Phi_{A}$ to the quasi-Keplerian phases $(\psi,\chi,\phi)$, again at 0PA order, is given in~\cite{Lynch:2024hco}. We note there are multiple conventions in the literature for the choice of origin for $\Phi_A$; Ref.~\cite{Lynch:2024hco} specifically adopts the convention that $\Phi_r=0,\pi$ correspond to periapsis ($\psi=0$) and apoapsis ($\psi=\pi$), and $\Phi_\theta=0,\pi$ correspond to maximum ($\chi=0$) and minimum $(\chi=\pi)$ inclination.

\end{appendices}

\printglossary[type=\acronymtype]


\bibliographystyle{plainone}
\bibliography{refs}

\end{document}